\documentclass[a4paper,12pt]{JHEP3}
\newcommand{\refeq}[1]{(\ref{#1})}

\usepackage{caption,graphicx,color}
\usepackage{amsmath,epsfig}
\usepackage{amssymb,amsfonts}
\usepackage{latexsym}
\usepackage[latin1]{inputenc}
\usepackage{float}
\usepackage{overpic}
\usepackage{slashed}
\usepackage{xcolor}
\let\normalcolor\relax
\usepackage{graphicx}
\usepackage{longtable}

\relax
\renewcommand{\theequation}{\arabic{section}.\arabic{equation}}

\def\be{\begin{equation}}
\def\ee{\end{equation}}

\newcommand{\ha}{{1 \over 2}}

\newcommand{\de}{\partial}
\newcommand{\bear}{\begin{eqnarray}}
\newcommand{\bea}{\begin{eqnarray}}
\newcommand{\eear}{\end{eqnarray}}
\newcommand{\eea}{\end{eqnarray}}

\newbox\pippobox

\def\II{\relax{\rm I\kern-.18em I}}

\def\e{\epsilon}

\def\m{\mu}
\def\n{\nu}

\def\g{\gamma}

\def\pa{\partial}

\def\sp{\;\;\;,\;\;\;}

\def\p{\partial}

\def\f{\varphi}

\def\hri#1#2{\href{http://arxiv.org/abs/#1}{[ArXiv:#1]#2}}
\def\hre#1#2{\href{http://arxiv.org/abs/#1/#2}{[ArXiv:#1/#2]}}

\def\hrj#1#2{\href{www.doi.org/#1}{#2}}
\def\hsp#1#2{\href{https://doi.org/#1}{#2}}

\title{The self-tuning of the cosmological constant and the holographic relaxion}

\author{Yuta Hamada$^\natural$, Elias Kiritsis$^\natural$$^\flat$, Francesco Nitti$^\natural$, Lukas T. Witkowski$^\dagger$
~\\
$^\natural$ \href{http://www.apc.univ-paris7.fr}{APC, AstroParticule et Cosmologie}, Universit\'e de Paris, CNRS/IN2P3, CEA/IRFU,
Observatoire de Paris,\\
 10, rue Alice Domon et L\'eonie Duquet, 75205 Paris
Cedex 13, France\\
~\\
$^\flat$ \href{http://hep.physics.uoc.gr}{Crete Center for Theoretical Physics}, Institute for Theoretical and Computational Physics,
Department of Physics, Voutes University Campus,\\
GR-70013, Vasilika Vouton, Heraklion, GREECE\\
~\\
$^\dagger$ \href{http://www.iap.fr/}{Institut d'Astrophysique de Paris}, GReCO, UMR 7095 du CNRS et de Sorbonne Universit\'e, 98bis boulevard Arago, 75014 Paris, France
}

\abstract{We propose a brane-world setup based on gauge/gravity duality that permits the simultaneous realisation of self-tuning of the cosmological constant and a stabilisation of the electroweak hierarchy. The Standard Model dynamics including the Higgs sector is confined to a flat 4-dimensional brane, embedded in a 5-dimensional bulk whose dynamics is governed by Einstein-dilaton-axion gravity. The inclusion of a dynamical bulk axion is new compared to previous implementations of the self-tuning mechanism.
Because of the presence of the axion,
the model generically exhibits a multitude of static solutions, with different values for the equilibrium position for the brane. Under mild assumptions regarding the dependence of brane parameters on bulk fields, a number of these solutions exhibit electroweak symmetry breaking with a  small Higgs mass as compared to the cutoff-scale of the brane theory. The realisation of self-tuning of the cosmological constant is generic and as efficient as in previous constructions without a bulk axion. Vacua with a small Higgs mass can sometimes be found, regardless of whether the brane theory depends explicitly on the bulk axion.
Because it is expected on general principles that the brane action will depend on the axion, the generation of solutions with a hierarchy is a robust feature.
}

\preprint{CCTP-2020-1\\ITCP-IPP-2020/1}

\begin{document}

\maketitle 


\section{Introduction and results}

The idea of naturalness in effective field theory emerged in the second half of the twentieth century, as one of the main guidelines for model building in the context of theories of fundamental physics. One of the main drives was the realization that the most complete theory of fundamental interactions, i.e.~the Standard Model of particle physics plus semiclassical gravity, suffers from (at least) two {\em naturalness problems,} i.e. the fact that some dimensionful parameters of the theory, which are sensitive to UV physics, are nevertheless much smaller than one would expect compared to other  mass scales in the theory.

The first is the {\em Electroweak hierarchy problem,} which does not involve (directly) gravitational physics and concerns the Higgs mass (equivalently, its vacuum expectation value). In a natural EFT, the latter should be of the order of the high-energy cut-off, i.e.~the energy scale where the theory breaks down (for example because new heavy degrees of freedom which had been integrated out at low energy have to be included).
The appearance of novel UV scales in the SM is guaranteed by the fact that the electromagnetic coupling is IR-free and the theory is not UV complete.
There are other good reasons to believe that a few such scales exist well above the Higgs mass scale of around 1 TeV. On the one hand, we have many hints for the existence  of physics beyond the Standard Model at a high energy scale (dark matter, inflation, neutrino masses if they are generated via a see-saw mechanism).
On the other hand, ultimately the need for a quantum description of  gravity at the Planck scale $M_p$ most likely needs new physics  at or below this scale.\footnote{The holographic realization of gravity via the AdS/CFT correspondence and its generalizations indicate that novel physics always appears {\em well-below} the Planck scale. The relevant scale is (most of the time) the string scale.}

The second  problem  concerns the {\em cosmological constant} (CC), a parameter  whose only effect can be felt when the theory is coupled to gravity. In a natural theory, the CC is expected to receive contributions from the vacuum energy of all quantum fields in the model, each contribution scaling like the respective mass-scale to the fourth power. However, the observed cosmological constant today (which in Einstein gravity is related to the four-dimensional curvature of the universe on very large scales) is measured to be many orders of magnitude below all known mass scale of particle physics. Unlike the electroweak hierarchy problem, we have  {\em direct} evidence for the existence of physics above the CC scale (namely, all of particle physics apart from Maxwell electromagnetism and neutrinos).

Although many attempts have been made to address one or the other of the two problems by introducing new physics, rarely both issues have been attacked at the same time, or using the same underlying mechanism.\footnote{There are a few notable exceptions, \cite{Dim}-\cite{Shafi}.} The aim of the present paper is to present a coherent framework that potentially addresses both hierarchy problems.

Recently, a class of models was put forward \cite{self-tuning} to address the cosmological constant problem in the context of holographic brane-worlds. In this framework, the Standard Model fields are confined on a 4-dimensional brane immersed in a 5-dimensional warped, non-compact bulk, similar to the original Randall-Sundrum (RS) brane-world model \cite{rs}.\footnote{See \cite{1912.12316} for recent work on theoretical obstacles to exactly localizing fields on 3-branes in 5-dimensional brane-worlds.} However,  the model proposed in \cite{self-tuning} departs from the RS model in several crucial ways:
\begin{enumerate}
\item The SM brane is not an ``end-of-space'' brane, but it is rather a defect in a geodesically complete bulk, which has an asymptotically Anti-de Sitter (AdS) region. In  the holographic language, the UV of the geometry is kept. This setup has a dual holographic interpretation in terms of a UV-complete holographic QFT dual to the bulk theory and a coupling to the SM realized on the brane, \cite{smgrav}. IT is also crucial for the success of the self-tuning mechanism.

\item The bulk theory contains one or more  scalar fields which have  a non-trivial profile in the vacuum (ground-state) solution. Their backreaction causes the bulk geometry to depart from AdS in the interior.

\item The localized brane action contains all terms allowed by the symmetries, up to second order in derivatives. In particular, it contains a localized Einstein-Hilbert term.

\end{enumerate}

As a consequence of these features,  the model displays a mechanism of {\em self-tuning} of the cosmological constant: the curvature observed on the brane is decoupled  from the vacuum energy of the brane-localized fields. In particular, for arbitrary values of the vacuum energy there generically exist solutions with a flat and stabilized brane. Moreover, thanks to the last property in the list above,\footnote{Even in the presence a brane-localized Einstein-Hilbert term it generally remains challenging to reproduce realistic gravitational interactions for a four-dimensional observer localized on a brane, as discussed in e.g.~\cite{Dubovsky:2003pn}.} a DGP-like mechanism of gravity quasi-localisation \cite{DGP} allows the four-dimensional observers on the brane to experience ordinary four-dimensional gravity in a range of scales.\footnote{Several works in the past displayed some, but not all, of the features listed above. Dilatonic brane-worlds were extensively studied, including in the context of self-tuning models \cite{st1,csaki}, but in these works the absence of a gravity-localisation mechanism and/or the presence of bulk singularities made these models untreatable. On the other hand, models like DGP or RS-DGP \cite{ktt2} without a bulk scalar are unsuitable for self-tuning.} In \cite{Ghosh:2018fbx} it was shown that the self-tuning mechanism is robust, in the sense that stabilized solutions with curved branes require a modification of the boundary conditions at the AdS boundary, and therefore belong to a different superselection sector than flat solutions.  A dynamical study of this model in the cosmological setting was initiated in \cite{Amariti:2019vfv}.

The self-tuning mechanism of  \cite{self-tuning} relies on the interplay between bulk and brane dilaton potentials. In that work,  only gravity  and the bulk dilaton where kept as dynamical fields, and the Standard Model fields where considered non-dynamical (they where ``integrated out''). In this work, we improve on that model by adding two new ingredients: the Higgs field on the brane, and the axion field in the bulk. Both are necessary to have a complete realistic model.

\paragraph{The brane Higgs sector.} In the full theory, the brane-localized Higgs field is  expected to also play an important role in the self-tuning dynamics: even at the classical level, the Higgs has a non-trivial brane-localized potential, and its vacuum expectation value enters the determination of the brane vacuum energy.\footnote{In orientifold realizations of the SM in string theory there are always two Higgs fields necessary in order to realize the symmetry breaking patters of the Standard Model, even in the absence of supersymmetry, \cite{A1,AK,ADKS}. These Higgses and the breaking are intertwined with anomalous U(1)'s that are always present, \cite{CIK}. We do not consider these subtleties in this paper.} In particular, the latter depends on whether the electroweak gauge group is in the broken or unbroken phase. Therefore, in order to find the correct self-tuning vacuum, it is necessary to minimize the potential for the dilaton and the Higgs field  at the same time.\footnote{In contrast,  the other SM fields can still be neglected for this purpose, as they do not take on a vacuum expectation value. There is an exception to this and this involves chiral symmetry breaking, but the correction for the self-tuning dynamics is negligible for our purposes.}

\paragraph{The bulk axion.} An extra bulk field other than the dilaton is universally present in holographic duals to large-$N$ gauge theories: it is the bulk axion field, dual to the gauge theory instanton operator $Tr[F\wedge F]$. This field enjoys a shift symmetry in the bulk, which  however may  be broken on the brane due to the coupling with the Standard Model.\footnote{The shift symmetry is also broken in the bulk by string theory instantons. Such a breaking is negligible at large $N$ as it is exponentially small, ${\cal O}(e^{-N})$.} The general bulk dynamics of Einstein-axion-dilaton theories (including axion backreaction) was recently discussed extensively in \cite{Hamada}. A peculiarity of the axion field is that the gauge theory coupling to which it is dual (namely the $\theta$-angle) is periodic. This implies the existence of several inequivalent bulk solutions corresponding to different branches of $\theta + 2\pi k$, which correspond to the same physical $\theta$-angle but different boundary conditions for the bulk axion.
This phenomenon is already known from gauge theory dynamics, \cite{witten1}, and has been seen in several related holographic contexts, \cite{witten,VQCD}.
Moreover it matches the analysis of the QCD chiral Lagrangian, \cite{witten1,VQCD}.

In this work,  we study the self-tuning and brane-stabilisation problem in the framework of \cite{self-tuning}, enriched by the dynamical  bulk axion and the SM Higgs field. We ask the question whether the electroweak  hierarchy problem can be resolved at the same time as the CC problem: do vacua with a small CC {\em and} a small  Higgs vacuum  expectation value (with respect to the high energy cut-off) exist, for generic model parameters? As we shall see, a positive answer relies on the existence of multiple, densely packed axion vacua, which gives rise to multiple flat extrema of the bulk-brane system, some of which lie in the region where the Higgs vev is small.

The idea of exploiting multiple axionic-like vacua (in four dimensions or in conjunction with extra dimensions) has been explored in the past, to solve either the cosmological constant problem, \cite{DV}, or the electroweak hierarchy problem like in the relaxion scenario \cite{relaxion}.
The latter had the feature, in addition to realizing the existence of vacua with small Higgs vev, of providing a dynamical mechanism (cosmological relaxation) for vacuum selection. Here we do not address this problem, which is of dynamical nature and we leave it for future work. Rather, we provide a proof of principle that a vacuum with a small Higgs mass and a self-tuned vacuum energy may generically exist in this class of models, given suitable (but non finely tuned) potentials.  For other related work where the two hierarchy problems are correlated, see   \cite{Dim}-\cite{Shafi}.

In the rest of this introductory section we summarize our setup and our main results.

\subsection{Setup and summary of results}

We consider an Einstein-axion-dilaton theory in the bulk, dual to a non-trivial holographic QFT.
Although we employ a single scalar and a single pseudoscalar (dual to an instanton density) our results generalize to the multiscalar case.

We add a codimension-$1$ brane on whose world-volume the SM fields are localized.
One of these fields is the Higgs scalar which will play a central role in our discussion.

  The bulk dynamics is described by the general two-derivative action which after field redefinitions reads,
\be
S_{bulk} = M_p^{3} \int d^{5}x \sqrt{-g} \left[R - {1\over 2}g^{ab}\de_a\f\de_b \f -\frac{1}{2}Y(\f)g^{ab}\de_a a\de_b a-V(\f)\right] + S_{GHY},
\label{IntroResult1}\ee
where $g_{ab}$ is the metric of the $5$-dimensional bulk space-time, $\f$ is the bulk dilaton field and $a$ is the bulk axion which only enters the bulk action via derivative terms. We shall consider the following ansatz for the bulk fields
\begin{align}
\label{IntroResult2} ds^2 =du^2 + e^{2 A(u)} \eta_{\mu \nu} dx^{\mu} dx^{\nu} \, , \quad \f=\f(u) \, , \quad a=a(u) \, ,
\end{align}
which is also employed in the description of holographic axionic RG flows \cite{Hamada}. The explicit Poincar\'e symmetry of the ansatz indicates that we are looking for ground state solutions.

Implementing the self-tuning mechanism of \cite{self-tuning},  we shall  seek solutions with a \emph{flat} $4$-dimensional brane embedded into a bulk described by \eqref{IntroResult1}. One way of achieving this is to embed the brane in such a way that it coincides with a constant-$u$-slice of the bulk geometry:
\begin{align}
\label{IntroResult3} \textrm{brane locus:} \quad u = u_0 \, , \quad \textrm{with} \quad \f_0 \equiv \f(u_0) \, , \ a_0 \equiv a(u_0) \, .
\end{align}
The brane has localized curvature terms on its world-volume. However, for a brane with a flat world-volume, only the cosmological constant term, denoted by $W_B$, will be non-vanishing in this sector,  as all terms involving the brane curvature vanish. In addition, the brane supports the SM fields. In this work, we wish to study the interplay between the self-tuning mechanism and the stabilisation of the electroweak hierarchy, and hence we leave the Higgs sector explicit. The rest of the SM fields are present but do not play a role in our arguments. Therefore, for our study, the relevant terms of the brane action are given by
\begin{align}
\nonumber S_{brane} = M_p^{3} \int d^4 x \sqrt{-\g}\bigg[ &- W_B(\f,a) -  Z_H(\f,a)|\de_\mu H|^2 \\
& - X_H(\f,a)|H|^2 -S_H(\f,a) |H|^4
\bigg] \, ,
\label{IntroResult4}
\end{align}
where $\gamma_{\mu \nu}$ is the induced metric on the brane, $W_B$ is the cosmological constant term (mentioned above) and $H$ is the Higgs doublet of the SM (in units of $M_p$). The quantities $X_H$ and $S_H$ correspond to the Higgs mass-squared and the Higgs quartic coupling, respectively and everything is a function of the two bulk scalars $\f,a$.
All the bulk scalars are dimensionless but the Higgs has dimensions of mass. Therefore $W_B, X_H^{-1}$ have dimensions of mass, while $Z_H,S_H$ have dimensions of (mass)$^{-3}$.

The precise functional form of the dependence on $\f, a$ is in principle calculable from a UV completion of the  model. This has been discussed in \cite{smgrav}, and this UV completion via bifundamental messenger fields determines the couplings between bulk fields to brane operators.
SM model quantum corrections then generate a localized action for the bulk fields, which in this case corresponds to $W_B$ as well as quantum corrections to the  functions $Z_H$, $X_H$ and $S_H$.

Calculating these is beyond the scope of the analysis in this work. Instead, here we shall  make educated guesses for these functions based on results from string compactifications.\footnote{An explicit dependence of at least one of the brane potentials $W_B$, $X_H$, $S_H$ on $a$ would correspond to a breaking of the axionic shift symmetry $a \rightarrow a + \textrm{const}$. While we exclude such a breaking in the bulk sector, here we permit this breaking as long as it only occurs in the brane sector of the theory. The reason is that as with the QCD axion, brane non-perturbative effects (that are not suppressed exponentially in $N$ will generically generate a potential for the axion on the brane.} Independent of the UV completion,  we can make a few observations regarding the brane potentials. In particular, if the theory on the brane has a UV cutoff given by the energy scale $\Lambda$, then we shall  expect quantum corrections due to fields on the brane to make the brane potentials UV sensitive to the UV cutoff as follows:\footnote{In the  UV completion of this model along the lines of \cite{smgrav} the scale $\Lambda$ is identified with the `messenger scale'.}
\begin{align}
\label{IntroResult5} W_B \sim {\Lambda^4\over M_p^3} \, , \qquad X_H \sim {\Lambda^2\over M_p^3} \, , \qquad Z_H,S_H \sim {\log\Lambda^2\over M_p^3}
\end{align}

The goal of this construction  is  to realise the self-tuning of the cosmological constant, while at the same time stabilising the electroweak hierarchy. In the context of this class of models, this implies the following:
\begin{enumerate}
\item \textbf{Self-tuning:} Self-tuning is realised successfully as long as our bulk-brane system exhibits a solution with a flat brane. Therefore by construction, the brane is flat, despite the presence of a non-vanishing cosmological constant $W_B$ and contributions from the Higgs sector, which can be of the order of the cutoff-scale. This is the essence of the self-tuning mechanism.
\item \textbf{Stable electroweak hierarchy:} Given a solution of the brane-bulk system,  we can calculate the corresponding Higgs mass on the brane, which (in both vacua with intact and broken electroweak symmetry) is given by
\begin{align}
\label{IntroResult6} m_h^2 \sim M^3_p |X_H(\f_0, a_0) | \, .
\end{align}
Here we define the Higgs mass to be low if it is small compared to the cutoff scale on the brane i.e.
\begin{align}
\label{IntroResult7} \frac{m_h^2}{\Lambda^2} \ll 1 \, .
\end{align}
However, note that from \eqref{IntroResult5} and \eqref{IntroResult6} it follows that for a generic self-tuning vacuum,  the condition \eqref{IntroResult7} is not automatically satisfied. A large electroweak hierarchy is only generated if in the self-tuning vacuum we also have that
\begin{align}
\label{IntroResult8} M^3_p |X_H(\f_0, a_0) |  \ll \Lambda^2 \, .
\end{align}
Not every solution will exhibit this property, and hence, in contrast to self-tuning of the cosmological constant, a large electroweak hierarchy is not a priori guaranteed. However, as we explain in more detail below, the setup described here typically exhibits a large number of vacua satisfying \eqref{IntroResult8} and hence a small Higgs mass (in addition to a large number of vacua with no significant hierarchy). The key to this is the presence of the bulk axion, and a holographic interpretation of the bulk solutions, as we now explain.
\end{enumerate}

An important property  of the type of brane-world model considered here is that the bulk geometry permits an interpretation in terms of holographic RG flow solutions, i.e.~the $5$-dimensional bulk solutions are dual to the RG flow of a particular $4$-dimensional gauge theory. The relevance of this for successful self-tuning has been thoroughly explored in \cite{self-tuning} and hence we refer readers to this work for details.

Here we focus on what is new compared to the setups considered in \cite{self-tuning} which is the existence of a non-trivial flow for a bulk axion and the Higgs dynamics. According to the standard holographic dictionary, the bulk axion $a$ is dual to the instanton density operator of the dual gauge theory. The coupling to this operator is known as the $\theta$-angle, whose RG running is then encoded in the bulk solution of the axion $a$. Part of the definition of the dual gauge theory (and thus our brane-world model) is the value of the $\theta$-angle at the UV fixed point of the RG flow, denoted by $\theta_{UV}$. This is a parameter we are free to choose and which is part of the definition of the model. In the dual geometry, a choice of $\theta_{UV}$ is equivalent to a choice of the value $a_{\star}$ of the bulk axion on the UV boundary of the geometry (here reached for $u \rightarrow - \infty$, i.e.
\be
\label{IntroResult9} a(u) \underset{u \rightarrow - \infty}{=} a_{\star} + {\cal O}(e^{4 u / \ell}) \, ,
\ee
where the ellipsis denotes subleading terms. The precise map between $a_\star$ and $\theta_{UV}$, however, is many-to-one and given by \cite{witten}:
\be
\label{IntroResult10} a_\star=c \, \frac{\theta_{UV}+2\pi k}{N_c} .
\ee
Here, $N_c$ specifies the number of colors of the dual gauge theory and $c$ is a dimensionless  constant whose value is determined by the precise implementation of the gauge-gravity correspondence. Most importantly, $k$ is an integer parametrizing different branches which exhibit the same value of $\theta_{UV}$, but different values of $a_\star$.

Thus, a model with a definite value of $\theta_{UV}$ will in fact correspond to a family of brane-worlds with different values of $a_\star$ related to $\theta_{UV}$ via \eqref{IntroResult10}. For every value of $a_\star$ we shall  obtain a different solution $a(u)$ for the axion flow, which will backreact differently on the geometry.  Solutions for different values of $a_\star$ will generically exhibit different values for the equilibrium position $u_0$ of the brane, and hence different $\f_0, a_0$. A model with a unique value $\theta_{UV}$ hence gives rise to a set of vacua (labelled by $k$), all with different values of $\f_0, a_0$. As the Higgs mass depends on $\f_0, a_0$ through $X_H$ as in \eqref{IntroResult6}, the various vacua will typically exhibit different values of the Higgs mass.

We should stress that as we work in a bottom-up setup, it is important to accommodate the constraints on axion actions as arising from string theory and further encapsulated in the form of swampland conjectures on theories with axions, \cite{ArkaniHamed:2006dz,Svrcek:2006yi,Rudelius:2015xta,Montero:2015ofa,Baume:2016psm}. They boil down to a constraint on the axion decay constant being sub-Planckian as well as having excursions in axion field space that are also sub-Planckian. However, in our setup the axion decay constant is field dependent and care is needed to assess the proper constraints. We will discuss them in detail in section \ref{constraint1}, following \cite{Hamada}.

To determine whether a particular model permits vacua with a large electroweak hierarchy one can proceed as follows. One can treat $a_\star$ as a free parameter and map the space of solutions for the equilibrium position of the brane $u_0$ as a function of $a_\star$. For every such equilibrium position one then records the values of $\f_0, a_0$, which then allows to calculate $X_H$. In this way one can extract $X_H$ as a function of $a_\star$.

A key point in the  space of solutions is the value of $a_\star$ (we denote it henceforth by $a_{\star,0}$) that leads to a vanishing effective Higgs mass, $X_H(a_{\star,0})=0$. If this happens, then we expect that around this value and in the regime in which $X_H<0$, we will have electroweak symmetry breaking with a small Higgs mass.
Then, for any value of $\theta_{UV}$, as long as there exist branches that satisfy\footnote{$c$ is an ${\cal O}(1)$ number that depends on the particular duality pair. It can be computed only in string theory dual pairs. For N=4 sYM,
$c={1\over 2\pi}$.}
\be
c \frac{\theta_{UV}+2\pi k}{N_c} \approx a_{\star,0}
\ee
these correspond to vacua with a low Higgs mass. Moreover, the steps with which the Higgs mass changes for these vacua is set by $1/N_c$. For large $N_c$ (as is assumed here) there will typically be many such solutions. It follows that a zero in $X_H(a_\star)$ is a sufficient condition for the existence of vacua with a low Higgs mass in our setup, i.e.~such vacua are guaranteed to exist if $X_H$ as a function of $a_\star$ exhibits at least one zero.

The goal of this work is then to check for the existence of such self-tuning vacua with large electroweak hierarchy, in models with several broadly generic choices for the brane potentials $W_B$, $X_H$ and $S_H$. In practice this is done by scanning the space of solutions as a function of $a_{\star}$ and identifying zeros of $X_H$ at the brane locus for specific values of $a_{\star}$ as explained above. If the backreaction of the axion flow on the geometry is sufficiently weak, this can be analysed partly analytically (sec.~\ref{small_axion}), but otherwise we turn to numerical methods (sec.~\ref{num}).

In situations where the axion backreaction on the bulk geometry is sufficiently ``small",  we can employ a probe approximation to assess how the presence of the axion affects a given self-tuning solution obtained without axion running. In this framework the modifications due to the axion can be calculated analytically and we display the analysis and the resulting formulae in sec.~\ref{small_axion}. The effect of the axion in this case is to slightly shift the brane equilibrium position leading to `small' changes in the quantities governing the physics of the brane.

For finite values of the axion data, we consider the following four distinct classes of models:
\begin{itemize}
\item \textbf{Case 1:} The brane potential $W_B$, $X_H$ and $S_H$ are functions of $\f$ only, and do not depend explicitly on $a$. See sec.~\ref{num1}.
\item \textbf{Case 2:} All brane potentials depend on $\f$ but in addition the brane potential $X_H$ (the Higgs-mass-squared) is taken to also depend linearly on $a$. See sec.~\ref{num2}.
\item \textbf{Case 3:} All brane potentials depend on $\f$ but in addition the brane cosmological constant $W_B$ has a periodic dependence on $a$. See sec.~\ref{num3}.
\item \textbf{Case 4:} All brane potentials depend on $\f$, $X_H$ also depends linearly on $a$ and $W_B$ also has a periodic dependence on $a$, i.e.~a combination of cases 2 and 3. See sec.~\ref{num4}.
\end{itemize}

\noindent Our results can be summarised as follows:

\begin{itemize}

\item In all four cases examined,  we find that the existence of a bulk axion does not destabilize or inhibit the holographic self-tuning mechanism for the cosmological constant. That is, for generic brane potentials, there typically exists at least one equilibrium position for the brane, as in the case without the axion field.

\item     In cases 2 and 4, i.e.~models where the Higgs-mass-squared parameter $X_H$ depends on the bulk axion field $a$ explicitly, we find that (for generic model parameters) $X_H$ as a function of $a_\star$ generically crosses zero and hence solutions with a small Higgs mass generically exist in these models.

\item In contrast, in cases 1 and 3, i.e.~models where the Higgs-mass-squared parameter $X_H$ does not depend on the bulk axion field $a$, this is not generically the case. Then,  a zero of $X_H$ as a function of $a_\star$ only occurs when the model parameters are chosen carefully and hence these models require a certain level of tuning to exhibit a significant electroweak hierarchy. Such a choice of parameters for case 1 is presented in sec.~\ref{num1}.

\end{itemize}

In conclusion we find that for several classes of brane data, the mechanism for the stabilisation of the electroweak hierarchy is viable and can appear in tandem with the self-tuning of the brane cosmological constant. This positive conclusion is however a first step towards obtaining a feasible and detailed model of the mechanism as we expand upon in the next subsection.

\subsection{Open questions and future work}
There are several open issues and future directions of our work:
\begin{itemize}
\item The analysis in this work constitutes a proof of principle that the self-tuning of the cosmological constant and a stable electroweak hierarchy can be achieved together in brane-world models based on axionic holographic RG flows. What we have not attempted is to propose a model that is quantitatively consistent with all current observations, e.g.~a model that reproduces the correct numerical value for the electroweak scale, which is therefore left for future work.
    One of the important constraints on such a model is that the function $W_B$, as well as the corrections to the other brane functions, should come from known SM corrections, in which the effective SM parameters are functions of the bulk fields along the lines described in \cite{zwirner,savas}.

\item  For our construction to be applicable to the existing universe, the interactions mediated by these fluctuations should reproduce four-dimensional Einstein gravity on the brane, at least over observable scales. It was observed that there exist two sets of propagating modes, with one corresponding to a spin-two mode associated to the 5d graviton realising a DGP-like scenario \cite{DGP}. Interestingly, both at short and long distances the graviton propagation is four-dimensional with the graviton exhibiting a mass. A five-dimensional phase may exist at intermediate distances if parameters are chosen accordingly. One of the main goals for future work is to construct an explcit  model in which self-tuning is effective and at the same time four-dimensional gravitational  physics is reproduced at the observed scales. 

\item An important question concerns the stability of the brane equilibrium position, and  under what circumstances fluctuations about an equilibrium position are neither tachyonic nor ghost-like.
 A key quantity controlling the gravitational coupling and the graviton mass is the induced Einstein term on the brane.
 For the holographic brane-world model of \cite{self-tuning} including a bulk dilaton, but without a bulk axion or brane Higgs field, such an analysis of fluctuations has been performed in that work.  Overall, it was observed that parametric regimes exist where all fluctuation modes are 
 non-tachyonic and not ghost-like. More specifically,  in  \cite{self-tuning} it was shown that if certain inequalities are obeyed by the various potentials evaluated at the brane, then stability is guaranteed.  It is plausible that  the key aspects of the analysis of \cite{self-tuning} may carry over to  setup including a bulk axion, whose presence will alter the aforementioned inequalities but not the general picture, as long as the bulk theory satisfies some positive energy conditions, e.g.~the NEC (which we always assume to be the case). 

The picture changes qualitatively, however, when we include the brane-localized  Higgs field. Even if we ensure that the latter has a positive kinetic term and a positive mass (in the EW-breaking vacuum) on the brane, bulk-brane mixing may result in the presence of an additional ghost or tachyon. In particular, a healthy brane Higgs can become ghost-like or tachyonic by linear mixing with the infinite tower of bulk KK modes in the scalar sector. This question is addressed in Appendix \ref{app:positivityanalysis} of this work. There, we study the Higgs-bulk mixing in a general model, and we formulate the conditions under which the effective four-dimensional Higgs field stays healthy after diagonalizing the kinetic and mass matrices. Again these constraints take the form of  inequalities, which this time involve sums or integrals over the tower of KK modes. In order to have a definite answer, one should evaluate these constraints on a specific model, which is beyond the scope of the present work. Nevertheless, we estimate the correction to the Higgs mass and kinetic term in simplified settings (two toy-models with a discrete and a continuous KK spectrum, respectively) and we find that, in each case, the effect is generically suppressed by powers of the four-dimensional Planck scale and/or the DGP-like transition length $r_c$. This  suggests that generically, in reasonable models,  the mixing of the Higgs with the KK tower does not lead to new instabilities (although a definite answer can only be provided by an explicit computation in a specific model).

\item
As discussed above, the phenomenology of our construction is highly sensitive to the dependence of the brane potentials $W_B$, $X_H$, $S_H$ on the bulk scalars $\f$ and $a$. For example, the existence of vacua with a small Higgs mass is favoured if $X_H$ depends on $a$ explicitly. Here we considered several simple functional forms for the brane potentials, but ideally this should be computed from a UV completion of our construction. More detailed knowledge regarding the functional form of $W_B$, $X_H$, $S_H$ would also help determine whether a large electroweak hierarchy is generic in our construction or whether it only occurs in certain corners of parameter space (i.e.~how much tuning is needed).

\item Even if a model exhibits a multitude of vacua with `small' Higgs mass, there typically also exist vacua where the Higgs mass is not small. For our brane-world scenario to reproduce the observed universe, we hence need to specify a dynamical mechanism that preferably populates (at late times) the vacua with small Higgs mass over those with no significant hierarchy.

Therefore, the next important question is how the vacuum realizing the light Higgs mass is selected in our world.
If the vacuum with a small Higgs mass minimizes the free energy of the total system, then the system evolves to this state after a sufficiently long time.   In the absence of the brane, it is well known that the minimum free energy occurs for minimal values of $k=0,1$, \cite{witten}. However,  the brane contributes  to the free energy and the minimization problem becomes complex, especially as it is affected by the scalar-dependent functions on the brane.

\item On the other hand, if the vacuum with a small Higgs mass does not minimize the free energy of the total system, this state could be realized as a metastable vacuum. Transitions to and from this state and rates are important in assessing the viability of this option, \cite{witten,Dubovsky:2011tu}.

\item We are therefore led to study the real time evolution of the bulk solutions as well as the brane along the lines studied in \cite{Kehagias:1999vr,Amariti:2019vfv}.
In our case, we have two effects that can happen in tandem.
The first is a semiclassical tunneling that interpolates between different $k$-bulk solutions.
Moreover, we also have the brane motion in a single bulk solution which will also be affected by the axion.

One of the relevant dynamical questions concerns the bulk motion of the brane that will generate the associated cosmology. This was studied in the absence of the axion in \cite{Kehagias:1999vr,Amariti:2019vfv}, in the probe approximation where this is solvable. What was found is that the setup corresponds to a brane moving in the radial bulk potential whose minimum (or minima) are at the places where the brane is flat and the brane cosmological constant cosmologically invisible. If the brane starts in a different bulk position it will move generating a non-trivial brane cosmology. This motion is affected, beyond the initial velocity and potential, by the presence of matter densities on the brane and  brane-bulk energy exchange, \cite{bb1,bb2}.

In the presence of a bulk axion we expect a similar behavior, but now the brane motions will also be affected by the axion.
It is important to find how the system may evolve to the metastable vacuum by studying the associated cosmology. At the same time, the lifetime of this vacuum should be long enough. An alternative possibility is to rely on anthropic arguments for the Higgs mass \cite{Agrawal:1997gf,Damour:2007uv,Donoghue:2009me,Hall:2014dfa,Meissner:2014pma}.

\item Our setup described in this paper has several similarities  to the standard relaxion scenario \cite{relaxion}.
These are discussed in our concluding section \ref{gauge_hierarchy}.

\end{itemize}

\section{The bulk theory and its dual QFT\label{bulk}}

As a bulk theory, we consider an Einstein-axion-dilaton theory in a $5$-dimensional bulk space-time, parameterized by coordinates $x^a\equiv (u,x^\mu)$ where $u$ is the holographic coordinate.
In the Einstein frame, the most general two-derivative action compatible with the axion shift symmetry is
\be
S_{bulk} = M_p^{3} \int d^{5}x \sqrt{-g} \left[R - {1\over 2}g^{ab}\de_a\f\de_b \f -\frac{1}{2}Y(\f)g^{ab}\de_a a\de_b a-V(\f)\right] + S_{GHY},
\label{A2}\ee
where $M_p$ is the bulk Planck scale, $g_{ab}$ is the bulk metric, $R$ is its associated Ricci scalar, $\f$ is the bulk scalar field, and $a$ is the bulk axion field.
$V(\f)$ is a bulk scalar potential, and $Y(\f)$ is a function controlling the axionic kinetic term.
$S_{GHY}$ is the Gibbons-Hawking-York term at the space-time boundary (e.g.~the UV boundary if the bulk is asymptotically AdS).

The  bulk field equations are given by:
\be
R_{ab} -{1\over 2} g_{ab} R = {1\over 2}\de_a\f\de_b \f +{Y\over 2}\de_a a\de_b a-  {1\over 2}g_{ab}\left( {1\over 2}(\pa \f)^2+{Y\over 2}(\pa a)^2 + V \right),
\label{FE1}\ee
\be
\de_a \left(\sqrt{-g} g^{ab}\de_b \f \right)- {\de V \over \de \f}-{Y\over 2}(\pa a)^2 =0 \sp  \de_a \left(\sqrt{-g} \,Y g^{ab}\de_b a \right)=0.
\label{FE2}\ee

We shall consider holographic RG flow geometries, which display $4$-dimensional Poincar\'e invariance and correspond therefore to vacuum states of the dual QFT.
In the domain-wall (or Fefferman-Graham) gauge,  the metric and scalar field are:
\be\label{FE7}
ds^2 = du^2  + e^{2A(u)} \eta_{\mu\nu}dx^\mu dx^\nu \, , \qquad \f = \f(u) \, , \qquad  a=a(u).
\ee
We take the coordinate $u$ to increase towards the IR region .
In this paper, we consider solutions which have an asymptotic AdS-like boundary for $u=-\infty\equiv u_{UV}$.
The bulk theory is dual  to a field theory with a UV conformal fixed point, deformed by a relevant operator dual to the dilaton and,  generically, a $\theta$-angle which is dual to the axion.
One important aspect of our analysis concerns the boundary conditions one should impose in the interior of the bulk geometry. There the metric scale factor generically vanishes at some coordinate value $u_{IR}$ (corresponding to the deep IR on the field theory side), which may be finite or infinite.

With \eqref{FE7}, the bulk equations of motion (\ref{FE1}, \ref{FE2}) become
\be
6\ddot A+\dot\f^2+Y \dot a^2=0,
\label{a1}\ee
\be
12\dot A^2-{\dot\f^2\over 2}-{Y \dot a^2\over 2}+ V =0,
\label{a2}\ee
\be
\ddot\f+4\dot A\dot \f-V'-{Y'\over 2}\dot a^2=0,
\label{a3}\ee
\be
\pa_u(Ye^{4A}\,\dot a)=0,
\label{a4}\ee
where a dot stands for a $u$-derivative while a prime stands for a $\f$-derivative.
Equation (\ref{a4}) can be solved as
\be
\dot a = {Q\over Y e^{4A}}
\label{a5}
\ee
with $Q$ an integration constant.
By substituting \eqref{a5} into (\ref{a1}, \ref{a2}, \ref{a3}), the remaining bulk equations become
\be
6\ddot A+\dot\f^2+{Q^2\over Y e^{8A}}=0,
\label{aa1}\ee
\be
12\dot A^2-{\dot\f^2\over 2}-{Q^2\over 2Y e^{8A}}+V(\f)=0,
\label{aa2}\ee
\be
\ddot\f + 4 \dot A\dot \f-V'-{Y'Q^2\over Y^2 e^{8A}}=0.
\label{aa3}\ee

\subsection{Constraints on the effective gravitational action\label{constraint1}}
The effective actions we use involve arbitrary functions of the scalar field $\f$. However, string theory puts constraints on such functions.
Especially for solutions in which scalars run towards the end of the moduli space, it is well known that they produce bulk metrics that are (mildly singular).
This is an effect that has been observed in the dimensional reduction of many well known solutions in string theory, (deformations of the AdS/CFT correspondence), \cite{pilch}.  It was observed that such singularities were artifacts of the dimensional reduction, and once the solutions were lifted to the higher dimensions, the solutions were regular.

Gubser introduced constraints on the lower-dimensional gravitational theory that imply that the higher-dimensional theories are regular, \cite{Gubser}.
Such constraints are conjectures in the general case, but they have been tested in many holographic examples, and its known that all holographic observables are finite despite the mild bulk singularities.\footnote{Our use of the word ``mild singularity" translates into  one that satisfies the Gubser criteria.}
For example, for a scalar potential that at large values of the scalar behaves as
\be
V\sim e^{b\f}
\label{add1}\ee
the Gubser criterion becomes
\be
b\leq \sqrt{2d\over d-1} \, ,
\label{add2}\ee
where $d$ is the dimension of the AdS boundary.
Eq.~(\ref{add2})  does not allow arbitrarily steep potentials in the regime of large $\f$.

The Gubser constraints were refined in \cite{iQCD,thermo,CGKKM}, where the notion of a repulsive singularity was introduced. As in general the mild singularity is resolved by the inclusion of the (missing) KK states, we have two possibilities.
\begin{enumerate}

\item The correlator in the lower dimension depends on the details of the singularity resolution.

\item The correlator in the lower dimension {\em does not} depend on the details of the singularity resolution.

\end{enumerate}

In  theories that we are in case 2 above, we call the singularity {\em repulsive}. Low energy observables in such cases can be computed reliably without resolving the singularity. The precise criteria to have a repulsive singularity were derived in \cite{iQCD,thermo,CGKKM} and for the example of a single scalar sharpen the Gubser bound in (\ref{add2}) to
\be
b\leq \sqrt{2(d+2)\over 3(d-1)} \, .
\label{add3}\ee
Therefore, theories that satisfy (\ref{add3}) have a well-defined holographic description and well-defined and finite correlation functions.
Moreover, the precise holographic dictionary, boundary conditions and holographic renormalization have been rigorously specified in \cite{Sked} for generic scalars and in \cite{Papa2} for the case of the Einstein-dilaton axion system we use in this paper. The scaling symmetries that appear when the dilaton runs to infinity were also interpreted in terms of the scaling symmetries of the higher-dimensional theories, \cite{GK}.

The addition of the axion pseudoscalar raises additional issues. It does not have a potential, but is constrained by the axion swampland conjectures, \cite{ArkaniHamed:2006dz,Svrcek:2006yi,Rudelius:2015xta,Montero:2015ofa,Baume:2016psm}.
They state that in cases where the axion kinetic term is constant, its scale $f_a^2$ must be at most as large as the Planck scale, $f\leq M_P$.
Also the field excursion of the axion must be (in Planck units as we use in this paper) smaller than one.
However, upon dimensional reductions these conditions are modified.
First, the axion kinetic term becomes field dependent, as is the case we consider here. Moreover, for large $\f$ the kinetic term coefficient, $Y(\f)$ behaves as
\be
Y(\phi)\sim e^{\gamma\f}\sp \gamma\geq \gamma_{min}\equiv {2d\over (d-1)b}-b\geq 0 \, ,
\label{add4}\ee
with $\gamma$ positive and this behavior seems to violate our constraints.

To convert our conditions on constraints on field dependent couplings we can dimensionally reduce the RR forms  and metric on a $N$-dimensional sphere to $d+1$ dimensions. In that case the scale factor of the sphere acquires a potential as in (\ref{add1}) with
\be
b=\sqrt{2(d+N-1)\over N(d-1)} \, ,
\label{add5}\ee
which is confining, \cite{GK}.
When $N=1$ one obtains a potential at the Gubser bound, (\ref{add2}), while the limit $N\to\infty$ gives a potential at the border of loosing confinement, \cite{GK,iQCD}. The repulsive constraint in (\ref{add3}) translates to $N\geq 3$.

The lower-dimensional axion acquires a kinetic function $Y(\f)$ as in (\ref{add4}) with
\be
\gamma=\sqrt{2N(d-1)\over d+N-1}={2\over b} \, ,
\label{add6}\ee
which automatically satisfies the inequality (\ref{add4}) if $d>1$.
The runaway $\f\to\infty$ solutions corresponds in $d+N$ dimensions to the volume of the N-sphere shrinking to zero, but without a curvature singularity.

Therefore, the constraint $f_a\leq M_p$ in the higher dimension translates to $b$ satisfying the Gubser bound in the lower dimension and\footnote{This is in fact a sufficient condition. In the higher dimension, the RR kinetic terms are further suppressed compared to the NS-NS kinetic terms with an extra factor of $g_s^2$. Via the holographic duality, this extra factor is matching the fact that the $\theta^2$ correction to the vacuum energy in the gauge theory is suppressed by a factor of $N_c^2$ compared to the leading term, \cite{witten,dissect}.}
\be
\gamma\leq {2\over b} \, .
\label{add7}\ee
We impose this inequality in all our models later on in section \ref{num} and in particular in equation (\ref{nfull0}).

We now proceed to the second constraint on the axion, namely that its field range in Planck units should be sub-Planckian. This implies that the variation of the axion, as it is normalized here should be smaller than one.
As shown in \cite{Hamada} the axion starts as $a=a_{UV}$ in the AdS boundary and then varies monotonically down to zero, as it evolves towards the IR end of the geometry.
This implies that the constraint translates to
\be
a_{UV}\lesssim 1\, .
\label{add8}\ee
It has been shown in \cite{Hamada} that the possible values of $a_{UV}$ that lead to a regular bulk solutions is restricted to the interval
\be
|a_{UV}|\leq a_{UV}^{max} \, ,
\label{add9}\ee
with the upper bound $a_{UV}^{max}$ being constrained by the detailed effective action and in particular, by the bulk dilaton potential $V(\f)$ and the axion kinetic term $Y(\f)$.
The bound derived analytically in \cite{Hamada} is
\be
a_{UV}^{max}\leq \int_{\f_{UV}}^{\infty}{d\f\over \sqrt{Y(\f)}} \, ,
\label{add10}\ee
where $\phi_{UV}$ is the value of the dilaton at the AdS boundary.
For potentials that in the IR behave as in  (\ref{add4}) the bound is determined essentially by the value of $\gamma$.
Moreover, in many examples analyzed numerically, we found that $a_{UV}^{max}$ is below the bound implied by (\ref{add10}), a fact that suggests that the bound in (\ref{add10}) can be improved.
In the cases we numerically analyse later on in this paper $a_{UV}^{max}<1$ and therefore in agreement with the axion swampland bounds.

Finally, the presence of $a_{UV}^{max}$,  implies that the number $n$ of distinct saddle points,  labelled by integers, is finite and of order ${\cal O}(N_c)$
\be
n=\Big\lfloor{N_c~a_{UV}^{max}\over 2\pi}\Big\rfloor \, ,
\ee
where where $\lfloor z\rfloor $ is the maximum integer smaller than or equal to the real number z.

\subsection{The first order formalism} \label{first_order}
Following \cite{Hamada}, we introduce three scalar functions of the bulk field $\f$, which we denote by $W(\f), S(\f)$ and $T(\f)$.
In terms of these scalar functions, the bulk equations of motion \eqref{aa1}--\eqref{aa3} reduce to the system of first order differential equations, as we shall show explicitly below.

The functions $W(\f), S(\f)$ and $T(\f)$ are defined as
\be
\dot A \equiv -{W(\f)\over 6},
\label{a6}\ee
\be
 \dot \f \equiv S(\f),
 \label{a7}
 \ee
 \be
 T(\f)\equiv {Q^2\over e^{8A}} .
 \label{a27} \ee
We immediately observe that $T\geq 0$ as $e^{4A}\geq 0$ by definition.

Using these definitions, it can be shown that the bulk equations of motion \eqref{aa1}--\eqref{aa3} can be written as the following set of first order differential equations in the $\f$ variable:
 \be
 S^2-W'S+{T\over Y}=0,
 \label{a8}\ee
 \be
 {T'\over T}={4\over 3}{W\over S},
 \label{a8-2} \ee
 \be
 {W^2\over 3}-{S^2\over 2}-{T\over 2Y}+V=0.
 \label{a9}\ee
The $u$-derivative of the axion field is given by
\be
\dot{a} = {\rm sign}(Q) {\sqrt{T}\over Y},
\label{a5-ii}\ee
from \eqref{a5}.
Note that,  as $Y\geq 0$ and $T\geq 0$, the sign of $Q$ determines the monotonicity properties of the axion evolution, which do not change along the flow.

The two equations in \eqref{a8}, \eqref{a8-2} are first order differential equations while the equation in \eqref{a9} is algebraic. Therefore, the solutions for $W$, $S$ and $T$ will depend on two integration constants. One of them can be taken to be  $Q$, which then enters in  the axion flow equation (\ref{a5}).  The second one will be denoted by $C_{UV}$, and can be shown to be related to the vev of the operator dual to $\f$.  Then, solving for $a$, $A$ and $\f$ by integrating \eqref{a5-ii}, \eqref{a6} and \eqref{a7} will introduce three further integration constants. However, the integration constant associated with $A$, just redefines the constant $Q$, and is hence not a physical parameter. Equivalently, it can be chosen so that the boundary metric has unit normalization, thus fixing the unit of measuring scales and other parameters such as $Q$.

We can compute asymptotic expressions for $W$, $S$, $T$ and hence $a$, $A$, $\f$ analytically both in the UV (near-boundary) and the IR region. Here we summarize the most important results. The reader can find the full analysis in \cite{Hamada}.

Consider a maximum of the scalar potential $V(\varphi)$, which we can always locate at $\f(u_{UV})=0$ by a shift in $\varphi$. As expected , a maximum of $V$ will be associated with a UV fixed point of a holographic RG flow. In the vicinity of that maximum, the bulk functions $V(\varphi)$, $Y(\varphi)$ can be expanded in a regular power series in $\varphi$,\footnote{This is the case in all known supergravity examples that are low energy limits of string theories.}
\be
V=-{12\over\ell^2}-{1\over2}{m^2\over\ell^2}\f^2 +\mathcal{O}(\f^3) \, ,
\quad
Y= Y_0 + Y_1\f + \mathcal{O}(\f^2) \, ,
\label{UV1}\ee
and we define
\be
\Delta_{\pm} \equiv 2 \pm \sqrt{4 - m^2\ell^2}.
\label{UV29}\ee
For a maximum,  $m^2>0$,  $2 < \Delta_{+}< 4$ and $0<\Delta_{-}< 2$.
The length scale $\ell$ is defined via (\ref{UV1}) as
\be
\ell^2 = -{12 \over  V(0)}\;\;\;.
 \ee
 It can be shown to correspond to the radius of the AdS space-time which the bulk space-time asymptotes to when approaching the boundary. The functions $W(\varphi)$, $S(\varphi)$ and $T(\varphi)$ can also be expanded in a series for small $\f$, but this type of series turns out to be a trans-series that contains also non-analytic powers.

The expansions for $W,S,T$ for small $\f$ can be found using similar techniques as in \cite{exotic}, \cite{curved}. The leading terms in this expansion are universal. As in the standard case of purely dilatonic flows, there are two branches for the solutions for $W,S,T$ depending on the coefficient of the leading $\f^2$ term in $W$, given either by ${\Delta_+\over 2}$ (plus-branch) or ${\Delta_-\over 2}$ (minus-branch) \cite{exotic}.
In the following, we focus on the minus branch solution, which will be relevant for our later applications.

The UV expansions for $W, S$ and $T$ on the minus-branch will contain two integration constants denoted by $C_{UV}$ and $q_{UV}$.
The first is related to the vev of the QFT operator dual to the dilaton $\f$.
The constant $q_{UV}$ determines the vev of the QFT operator dual to the axion and is related to $Q$ introduced in \eqref{a5}. The precise relation will be given later, in \eqref{aUV28-main}.
Collecting the universal terms and the leading terms containing $C_{UV}$ and $q_{UV}$ the near-UV expansions of $W$, $S$ and $T$ on the minus branch are given by:
\be
W_{-}(\f) = {6 \over \ell} + {\Delta_- \over 2\ell} \f^2 + \cdots+ {C_{UV}}|\f|^{4\over\Delta_-} + \cdots +{q_{UV} \over 8 Y_0}|\f|^{8\over\Delta_-} + \cdots ,
\label{w-}\ee
\be
S_{-}(\f) = {\Delta_- \over \ell} \f+ \cdots+ {4C_{UV} \over\Delta_- }|\f|^{{\Delta_+\over\Delta_-}} + \cdots +  {q_{UV} \, Y_1\over 2Y_0^2(4+\Delta_-)}|\f|^{8\over\Delta_-}
+ \cdots
\label{s-}\ee
\be
T_{-}(\f) = {q_{UV}\, |\f|^{8\over\Delta_-}} \left[ 1 + \cdots   -{32C_{UV} \over (\Delta_+-\Delta_-)\Delta_-^2} |\f|^{\Delta_+ - \Delta_-\over\Delta_-} + \cdots \right]
\label{t-}.\ee
Further details regrading the UV expansion can be found in section 4.1 and appendix A of \cite{Hamada}.

Next, we discuss the asymptotic behavior of the solutions to \eqref{a8}--\eqref{a9} in the deep IR region.
In \cite{Hamada}, the regularity of general axionic flows was studied.
If the flow ends at a finite end-point, $\f_{end}$, then regularity requires that $Y(\f)$ diverges at $\f_{end}$. This is not permissible in string theory, though.
Therefore, regular axionic flows exist only when $\f$ runs to the boundaries of its space,\footnote{This is the behavior in top-down holographic theories like the Witten realization of QCD once it is dimensionally reduced to 5 dimensions, \cite{witten2}. It is also the behavior in Improved Holographic QCD (IHQCD), \cite{iQCD}, a bottom-up holographic theory constructed to emulate the dynamics of YM in four dimensions. Moreover, it is also  the behavior in V-QCD, \cite{VQCD}, which emulates the dynamics of QCD in the Veneziano limit.} i.e.~as  $\f \rightarrow \pm \infty$.  One may expect that a `mild-enough' singularity in this regime can be resolved by KK or stringy states as advocated by Gubser, \cite{Gubser}.

In the following, we choose the IR to be reached for $\f\to + \infty$.
Then, motivated by top-down results from string theory, we assume that for large dilaton-values $V$ and $Y$ can be approximated by exponentials in $\f$, i.e.
\be
V\simeq -{V_\infty\over\ell^2}e^{b\f}\sp
Y\simeq Y_\infty e^{\g \f},
\label{a32}\ee
with $V_\infty, Y_\infty, b$ and $\g$ constant. The corresponding solutions for $W$, $S$, $T$ in this regime are \cite{Hamada}
\be
W=
W_\infty \,e^{{b\over2}\f}
	-{D_{IR}\over2}{ e^{-\left({b\over2}+\g-{8\over 3b}\right)\f} \over {b\over2}+\g-{4\over 3b}}
+\cdots,
\label{sub8-2}\ee
\be
S=
{b\over2}W_\infty\, e^{{b\over2}\varphi}
	-{D_{IR}\over2}{{b\over2}+\g \over {b\over2}+\g-{4\over 3b}}
	e^{-\left({b\over2}+\g-{8\over 3b}\right)\f}
	+\cdots,
\label{sub9}
\ee
\be
T=
{b\over2}D_{IR} W_\infty Y_\infty  \,e^{{8\over 3b}\f}+\cdots,
\label{sub10}\ee
with
\be
W_\infty=
\sqrt{8V_{\infty} \over {8 \over 3}-b^2}\, ,
\label{a34}\ee
and $D_{IR}$ an integration constant which is related to $q_{UV}$ as
\be
{q_{UV}\over D_{IR}}
=\lim_{\f(u_{UV})\to0,\, \f(u_{IR})\to\infty}
{{b\over2}W_\infty Y_\infty e^{{8\over 3b}\f(u_{IR})} \over |\f(u_{UV})|^{8/\Delta_-} \exp\left({4\over 3}\int^{\f(u_{IR})}_{\f(u_{UV})}d\f\,{W\over S}\right)}.
\label{sub14}\ee
If the asymptotic form of $V$ satisfies the Gubser bound, \cite{Gubser,thermo}, which here corresponds to
\be
b\leq \sqrt{8 \over 3} \, ,
\label{a41}\ee
then the solutions, although singular, are expected to have a resolvable singularity.
We also require
\be
\gamma\geq {8\over 3b}-b={8-3b^2\over 3b},
\label{sub11}\ee
for the validity of the expansion.
Note that the backreaction due to the axion flow on $W$ and $S$ enters only at subleading order in the IR.
There exists another solution in which the axion field backreacts at the leading order.
However, as discussed in \cite{Hamada}, this solution does not have a holographic interpretation, and therefore we discard it. The reader can find all the details of the IR expansion in section 4.2 and appendix B of \cite{Hamada}.

Last, we review the holographic interpretation of the various integration constants appearing in the solutions.
As was argued below equation \eqref{a5-ii}, the physical system in question is characterized by four physical integration constants. As we shall  describe below, these correspond to two pairs of the form (source, vev) for the operators dual to the dilaton and axion fields, respectively.

The axion bulk profile is characterized by the two coefficients $a_\star$ and $Q$,  which enter as the integration constants of the second order axion equation of motion and control the leading and subleading terms in the near-boundary expansion,
\be \label{intro1}
a(u) = a_\star  + Q\,{\ell \over 4 Y_0} e^{4 u /\ell} + \cdots
\qquad u \to -\infty,
\ee
In the holographic dictionary,  $a_\star$ is related to the value of the $\theta$-term in the UV field theory (modulo 2$\pi$ shifts)  and $Q$ is proportional to the vacuum expectation value of the corresponding topological density operator $O_a(x)$.
More precisely, the relation between the source $a_\star$ and the UV $\theta$-angle $\theta_{UV}$ in the dual QFT is
\be
a_\star=c \, {\theta_{UV}+2\pi k\over N_c}
\label{i2}\ee
where $N_c$ is the number of the color in the dual QFT, $\theta_{UV}\in[0,2\pi)$, $k\in \mathbb{Z}$ and $c$ a dimensionless number of ${\cal O}(N_c^0)$. In instances where the dual geometry including the compact internal manifold is known, the parameter $c$ can in principle be computed. For example, in the conventional IIB normalization of the RR axion, $c=1$.
The expectation value of the operator $O_a$, dual to the axion  is
\be
\langle O_a\rangle = Q {M_p^3 \over N_c}.
\label{UV26-2}\ee
Due to the exact axion shift symmetry of the bulk Lagrangian, of the two parameters $a_\star$ and $Q$, only $Q$ enters non-trivially in the non-linear equations for the metric and dilaton \eqref{aa1}--\eqref{aa3}.
Therefore, seemingly, $a_\star$ remains a free parameter. This, however, would go against the expectation from holography where one does not expect any additional freedom in the interior. Instead, a solution should be completely fixed by the choice of boundary sources plus some universal requirement in the IR. In \cite{Hamada} we therefore proposed such an IR condition in the form of the requirement that the axion field should vanish at the IR endpoint,
\be \label{a10}
a (u_{IR}) = 0.
\ee
This axion regularity condition \eqref{a10} leads to a relation between the axion source $a_\star$ to the axion vev $Q$, and hence no free parameter remains. The condition \eqref{a10} is motivated by top-down string theory constructions, where the axion is a form field component along an  internal cycle, which shrinks to zero-size in the IR as in \cite{witten}.
Single-valuedness then demands that the axion field vanishes at such IR end-points.
Combining \eqref{a10} with \eqref{a5-ii}, the axion source is expressed as
\be
a_\star = {\rm sign}(Q)\int^{\f(u_{UV})}_{\f(u_{IR})} d\f~{\sqrt{T}\over YS},
\label{UV8}\ee

Similarly, the dilaton bulk profile in the minus branch is characterized by two integration constants $\f_-$ and $C$, which also control the leading and subleading terms in the near-boundary expansion:
\be
\f=
\f_- \ell^{\Delta_-}e^{\Delta_- {u/\ell}}
+4C_{UV}\,{\left(|\f_-|\,\ell^{\Delta_-}\right)^{{\Delta_+\over\Delta_-}}\over(\Delta_+-\Delta_-)\Delta_-}e^{\Delta_+{u/\ell}} + \cdots
\qquad u \to -\infty,
\label{intro3}\ee
where $\f_-$ determines the UV coupling constant of the scalar operator $O(x)$ dual to $\f$,
and $C_{UV}$ is related to its vev. The vev of $O(x)$ is given by
\be \label{vev-C}
\langle O\rangle = C_{UV} \, (M_p \ell)^{3}{4\over \Delta_-}|\f_-|^{\Delta_+\over\Delta_-}.
\ee
Again, one can show that the IR regularity condition \eqref{a10} leads to a relation between $C_{UV}$ and $a_{\star}$. For completeness, we also recall that the integration constants $Q$ and $q_{UV}$ are related as
\be \label{aUV28-main}
Q^2=q_{UV} \, {1\over\ell^2} \left(\ell |\f_-|^{1/\Delta_-}\right)^{8} \, .
\ee
For details we once more refer readers to  \cite{Hamada}.

\section{The brane theory and its couplings to the bulk fields}

Given the bulk system discussed in the previous section, we now introduce a co-dimension-1 brane at a fixed value $u=u_0$ in the bulk. The world-volume of this brane is taken to model
our universe and correspondingly we assume that SM fields are localized on this brane.
At the two derivative level, the brane action is
\be
S_{brane} = S_g+S_{SM},
\label{A3}\ee
where
\begin{align}
 \nonumber S_g=M_p^{3} \int d^4 x \sqrt{-\g}\bigg[- W_B(\f,a) &- {1\over 2} Z_B(\f,a) \gamma^{\mu\nu}\de_\mu\f\de_\nu \f \\
 & -{1\over 2}Y_B(\f,a) \gamma^{\mu\nu}\de_\mu a\de_\nu a + U_B(\f,a) R_B \bigg],
\label{A4}\end{align}
and
\begin{align}
\nonumber S_{SM}=M_p^{3} \int d^4 x \sqrt{-\g}\bigg[ &-  T_H(\f,a)|\de_\mu H|^2 - X_H(\f,a)|H|^2 \\
&-S_H(\f,a) |H|^4
+ U_H(\f,a) R_B |H|^2  +\ldots
\bigg].
\label{A5}\end{align}
The ellipsis represent  omissions corresponding to higher-dimensional terms involving the Higgs field, higher curvature terms  as well as higher derivative  terms for the other SM fields.
Here, $\gamma_{\mu\nu}$ is the induced metric on the brane, which from \eqref{FE7} is given by
\be
\g_{\mu\nu} = e^{2A(u)}\eta_{\mu\nu},
\label{FE3-5}\ee
and $R_B$ is the corresponding scalar curvature.
$H$ is Higgs field of the SM (in units of $M_p$), and $W_B, Z_B, Y_B, U_B, X_H, S_H, T_H$ and $U_H$ are scalar potentials generated by quantum corrections of the brane-localized fields. In particular, $W_B(\f,a)$ is the ``cosmological constant" on the brane.
From the ansatz \eqref{FE7}, we can set the kinetic term of the Higgs field to be
\be
T_H=M_p^{-1} \, ,
\ee
without loss of generality.\footnote{
With our choice of ansatz in \eqref{FE7}, $\f$ and $a$ do not depend on $x^\mu$. Therefore, the rescaling of the Higgs field does not generate terms including $\p_\mu T_H$.
}

The brane separates the bulk into two parts, denoted by ``$UV$'' ($u<u_0$, which contains the conformal AdS boundary region or more generally, in non-asymptotically  AdS solutions,  the region where the volume form becomes infinite) and ``$IR$'' ($u>u_0$, which may contain the AdS Poincar\'e horizon, or a good singularity, or a black hole horizon etc.).

\section{Brane-bulk interactions: the Israel matching conditions}
\label{sec:junc}
We denote the bulk solutions and scalar functions in the UV and IR regions by
\be
\big(A(u), \f(u), a(u)\big) \equiv
\begin{cases}
\big(A_{UV}(u), \f_{UV}(u), a_{UV}(u)\big)
\quad \text{for $u<u_0$}
\\ \\
\big(A_{IR}(u), \f_{IR}(u), a_{IR}(u)\big)
\quad \text{for $u>u_0$}
\end{cases},
\label{FE1-2}\ee
and
\be
\big(W(\f(u)), S(\f(u)), T(\f(u))\big) \equiv
\begin{cases}
\big(W_{UV}(\f(u)), S_{UV}(\f(u)), T_{UV}(\f(u))\big)
\quad \text{for $u<u_0$}
\\ \\
\big(W_{IR}(\f(u)), S_{IR}(\f(u)), T_{IR}(\f(u))\big)
\quad \text{for $u>u_0$}
\end{cases}.
\label{FE1-3}\ee
Both sets $(W_{UV}, S_{UV}, T_{UV})$ and $(W_{IR}, S_{IR}, T_{IR})$ are solutions to the bulk equations \eqref{a8}--\eqref{a9}.
The integration constant $Q$ will in principle differ in the UV and IR regions as we have
\be
\dot{a}\equiv
\begin{cases}
\dfrac{Q_{UV}}{Y\,e^{4A_{UV}}} = {\rm sign}(Q_{UV}) \dfrac{\sqrt{T_{UV}}}{Y}
\quad \text{for $u<u_0$}
\\ \\
\dfrac{Q_{IR}}{Y\,e^{4A_{IR}}} = {\rm sign}(Q_{IR}) \dfrac{\sqrt{T_{IR}}}{Y}
\quad \text{for $u>u_0$}
\end{cases}.
\label{FE1-4}\ee
In the following it will be convenient to define the jump of a quantity $X$ across the brane by
\be
\Big[ X\Big]^{IR}_{UV}
\equiv \lim_{\epsilon\to0+}\big(X(u_0+\epsilon) - X(u_0-\epsilon) \big).
\label{FE1-5}\ee

The solutions in the UV and IR regions are then to be matched at the locus of the brane. The relevant conditions are known as Israel matching conditions and are given by the following:
\begin{enumerate}
\item Continuity of the metric and scalar fields:
\be
\Big[g_{ab}\Big]^{IR}_{UV} = 0,   \qquad \Big[\f\Big]^{IR}_{UV} =0\sp  \qquad \Big[a\Big]^{IR}_{UV} =0.
\label{FE3}\ee

For later convenience, we can define
\be
A_0\equiv A(u_0)=A_{UV}(u_0)=A_{IR}(u_0),
\label{FE3-2}\ee
\be
\f_0\equiv \f(u_0)=\f_{UV}(u_0)=\f_{IR}(u_0),
\label{FE3-3}\ee
\be
a_0\equiv a(u_0)=a_{UV}(u_0)=a_{IR}(u_0).
\label{FE3-4}\ee
\be
Y_0\equiv Y(\f_0).
\label{FE3-6}\ee
Only $\f_0$ and $a_0$ (not $u_0$) are gauge-invariant quantities.\footnote{By gauge invariant, we mean invariant under bulk diffeomorphisms.}

\item Discontinuity of the extrinsic curvature and normal derivatives of $\f$ and $a$:
\be\label{FE4}
\Big[K_{\mu\nu} - \gamma_{\mu\nu} K \Big]^{IR}_{UV} =   {1\over \sqrt{-\gamma}}{\delta S_{brane} \over \delta \gamma^{\mu\nu}}  ,
\ee
 \be
 \Big[n^a\de_a \f\Big]^{IR}_{UV} =- {1\over \sqrt{-\gamma}} {\delta S_{brane} \over \delta \f} ,
 \qquad \Big[n^a\de_a a\Big]^{IR}_{UV} =- {1\over \sqrt{-\gamma}} {\delta S_{brane} \over \delta a} ,
\label{FE4-2}\ee
where $K_{\mu\nu}$ is the extrinsic curvature of the brane, $K = \gamma^{\mu\nu}K_{\mu\nu}$ is its trace, and $n^a$ is a unit normal vector to the brane,  oriented towards the $IR$.
\end{enumerate}
Using the  form of the brane action in \eqref{A4} and \eqref{A5}, equations (\ref{FE4}, \ref{FE4-2}) are given explicitly by:
\begin{align}
\nonumber \Big[K_{\mu\nu} - \gamma_{\mu\nu} K \Big]^{IR}_{UV} = \ &
\ha \hat{W}_B(\f,a,H) \gamma_{\mu\nu} + \hat U_B (\f,a,H) G^B_{\mu\nu}-(\nabla_{\m}\nabla_{\n}-\gamma_{\m\n}\square)\hat U(\f,a,H) \\
\nonumber &-Z_B \left(\de_\mu \f \de_\nu \f - \ha \g_{\mu\nu} (\de \f)^2 \right) -Y_B \left(\de_\mu a \de_\nu a - \ha \g_{\mu\nu} (\de a)^2 \right) \\
\label{FE5} & - \Big( \left(\de_\m H \de_\nu H^*+c.c.\right) - \g_{\mu\nu} |\de H|^2 \Big)
 \Big|_{u=u_0},
\end{align}
\be
\Big[n^a\de_a \f\Big]^{IR}_{UV}  =
\left.
{\de \hat W_B \over \de \f}
- {\de \hat U_B \over \de \f} R_B
+ \ha {\de Y_B \over \de \f}(\de a)^2
- {1\over \sqrt{-\g}}\de_\mu \left( Z_B \sqrt{-\g} \g^{\mu\nu}\de_\nu \f \right)
\right|_{u=u_0},
\label{FE6}\ee
\be
\Big[n^a\de_a a\Big]^{IR}_{UV}  =
	\left.
	{\de \hat W_B \over \de a} - {\de \hat U_B \over \de a} R_B
	+ {1\over2} {\de Z_B \over \de a} (\de \f)^2
	- {1\over \sqrt{-\g}} \de_\mu \left( Y_B \sqrt{-\g} \g^{\mu\nu} \de_\nu a \right)
	\right|_{u=u_0}
\label{FE6-2}\ee
where $G^B_{\mu\nu}$ is the Einstein tensor constructed from $\g_{\mu\nu}$, and
\be
\hat W_B(\f, a, H)=W_B(\f,a)+X_H(\f,a)|H|^2+S_H(\f,a)|H|^4,
\label{FE8}\ee
\be
\hat U_B(\f, a, H)=U_B(\f,a) + U_H(\f,a)|H|^2.
\label{FE8-2}\ee

In the following we rewrite the conditions (\ref{FE5}, \ref{FE6}, \ref{FE6-2}) for the physical system at hand. For one, in this work we shall  be exclusively interested in situations with a constant Higgs field on the brane world-volume,
\be
H = {\rm const.}\quad .
\label{FE7-2}\ee
Further, for the ansatz of the bulk geometry and the chosen brane embedding, the extrinsic curvature and normal derivatives of $\f$ and $a$ can be written as:
\be \label{FE13}
K_{\mu\nu} = \dot{A} \, e^{2A} \eta_{\mu\nu}, \qquad
K_{\mu\nu} - \gamma_{\mu\nu} K = -3 \dot{A} \, e^{2A}\eta_{\mu\nu}, \qquad
n^a\de_a \f = \dot{\f}, \qquad
n^a\de_a a = \dot{a}.
\ee
Then, the matching conditions can be cast in a gauge-invariant form using the scalar functions $(W,S,T)$ on each side of the brane: making use of the expressions (\ref{a6}, \ref{a7}, \ref{a5-ii}) for $\dot{A}$, $\dot{\f}$ and $\dot{a}$, as well as (\ref{FE13}), equations  (\ref{FE5}, \ref{FE6}, \ref{FE6-2}) become conditions specifying the discontinuities in the scalar functions across the brane in terms of brane-localized terms:
\be
\Big[ W\Big]^{IR}_{UV} = \hat W_B(\f_0,a_0,H)  \label{match1}\\
\ee
\be
\Big[ S \Big]^{IR}_{UV}
={\de \hat W_B \over \de\f}(\f_0,a_0,H) \label{match2},
\ee
\be
	{\left[{\rm sign}(Q)  \sqrt{T} \right]^{IR}_{UV} \over Y_0}
	  ={\de \hat W_B \over \de a}(\f_0,a_0,H), \label{match3}
\ee
From \eqref{A5} and \eqref{FE3-5}, the equation of motion for the SM Higgs field on the brane is
\be
 \left(X_H(\f_0,a_0) + 2S_H(\f_0,a_0)|H|^2 \right) H
=0,
\label{match4}\ee
where we also used \eqref{FE7-2}.
This leads to two solutions $H_{\rm min}$ for the Higgs field:
\be
|H_{\rm min}|^2 =
\begin{cases}
\phantom{aaa}0
\qquad\qquad \text{},
\\ \\
-\dfrac{X_H}{2S_H}
\qquad \text{for $X_H<0$},
\end{cases}
\label{match4-2}\ee
where we assume positivity of $S_H$.
The physical Higgs mass squared $m_H^2$ in the two cases is given by
\be
m_H^2 =
\begin{cases}
M_p X_H,  \quad \quad \text{for $X_H \geq 0$}, \\ \\
-2M_p  X_H, \quad \quad \text{for $X_H<0$},
\end{cases}
\label{nth2}\ee
from \eqref{A5}, where we have also used $T_H=M_p^{-1}$.

When we choose $|H|^2=|H_{\rm min}|^2$, the three conditions (\ref{match1}), (\ref{match2}) and (\ref{match3}) can be written as
\be
\Big[ W \Big]^{IR}_{UV}  = \hat W^{eff}_B(\f_0, a_0)
\label{match1-2}\ee
\be
\Big[ S \Big]^{IR}_{UV}
= {\de \hat W^{eff}_B \over \de\f}(\f_0, a_0) ,
\label{match2-2}\ee
\be
{\left[{\rm sign}(Q)\sqrt{T}\right]^{IR}_{UV} \over Y_0}
	  ={\de \hat W_B^{eff} \over \de a}(\f_0,a_0),
\label{match3-2}\ee
with
\be
\hat{W}_B^{eff}(\f_0, a_0) \equiv
\hat{W}_B(\f_0, a_0,  H_{\rm min})
\label{match5}\ee

To summarize, the full system of bulk and brane field equations boils down to the bulk equations, (\ref{a8}, \ref{a8-2}, \ref{a9}, \ref{a5-ii}) and the three matching conditions  (\ref{match1-2}, \ref{match2-2}, \ref{match3-2}).
Once we impose the IR regularity conditions, the matching conditions (\ref{FE3}, \ref{match1-2}, \ref{match2-2}, \ref{match3-2}) fix the subleading (vev) boundary conditions on the UV side.

In the following, we explore to what extent the relations between the various integration constants and the bulk solutions are affected by the presence of the brane. For one, as both the integration constants $Q_{UV}$ and $q_{UV}$ are properties of the solution on the UV side alone, the relation between them is unaffected by the brane and hence still given by \eqref{aUV28-main}, which we reproduce here:
\be
Q_{UV}={\rm sign}(Q_{UV}) {\sqrt{q_{UV}}\over \ell}\, \left(\ell |\f_-|^{1\over \Delta_-}\right)^4 \, .
\label{Q18}
\ee
On the other hand in \eqref{UV8} the integration constant $a_{\star}$ was defined in terms of an integral over the whole bulk solution from IR to UV. This expression will be modified in the presence of the brane due to the condition \eqref{match3-2}, to become:
\begin{align}
\nonumber a_\star &=-Q_{UV} \int_{u_{UV}}^{u_0} {du\over Y e^{4A}}-Q_{IR} \int_{u_0}^{u_{IR}} {du\over Y e^{4A}}
 \\
\nonumber &=-\left( {\rm sign}(Q_{UV}) \int^{\f_0}_{\f(u_{UV})} + {\rm sign}(Q_{IR}) \int^{\f(u_{IR})}_{\f_0}\right) {\sqrt{T}\over Y S}d\f
\\
&=-Q_{IR} \int_{u_{UV}}^{u_{IR}} {du\over Y e^{4A}}
+{\partial \hat{W}^B\over\partial a} Y_0 e^{4 A_0} \int_{u_{UV}}^{u_0} {du\over Y e^{4 A}}. \label{Q19}
\end{align}
In the last line, the first term is unchanged compared to the case without the brane. The second term appears because of the junction condition \eqref{match3-2}.
To arrive at the above we used equation \eqref{match3-2}, \eqref{a5} and \eqref{a5-ii} to write:
\be
{ {\rm sign}(Q_{IR})\sqrt{T_{IR}} - {\rm sign}(Q_{UV})\sqrt{T_{UV}} \over Y_0}={Q_{IR}-Q_{UV}\over Y_0 \, e^{4 A_0}}={\partial \hat{W}_B^{eff} \over \partial a}(\f_0, a_0).
\label{Q7}\ee

As we will see,  the full system (\ref{a8}, \ref{a8-2}, \ref{a9}, \ref{a5-ii}, \ref{match1-2}, \ref{match2-2}, \ref{match3-2}) permits solutions for generic choices of brane potentials, up to mild assumptions stated below. The solutions can be obtained analytically in the case of small axion backreaction (section \ref{small_axion}), otherwise we resort to numerical methods (section \ref{num}).

In addition,  we require
\begin{align}
\nonumber W_{UV}(\f_0) &= W_{IR}(\f_0) - \hat{W}_B^{eff}(\f_0, a_0) >0 \, ,\\
S_{UV}(\f_0) &= S_{IR}(\f_0)- {\p \hat{W}_B^{eff}(\f_0, a_0) \over \p \f_0} >0 \, ,
\label{nfull8}\end{align}
which can be regarded as a set of mild constraints on the bulk potentials.
The first condition comes from the requirement that the scale factor $A$ asymptotes to $+\infty$ when approaching the UV boundary for $u\to u_{UV}$. The second condition comes from the requirement that the $\f$ asymptotes  to $\f_{UV}=0$ for $u\to u_{UV}$.
These conditions also apply in the case of absence of a running bulk axion, see \cite{self-tuning}.

Using the equations of the full system, we can calculate the renormalized on-shell action $S_{\text{on-shell}}^{\text{ren}}$, free energy $F$ and the topological susceptibility $\chi$.
Here we collect the results, which are given by:
\be
S_{\text{on-shell}}^{\text{ren}}=
M_p^{3} V_4 \, \ell^{3} |\f_-|^{4\over\Delta_-} (C_{UV}(q_{UV})-C_{UV, ct}),
\label{f7-main}\ee
\be
F\left(\f_-,\theta_{UV}\right)=-{\rm Min_ {k\in\mathbb{Z}}}~S_{\text{on-shell}}^{\text{ren}}
\label{f13-main}\ee
\be
\chi =
\left.- {\rm Min}_k |\f_-|^{4\over\Delta_-}{(M_p \ell)^{3}\over N_c^2}{\partial ^2 C_{UV} (a_{\star,k})\over \partial {a_{\star,k}}^2}\right|_{a_{\star,k} = {\theta_{UV}+2\pi k\over N_c}}
\label{f9-main}\ee
with $V_4$ is the $4$-dimensional space-time volume, $C_{UV, ct}$ the free parameter corresponding to the choice of the renormalization scheme, and $C_{UV}(q_{UV})$ the integration constant setting the vev of the operator dual to $\f$, which depends on $q_{UV}$ (or $a_\star$) in virtue of the IR regularity condition \eqref{a10}.
The integer $k$ labels the various oblique holographic vacua of the gauge theory.
We have introduced a minimization over $k$ to take into account the many-to-one relation between $a_\star$ and $\theta_{UV}$ \eqref{i2}.
The details of the computation are given in Appendix \ref{Free}.

\section{Solutions in the small axion backreaction approximation}\label{small_axion}

In this section we study solutions of the brane-bulk system analytically by assuming the axion backreaction to the whole system is small.
Concretely, we consider a small perturbation around the trivial axion solution  $a_\star = q_{UV} = Q_{UV} = Q_{IR}=0$, which will also be referred to as the `probe limit'.

We shall calculate the leading corrections in $q_{UV}$ to various quantities.
First, we clarify the relation among the various axion-related integration constants $q_{UV}, Q_{UV}$ and $Q_{IR}$. Recall the relation between $Q_{UV}$ and $q_{UV}$
\be
\ell Q_{UV} ={\rm sign}(Q_{UV}) \left(\ell |\f_-|^{1\over \Delta_-}\right)^4 \sqrt{q_{UV}}
\label{Q55}\ee
from \eqref{Q18}. Note that, for a given $q_{UV}$ one has the freedom to choose ${\rm sign}(Q_{UV})$.
Then, in the probe limit one can show that $Q_{IR}$ and $q_{UV}$ are related as (see Appendix \ref{small_axion_detail} for details)
\begin{align}
\nonumber \ell Q_{IR} &=
 \frac{{\rm sign}(Q_{UV}) \left(\ell |\f_-|^{1\over\Delta_-}\right)^4}{1 +  Y_0 e^{4A_0} {\partial^2 \hat{W}_B^{eff} \over\partial a^2} \int^{u_{IR}}_{u_0} {du \over Y e^{4A}}} \sqrt{q_{UV}}
 + {\cal O} (q_{UV}) \\
\label{Q52} &=\frac{\ell Q_{UV} }{1 +  Y_0 e^{4A_0} {\partial^2 \hat{W}_B^{eff} \over\partial a^2} \int^{u_{IR}}_{u_0} {du \over Y e^{4A}}}
+ {\cal O} (q_{UV}) ,
\end{align}
where the argument of ${\partial^2 \over\partial a^2} \hat{W}_B^{eff}$ is suppressed for simplicity.
From \eqref{Q52}, we observe that the sign of $Q_{IR}$ is same as that of $Q_{UV}$ as long as ${\partial^2 \over\partial a^2} \hat{W}_B^{eff} \geq0$.

Expanding in powers of $q_{UV}$, the expansion coefficients are defined as follows:
\be
W=W^{(q0)}+ q_{UV} W^{(q1)} + {\cal O}(q_{UV}^2),
\label{Q1}\ee
\be
S=S^{(q0)} + q_{UV} S^{(q1)} + {\cal O}(q_{UV}^2),
\label{Q2}\ee
\be
T= q_{UV} T^{(q1)} + {\cal O}(q_{UV}^2).
\label{Q3}\ee
\be
\f_0=\f_0^{(q0)} + q_{UV} \f_0^{(q1)} + {\cal O}(q_{UV}^2),
\label{Q4}\ee
\be
C_{UV}=C_{UV}^{(q0)} + q_{UV} C_{UV}^{(q1)} + {\cal O}(q_{UV}^2).
\label{Q23}\ee
The leading axion backreaction effects (the quantities with the superscript $(q1)$) can be expressed in terms of the unperturbed quantities (with the superscript $(q0)$).
The calculation is straightforward, but lengthy. Therefore, the computation is relegated to Appendix \ref{small_axion_detail}, and here we focus on the result and its consequences for the sign of the Higgs mass parameter $X_H$.
The readers can find the expressions for $W^{(q1)}, S^{(q1)}, T^{(q1)}, \f_0^{(q1)}$ and $C_{UV}^{(q1)}$ in \eqref{Q37}, \eqref{Q38}, \eqref{Q39}, \eqref{Q51} and \eqref{Q49}, respectively.\footnote{Apart from the probe limit discussed here, there are other configurations of the brane-bulk system that are amenable to an analytical study. For example, this is the case when the brane is located in the asymptotically UV or IR region of the bulk and the corresponding analysis is recorded in Appendix \ref{brane_UV_IR}.}

Using the above coefficients of the small-$q_{UV}$-expansion, the axion field values at the brane position $a_0$ and axion source $a_\star$ can be expressed as
\be
a_0= -\sqrt{q_{UV}}\,{\rm sign}(Q_{IR})\int^{\f_{IR}}_{\f_0^{(q0)}} {\sqrt{T^{(q1)}}\over S^{(q0)} Y}d\f
+ {\cal O}(q_{UV})
\equiv g_0 \sqrt{q_{UV}} + {\cal O}(q_{UV}) ,
\label{Q32}\ee
\be
a_\star=
\left(
- {\rm sign}(Q_{UV})\int^{\f_0^{(q0)}}_{\f(u_{UV})} d\f{\sqrt{T^{(q1)}}  \over S^{(q0)} \,Y }
- {\rm sign}(Q_{IR}) \int^{\infty}_{\f_0^{(q0)}} d\f{\sqrt{T^{(q1)}} \over S^{(q0)} \,Y }
 \right) {\sqrt{q_{UV}}}
+ {\cal O}(q_{UV})
\label{Q50-1}\ee
$$
\equiv g_\star \sqrt{q_{UV}} + {\cal O}(q_{UV}) .
$$

The $q_{UV}$-expansion of the Higgs mass parameter $X_H$ reads
\begin{align}
X_{H} &= \left. X_H \right|_{a=0} + \left({\p^2 X \over \p a^2} g_0^2 + {\p X_H \over \p \f}\f_0^{(q1)}\right) q_{UV} + {\cal O}(q_{UV}^2)
\label{N6} \\
\nonumber &=
\left. X_H \right|_{a=0} + \left({\p^2 X_H \over \p a^2} {g_0^2 \over g_\star^2}+ {\p X_H \over \p \f} {\f_0^{(q1)}\over g_\star^2}\right)
\left({\theta_{UV}+2\pi k \over N_c}\right)^2
+{\cal O}\left( \left({\theta_{UV}+2\pi k \over N_c}\right)^4 \right),
\end{align}
where we used \eqref{i2} (with $c$, defined in \eqref{i2} to be equal to $1$)\footnote{If $c$ is not $1$, then $a_\star^{\textrm{max}}$ should be replaced by $a_\star^{\textrm{max}}/c$} and \eqref{Q50-1}. In the second line, we assumed CP invariance, i.e.
\be
{\p X\over \p a} (\f^{(q0)}_0,0) =0 \, .
\label{N8}\ee
Also, functions on the right hand side of \eqref{N6} are evaluated at $\f_0=\f_0^{(q0)}, \, a_0=0$.

As one can observe from \eqref{N6}, the different solution labelled by $k=0,1,\cdots,n$ differ in the value of the Higgs mass.
From \eqref{i2} we obtain the number of distinct solutions $n$ as
\be
n=\left\lfloor{N_c a_\star^{\textrm{max}}\over 2\pi} - \theta_{UV} \right\rfloor + 1,
\label{nfull10}\ee
where we take $c=1$ and $\lfloor z \rfloor$ is the maximum integer smaller than or equal to the real number $z$. Here, $a_\star^{\textrm{max}}$ is the maximum value of the axion source observed in \cite{Hamada} (see also section \ref{num}).
For large $N_c$, $n$ is a large number, which compares well to a similar number emerging from the chiral Lagrangian, see section 5 of \cite{VQCD}. This multitude of saddle points opens up the following phenomenologically interesting situation.
Suppose that we find the branch $k=k_1$ which realizes $\left. X_H\right|_{k=k_1}\approx0$ in the region where the small axion backreaction is valid, i.e. $(\theta_{UV}+2\pi k_1)/N_c\ll1$. If this is the case, around this saddle point, we can find many other branches realizing $X_H<0$ with Higgs mass which is smaller than any other scales characterizing the brane-bulk system.
The existence of a saddle point with a small Higgs vev therefore arises as a consequence of the multiplicity (and density) of axionic saddle points, similar to the case of the `relaxion' proposal for solving the EW hierarchy problem \cite{relaxion}.

 In this section, we argued that a small electroweak scale can be realized assuming that the small axion backreaction approximation is applicable. In the next section, we will go beyond the probe limit, and solve the full system numerically.  We shall show that  solutions with a small Higgs vev persist beyond the probe limit.

\section{General numerical solutions}\label{num}
In this section, we explore numerical solutions of the brane-bulk system.
Throughout this section, we work with the following choice for the bulk functions $V$ and $Y$:
\be
V=
-{1\over\ell^2}
\left[12+\left(\frac{(4-\Delta_-)\Delta_-}{2} -b^2V_\infty\right)\f^2+4V_\infty \sinh^2\left(b\f\over2\right)\right],
\quad
Y=Y_\infty  e^{\gamma\f}.
\label{Num1}\ee
The bulk potential $V$ has an AdS maximum at the origin $\f=0$, and does not have any other extrema. Therefore, the solution in the bulk extends reaches the boundary of field space, $\f \to \pm\infty$. For definiteness, we consider solutions in which $\f >0$.

For large dilaton values the potential asymptotes to
\be
V \overset{\f \rightarrow +\infty}{\longrightarrow}
-{1\over \ell^2}V_\infty e^{b\f}+\mathcal{O}(\f^2).
\label{Num2}\ee
This choice of bulk functions is the same as in our previous work \cite{Hamada} where axionic RG flow solutions without a brane were studied, and is motivated from top-down string-generated supergravity examples.

In the following, in all numerical examples the parameters in \eqref{Num1} are chosen as
\be
\Delta_-=1.2,
\quad b=1.3, \quad \gamma=1.5, \quad V_\infty=1, \quad Y_\infty =1.
\label{nfull0}\ee
Without loss of generality we also set
\be
\ell=1.
\label{Num0}\ee
The set of parameter values is consistent with the bound on $\g$ in \eqref{sub11} and the Gubser bound on $b$ in \eqref{a41}. The condition \eqref{sub11} reads
\be
\gamma\geq {8-3b^2\over3b}=\frac{293}{390}\simeq0.75 \, ,
\label{Num10}\ee
while the Gubser bound \eqref{a41} is given by
\be
b\leq \sqrt{8\over 3} =2\sqrt{2/3}\simeq1.63.
\label{Num11}\ee
These are satisfied by the choice \eqref{nfull0}.

To set up the numerical study, we also need to specify the brane potentials $W_B$, $X_H$ and $S_H$.\footnote{Recall that we are considering a flat brane and hence $R_B=0$. As a result, the terms multiplying the brane potentials $U_B$ and $U_H$ in \eqref{A4}, \eqref{A5} are absent and we can refrain from specifying $U_B$ and $U_H$.} In the following, we shall  consider four different choices for the brane potentials, which will be discussed in sections \ref{num1}, \ref{num2}, \ref{num3} and \ref{num4}. Note that, from \eqref{A5}, the mass dimensions of $W_B, X_H$ and $S_H$ are $1, 2$ and $3$, respectively.

To solve numerically, we impose boundary conditions on the IR end of the flow and then evolve the solutions towards the UV. As the IR is only reached for $\f \rightarrow \infty$, in practice boundary conditions are implemented at a finite, but sufficiently large value of $\f$, where the bulk potential is well-approximated by the leading exponential in \eqref{Num2}. The appropriate boundary conditions for $W_{IR}$, $S_{IR}$ and $T_{IR}$ are then given by \eqref{sub8-2}, \eqref{sub9}, \eqref{sub10}.

Then, we evolve the expressions for $W_{IR}$, $S_{IR}$ and $T_{IR}$ from the IR towards the UV until we encounter the locus $\f=\f_0$ where the brane is located. This can be found using only the IR solutions $W_{IR}$, $S_{IR}$ and $T_{IR}$ as well as the brane potentials and the value of $\f$ for which the following condition is satisfied:
\be
{1\over3}\left(W_{IR} - \hat{W}_B^{eff}\right)^2
-{1\over2}\left( S_{IR} - {\p \hat{W}_B^{eff}\over \p \f} \right)^2-
\label{nfull5}\ee
$$
\left.
-{1\over2 Y}\left({\rm sign}(Q_{IR}){\sqrt{T_{IR}}} - Y{\p \hat{W}_B^{eff}\over \p a}\right)^2
+V
\right|_{\f=\f_0}=0,
$$
This is just the equation of motion \eqref{a9} for the scalar functions on the UV side, where we substituted for $W_{UV}$, $S_{UV}$, $T_{UV}$ using the matching conditions \eqref{match1-2}, \eqref{match2-2}, \eqref{match3-2}.
Since the scalar functions $W_{IR}$, $S_{IR}$, $T_{IR}$ also satisfy the equation of motion \eqref{a9}, the condition \eqref{nfull5} can be rewritten as
\be
-{2\over3}\left(W_{IR} \hat{W}_B^{eff} - {1\over2} \left(\hat{W}_B^{eff}\right)^2\right)
+\left(  S_{IR} {\p \hat{W}_B^{eff}\over \p \f} - {1\over2} \left({\p \hat{W}_B^{eff}\over \p \f}\right)^2\right)+
\label{nfull6}\ee
$$
\left.+\left({\rm sign}(Q_{IR}){\sqrt{T_{IR}}\over Y} {\p \hat{W}_B^{eff}\over \p a} - {Y\over2} \left( {\p \hat{W}_B^{eff}\over \p a}\right)^2\right)\right|_{\f=\f_0}
=0.
$$
This equation has generically multiple solutions corresponding to multiple possible positions for the brane.

For a given solution, we then calculate the Higgs mass numerically. One priority of this analysis will be to determine whether solutions with a low Higgs mass arise in the fully backreacted setup considered here. The analysis will be performed for the following four choices of brane potentials:
\begin{itemize}
\item The brane potentials only depend on the bulk scalar $\f$, but not on the axion $a$ (Section \ref{num1}). The explicit form of the brane potentials is given in \eqref{nfull1}.

\item The Higgs mass function $X_H$ depends linearly on the axion $a$, while the other brane functions do not depend on $a$. This is motivated by the original relaxion scenario \cite{relaxion}, and by the stringy constructions of (rel)axion monodromy which exhibit an axion-Higgs coupling \cite{Ibanez:2015fcv} (Section \ref{num2}). The explicit form of the brane potentials is given in \eqref{nth1}.

\item The brane cosmological constant $W_B$ depends on $\cos(a)$, which is motivated by the standard instanton-generated potential in the dilute gas approximation (Section \ref{num3}). The other brane functions do not depend on $a$. The explicit form of the brane potentials is given in \eqref{nf1}.

\item The Higgs mass function depends linearly on $a$, and the brane cosmological constant depends on $\cos(a)$ (Section \ref{num4}). The Higgs self coupling $S_H$ does not depend on $a$.
This is a combination of the ans\"atze in \ref{num2} and \ref{num3}.
The explicit form of the brane potentials is given in \eqref{nfive1}.
\end{itemize}

 \begin{figure}[t]
 \begin{center}
  \includegraphics[width=.45\textwidth]{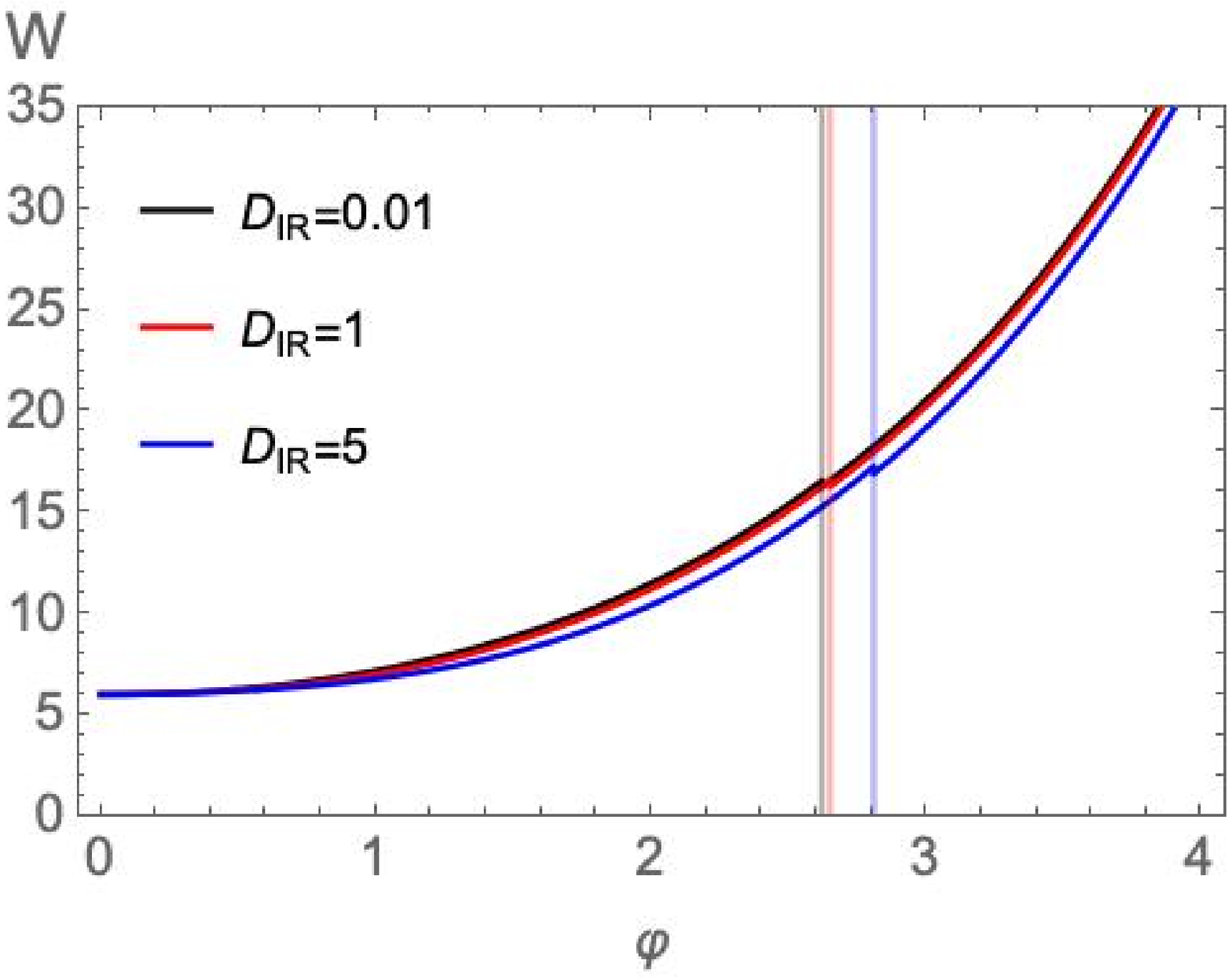}
   \includegraphics[width=.45\textwidth]{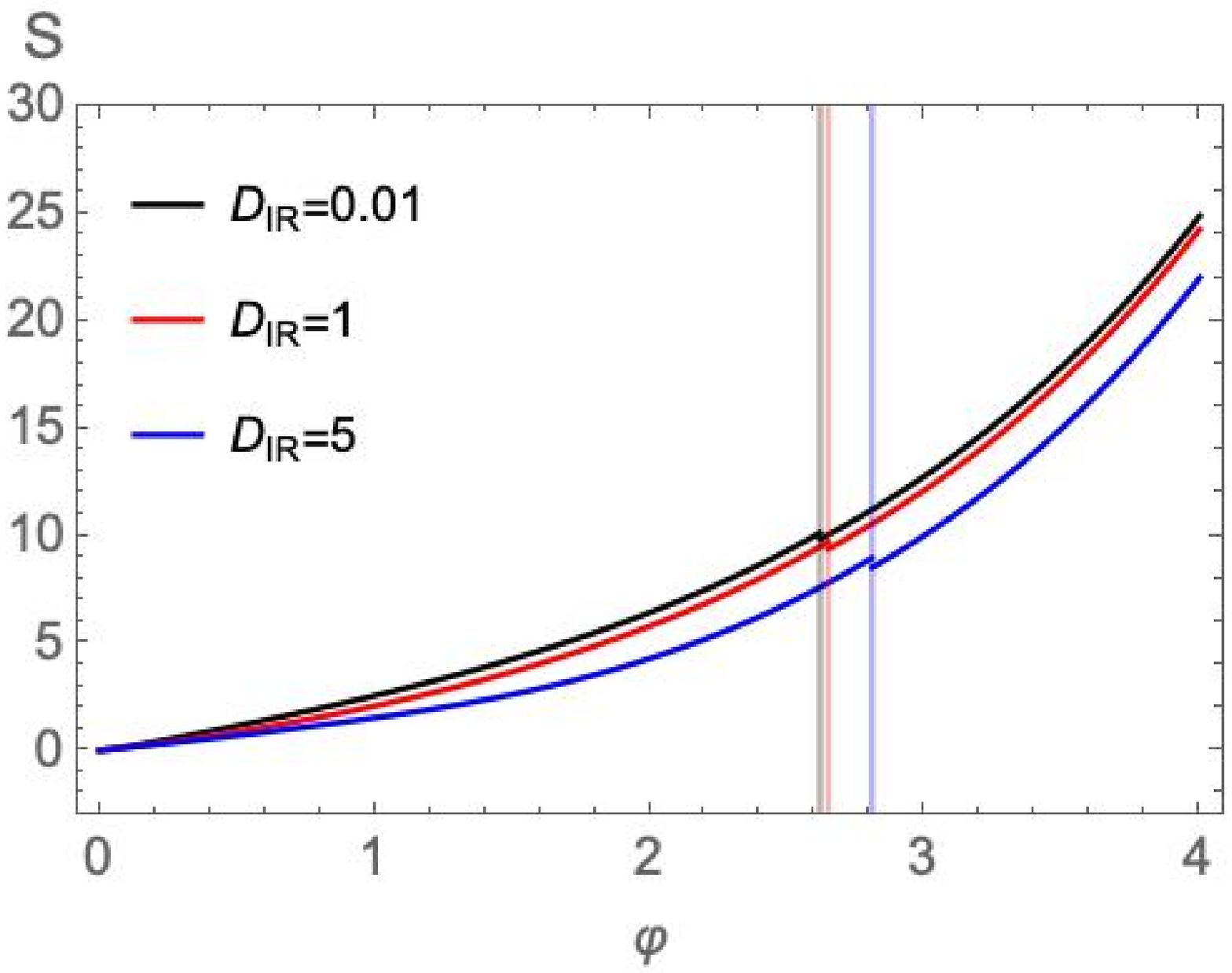}
 \end{center}
 \begin{center}
 \includegraphics[width=.5\textwidth]{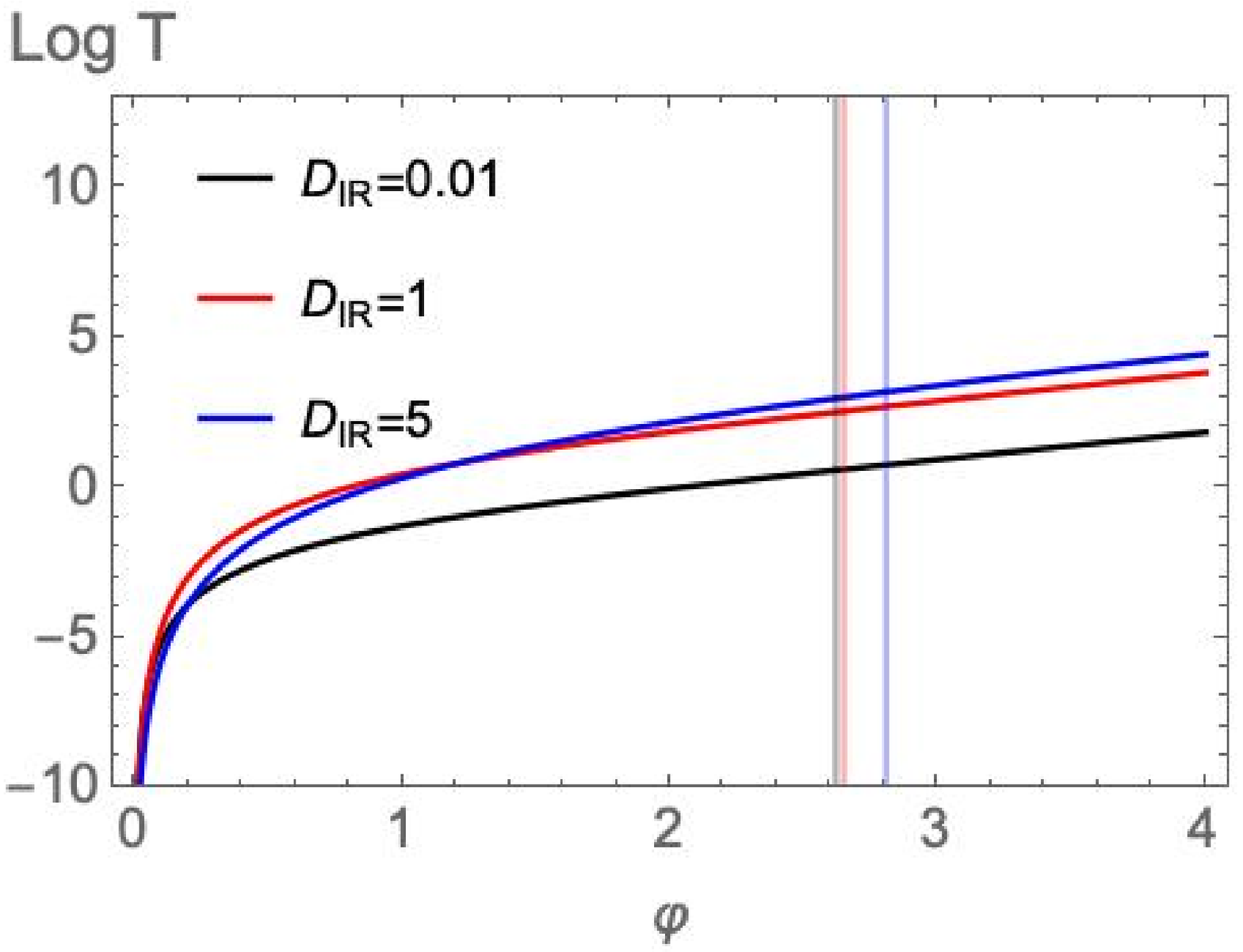}
  \end{center}
   \caption{Plots of $W$, $S$, $T$ vs.~$\f$ for a model with bulk functions \protect\eqref{Num1}, brane functions \protect\eqref{nfull1},  model parameters \protect\eqref{nfull0} with \protect\eqref{nfull9}, and $D_{IR}=0.01, 1, 5$. The vertical line represents the brane position, which is determined by solving \protect\eqref{nfull7} and \protect\eqref{nfull8}.
   \textbf{Top row, left:} Plot of $W(\f)$.  \textbf{Top row, right:} Plot of $S(\f)$.
\textbf{Bottom row:} Plot of $\log T(\f)$.
}
  \label{fig18}
 \end{figure}

 \begin{figure}[t]
 \begin{center}
  \includegraphics[width=.44\textwidth]{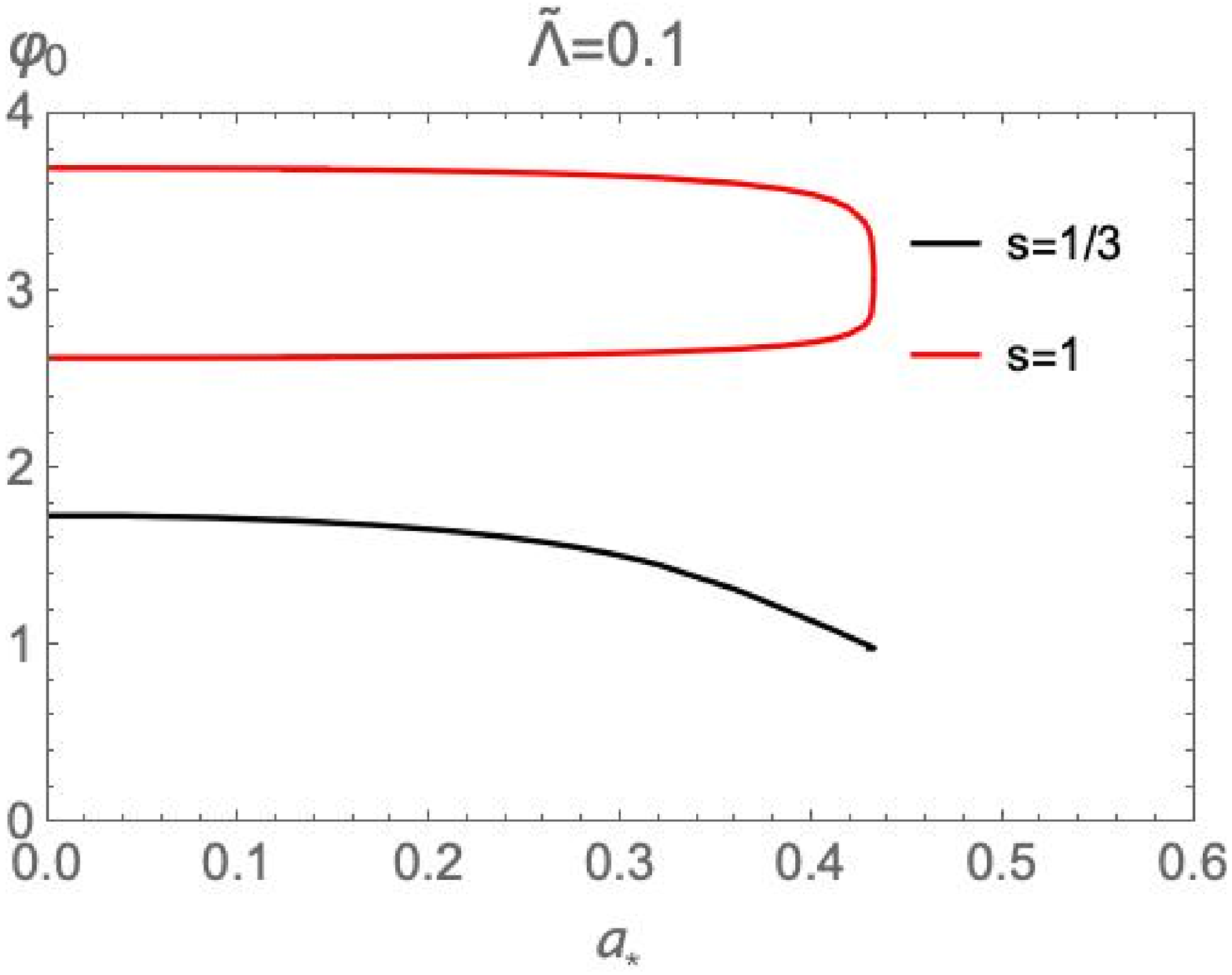}
   \includegraphics[width=.5\textwidth]{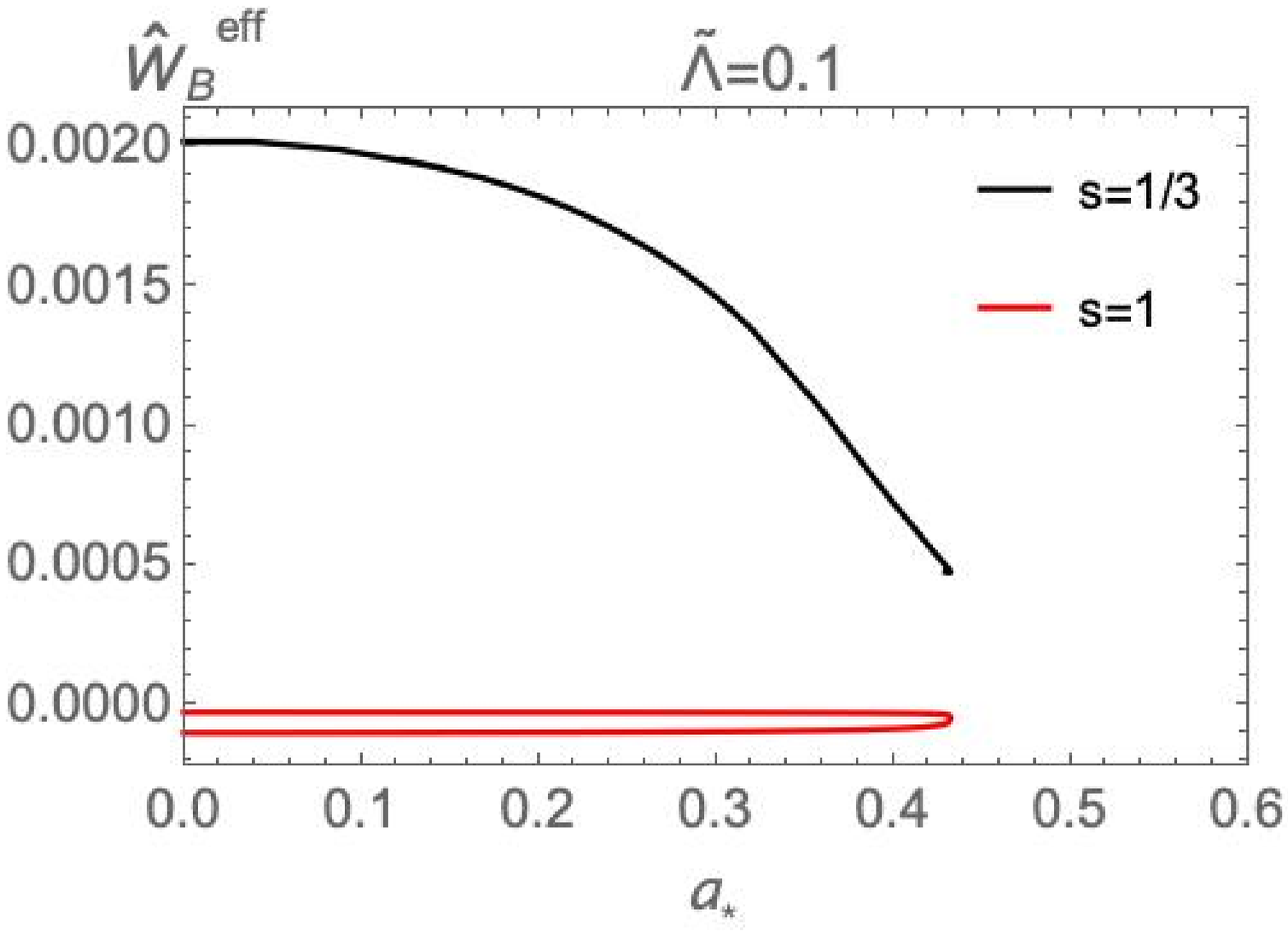}
 \end{center}
 \begin{center}
 \includegraphics[width=.45\textwidth]{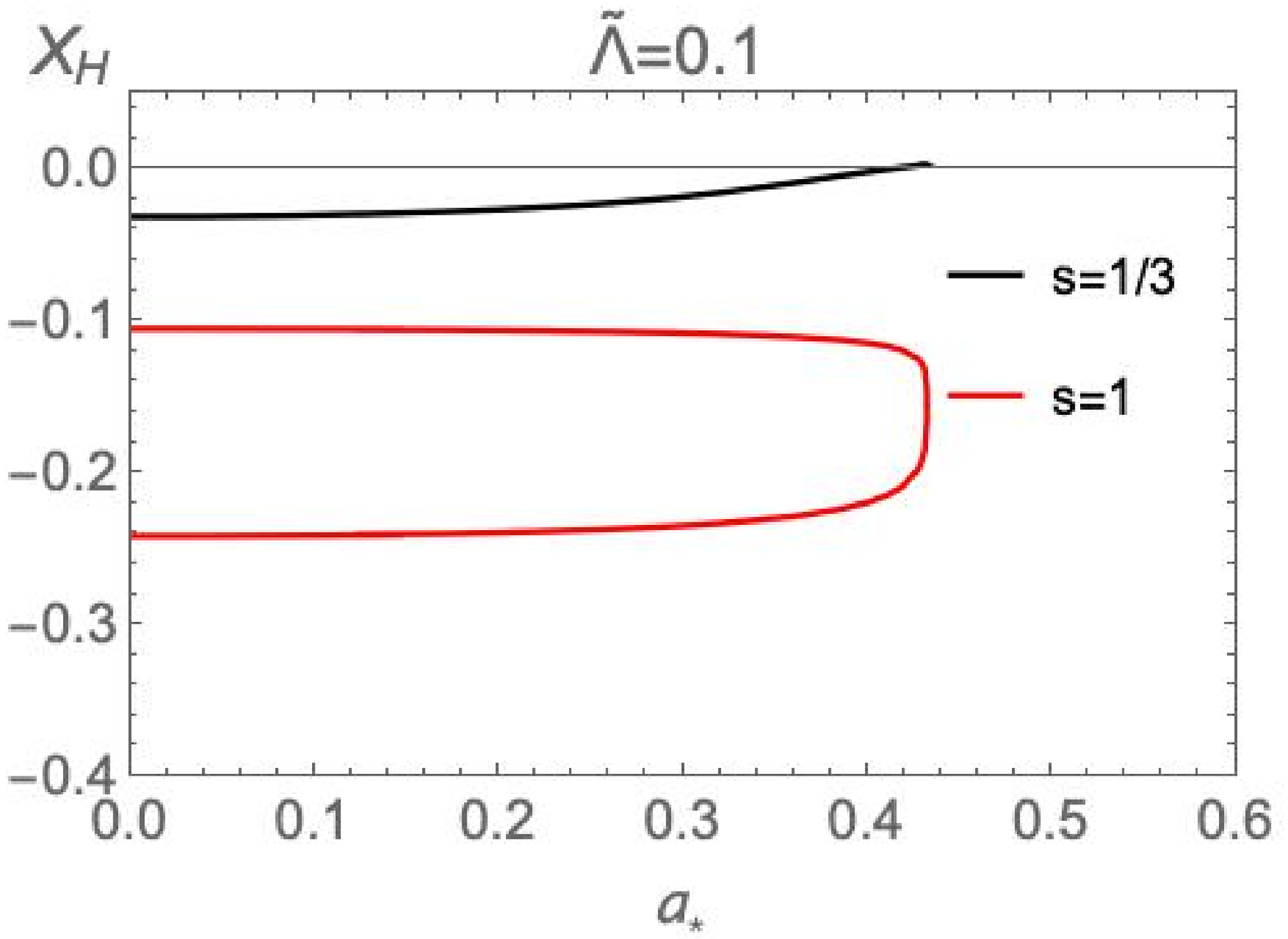}
 \includegraphics[width=.45\textwidth]{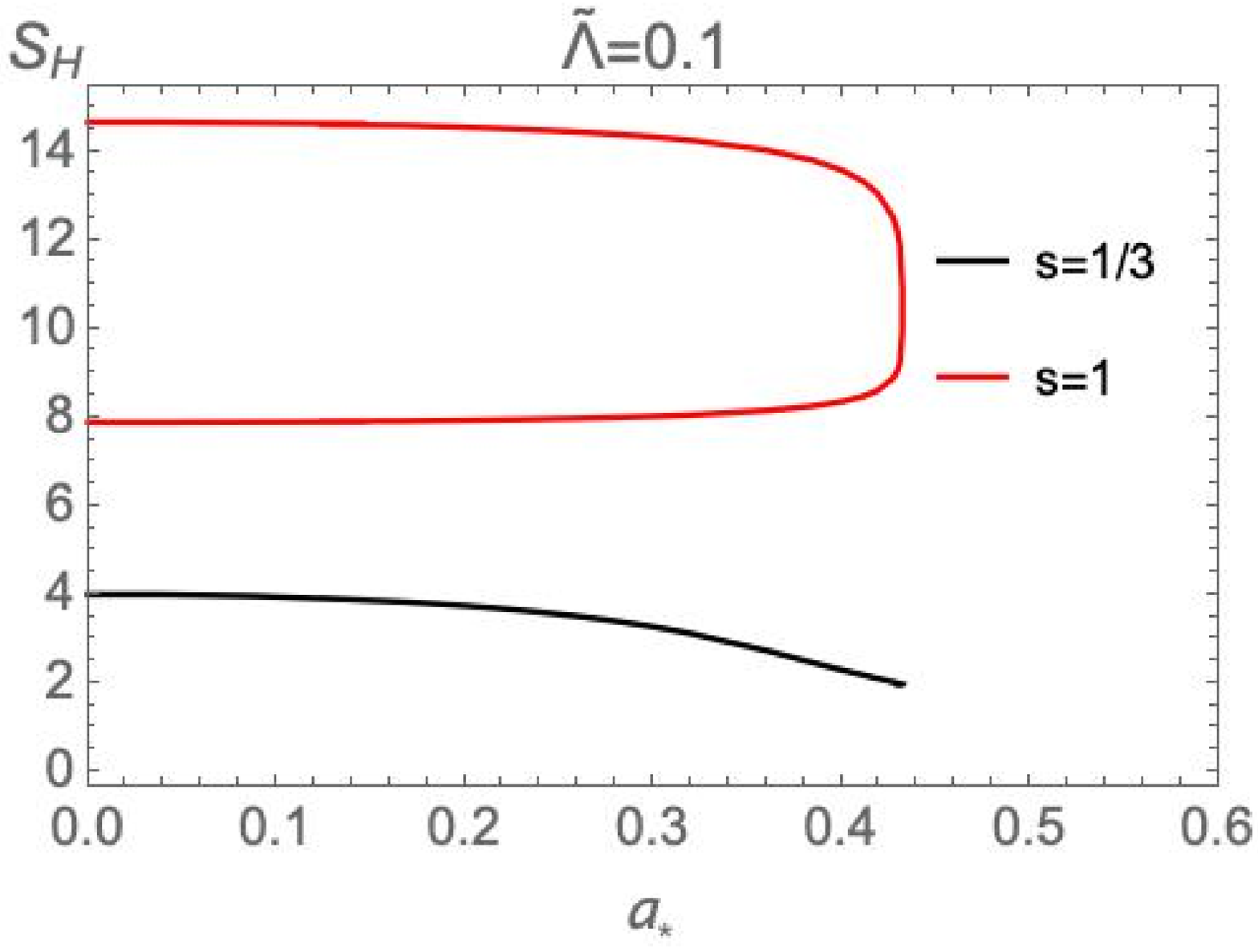}
  \end{center}
   \caption{
    Plots of $\f_0, \,\hat{W}_B^{eff}, \, X_H$ and $S_H$ vs.~$a_\star$ for a model with bulk functions \protect\eqref{Num1} and brane functions \protect\eqref{nfull1}.
 The bulk parameters are \protect\eqref{nfull0}, and the brane parameters are \protect\eqref{nfull9}, $\tilde{\Lambda}=0.1$, and $s=1/3, 1$.
\textbf{Top row, left:} Plot of $\f_0$.  \textbf{Top row, right:} Plot of $\hat{W}_B^{eff}(\f_0,a_0)$.
\textbf{Bottom row, left:} Plot of $X_H(\f_0,a_0)$.  \textbf{Bottom row, right:} Plot of $S_H(\f_0,a_0)$.
}
  \label{fig22}
 \end{figure}

  \begin{figure}[t]
 \begin{center}
  \includegraphics[width=.44\textwidth]{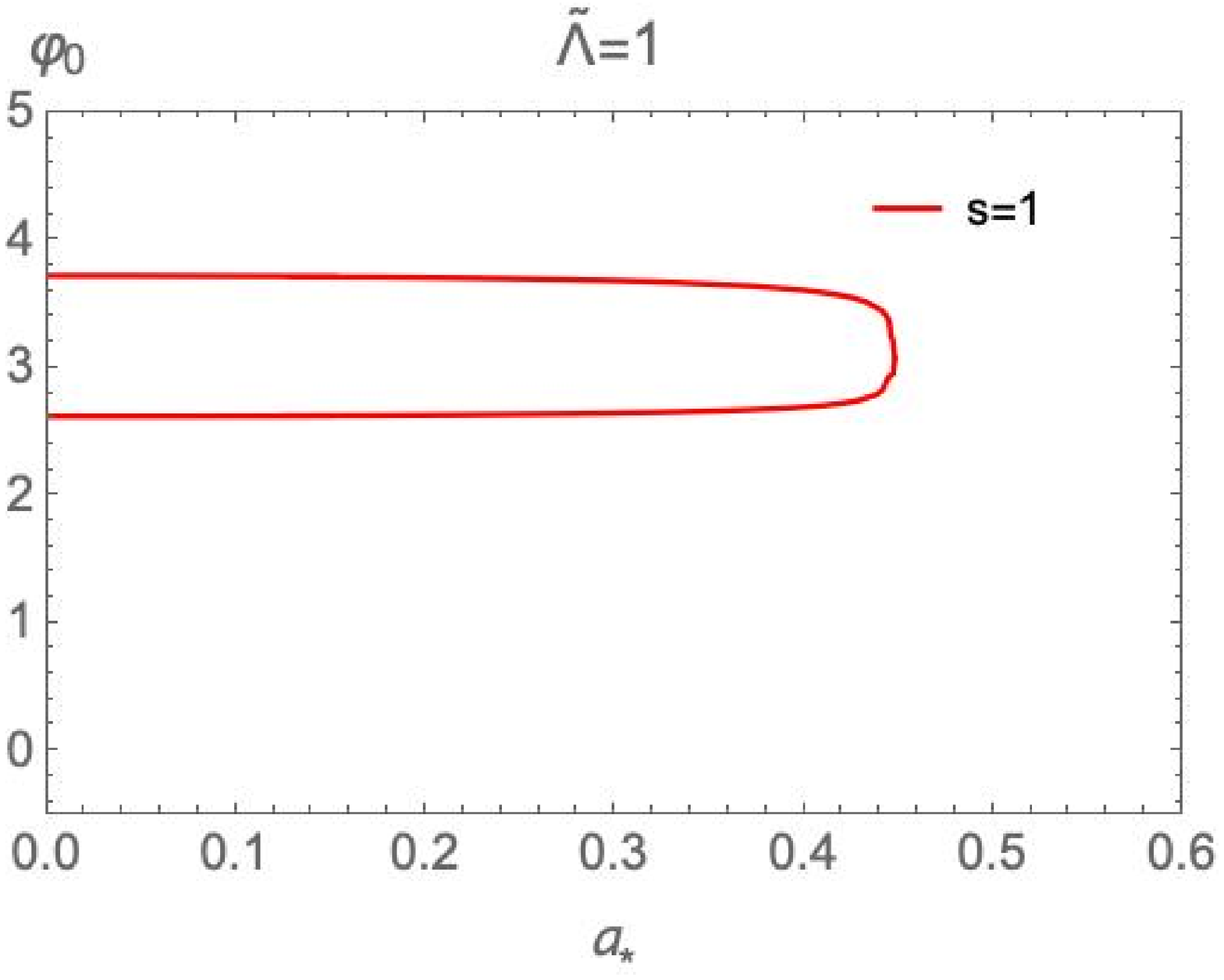}
   \includegraphics[width=.5\textwidth]{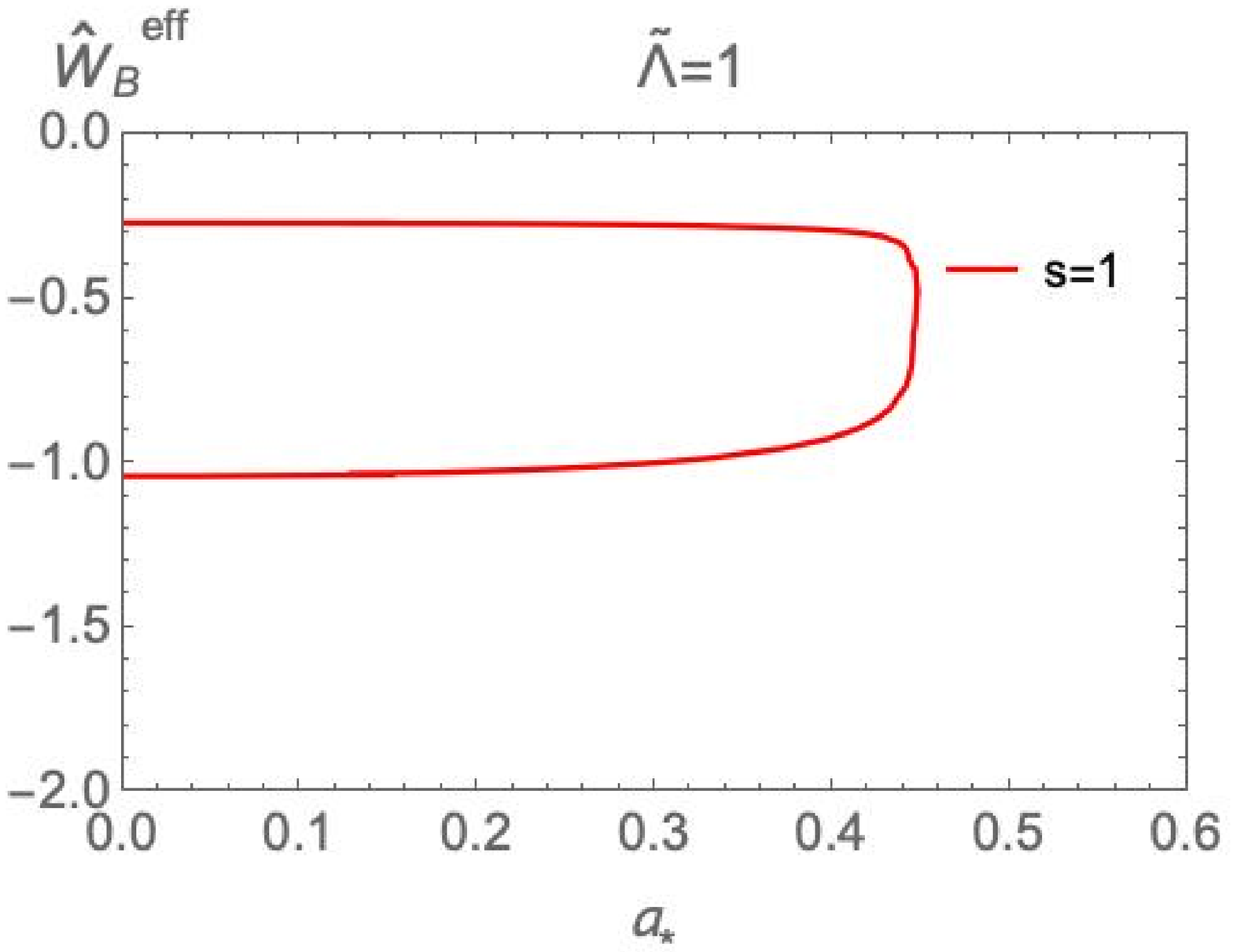}
 \end{center}
 \begin{center}
 \includegraphics[width=.45\textwidth]{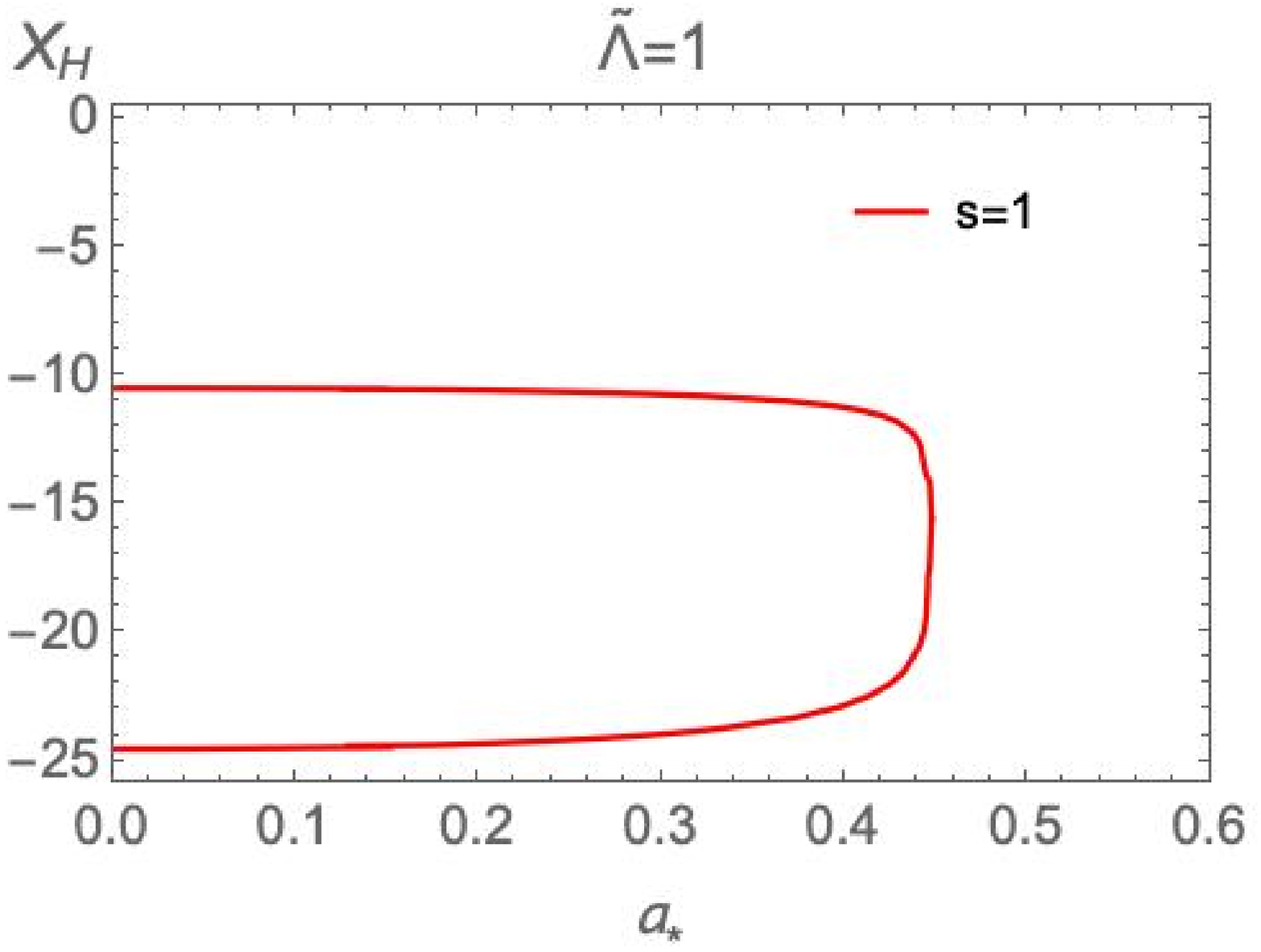}
 \includegraphics[width=.45\textwidth]{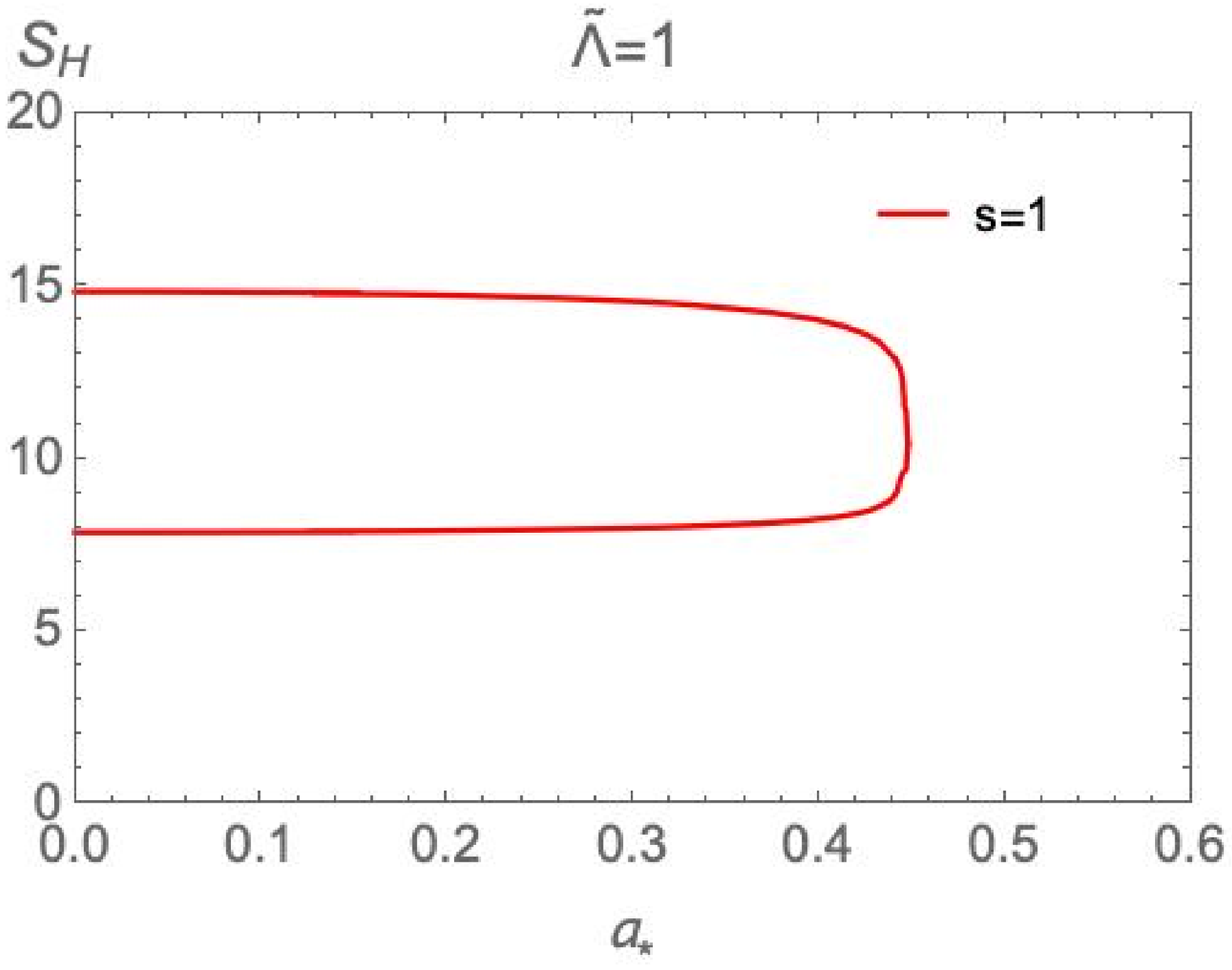}
  \end{center}
   \caption{
    Plots of $\f_0, \,\hat{W}_B^{eff}, \, X_H$ and $S_H$ vs.~$a_\star$ for a model with bulk functions \protect\eqref{Num1} and brane functions \protect\eqref{nfull1}.
 The bulk parameters are \protect\eqref{nfull0}, and the brane parameters are \protect\eqref{nfull9}, $\tilde{\Lambda}=1$, and $s=1$.
\textbf{Top row, left:} Plot of $\f_0$.  \textbf{Top row, right:} Plot of $\hat{W}_B^{eff}(\f_0,a_0)$.
\textbf{Bottom row, left:} Plot of $X_H(\f_0,a_0)$.  \textbf{Bottom row, right:} Plot of $S_H(\f_0,a_0)$.
}
  \label{fig23}
 \end{figure}

 \begin{figure}[t]
 \begin{center}
  \includegraphics[width=.44\textwidth]{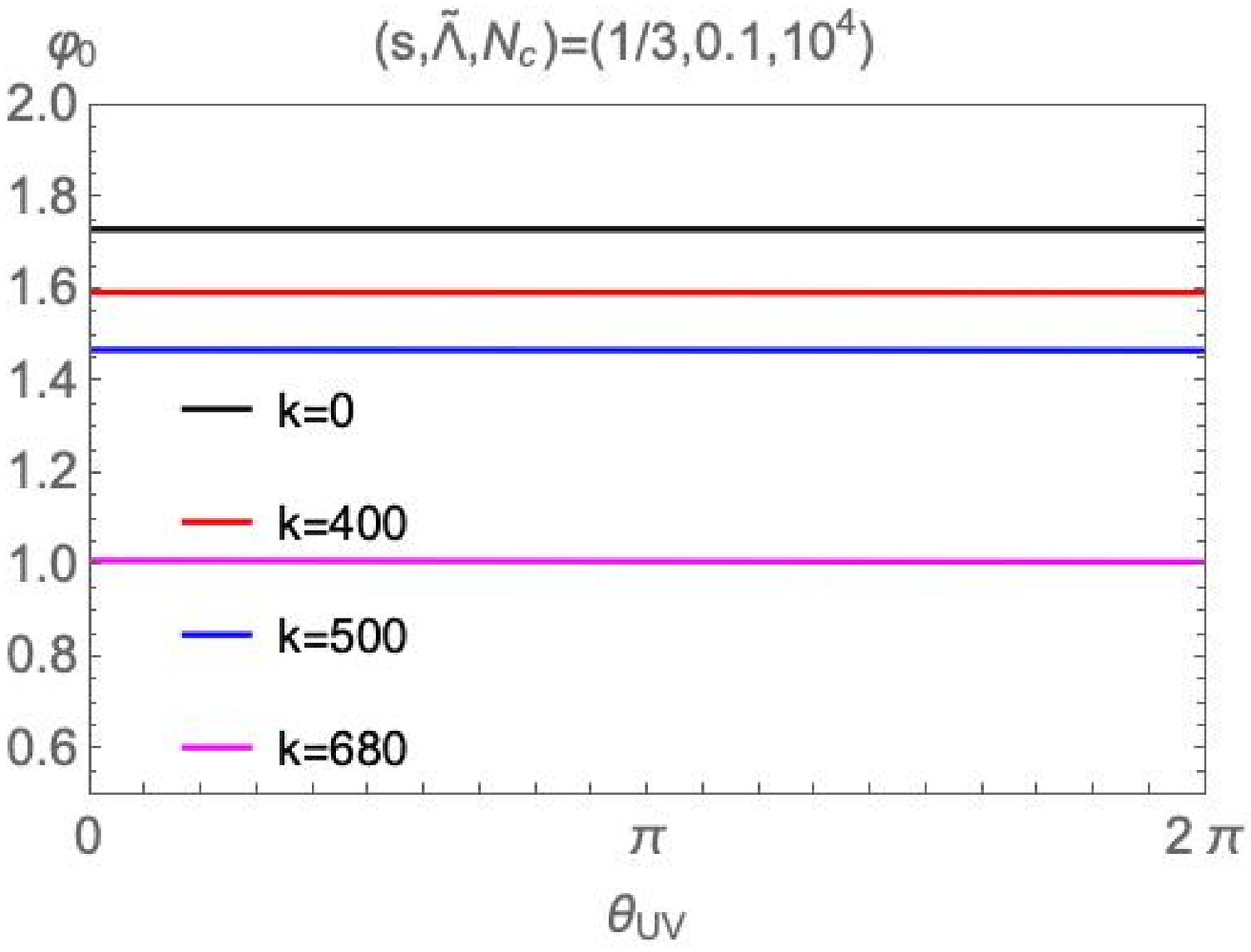}
   \includegraphics[width=.5\textwidth]{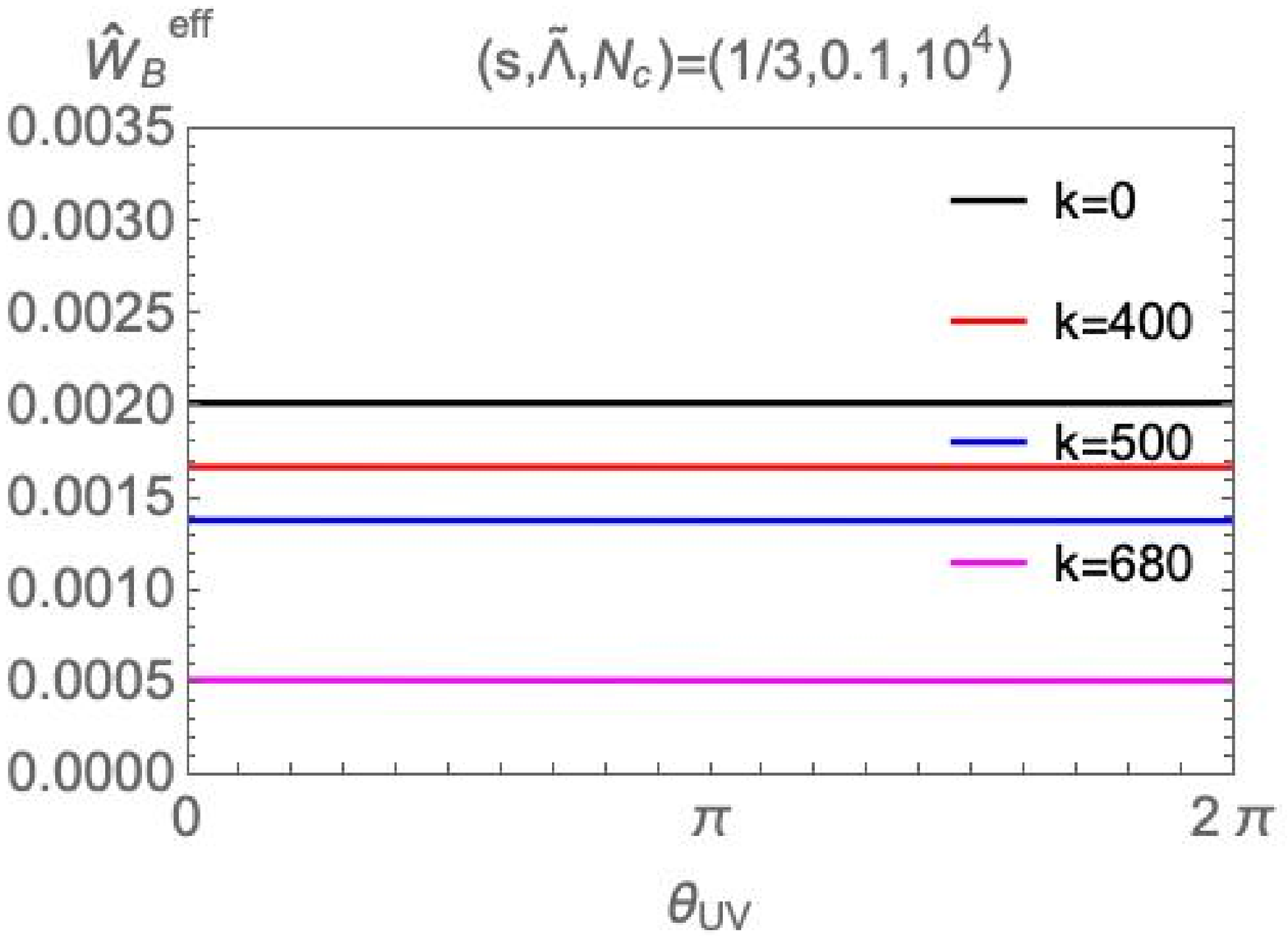}
 \end{center}
 \begin{center}
 \includegraphics[width=.45\textwidth]{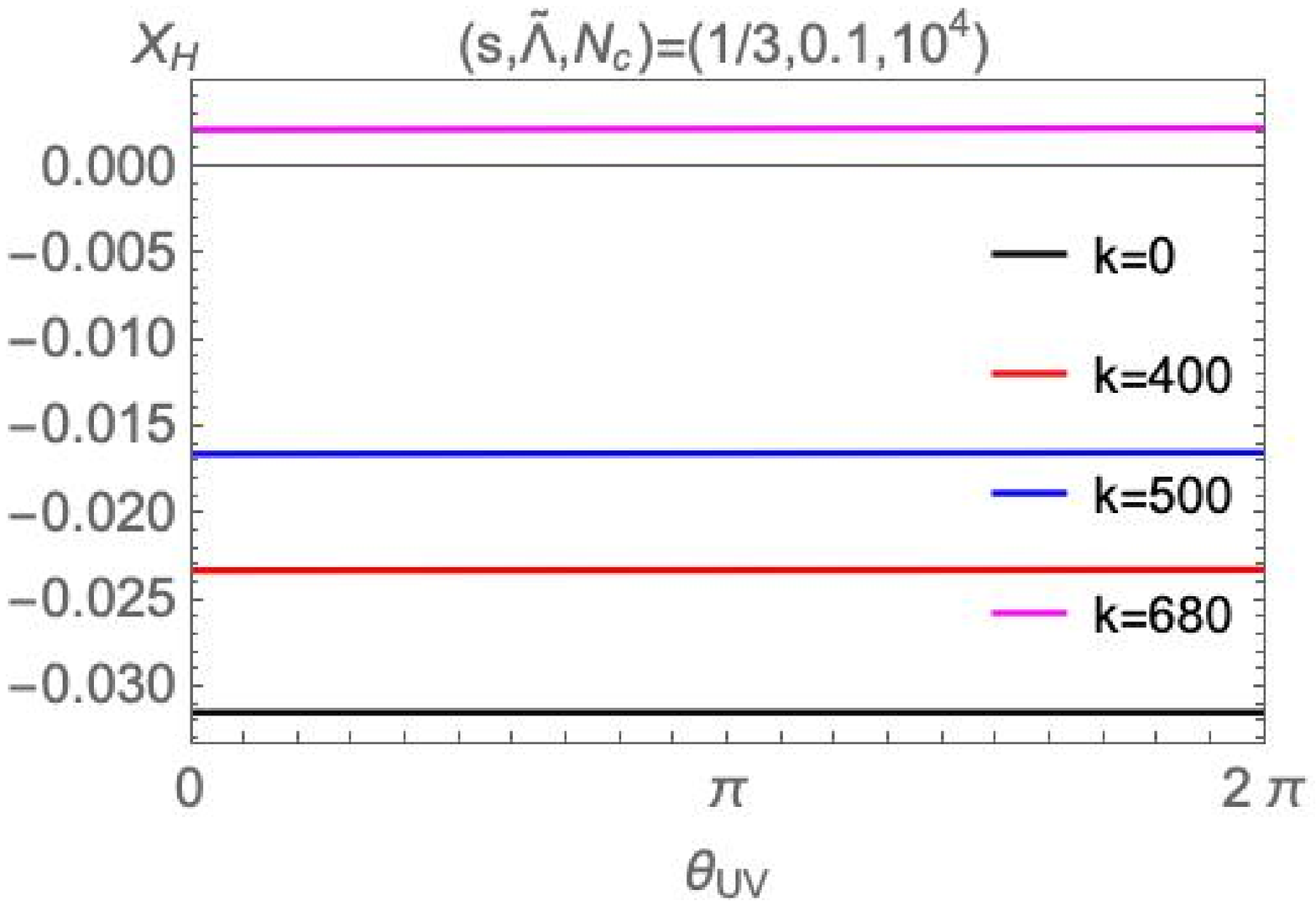}
 \includegraphics[width=.45\textwidth]{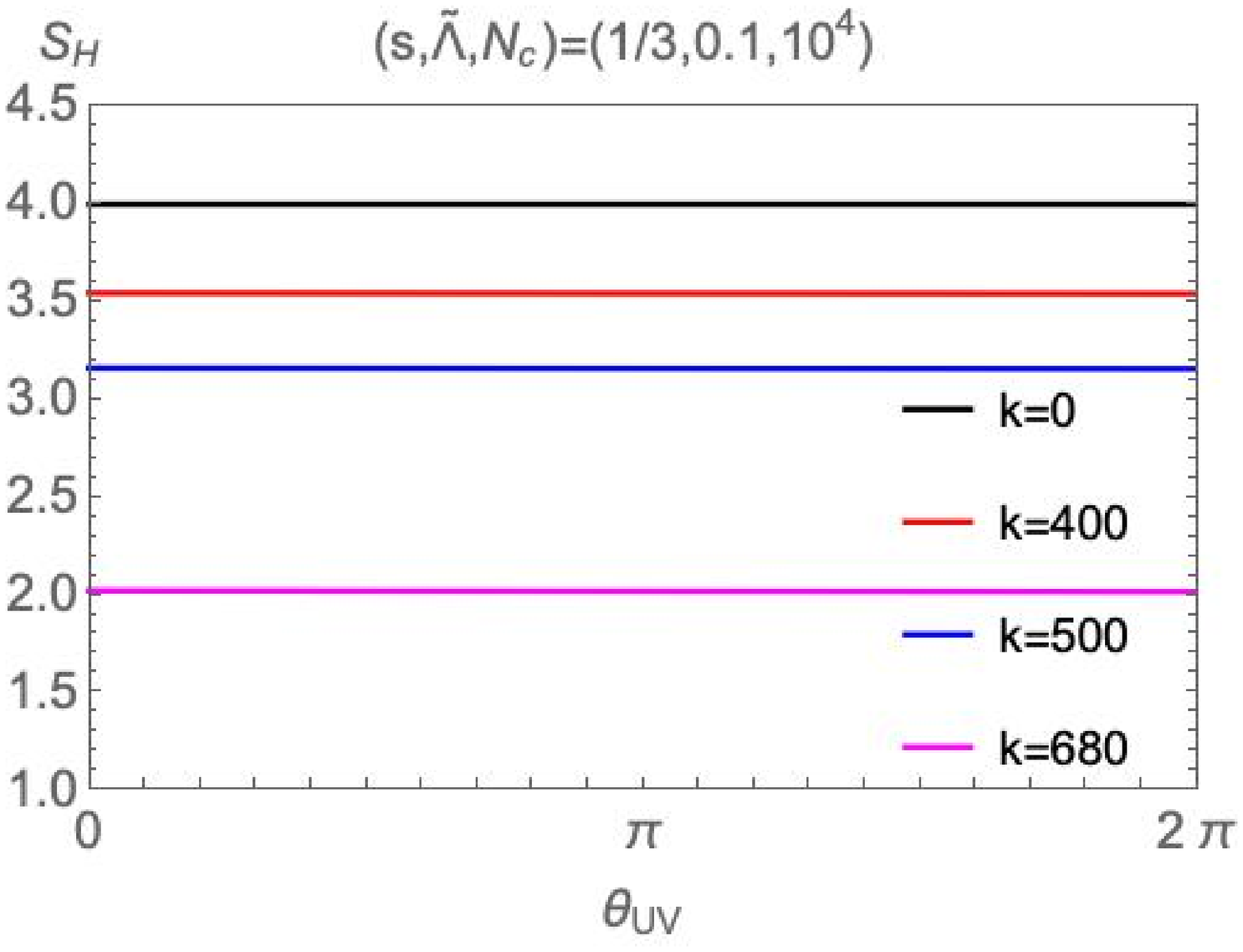}
  \end{center}
   \caption{
    Plots of $\f_0, \,\hat{W}_B^{eff}, \, X_H$ and $S_H$ vs.~$\theta_{UV}$ for a model with bulk functions \protect\eqref{Num1} and brane functions \protect\eqref{nfull1}. We take $s=1/3$, $\tilde{\Lambda}=0.1$ and $N_c=10^4$. Other parameters are \protect\eqref{nfull0} and \protect\eqref{nfull9}. Only the $k=0, 400, 500, 680$ branches are shown for the illustration.
\textbf{Top row, left:} Plot of $\f_0$.  \textbf{Top row, right:} Plot of $\hat{W}_B^{eff}(\f_0,a_0)$.
\textbf{Bottom row, left:} Plot of $X_H(\f_0,a_0)$.  \textbf{Bottom row, right:} Plot of $S_H(\f_0,a_0)$.
}
  \label{fig24}
 \end{figure}

 \begin{figure}[t]
 \begin{center}
   \includegraphics[width=.5\textwidth]{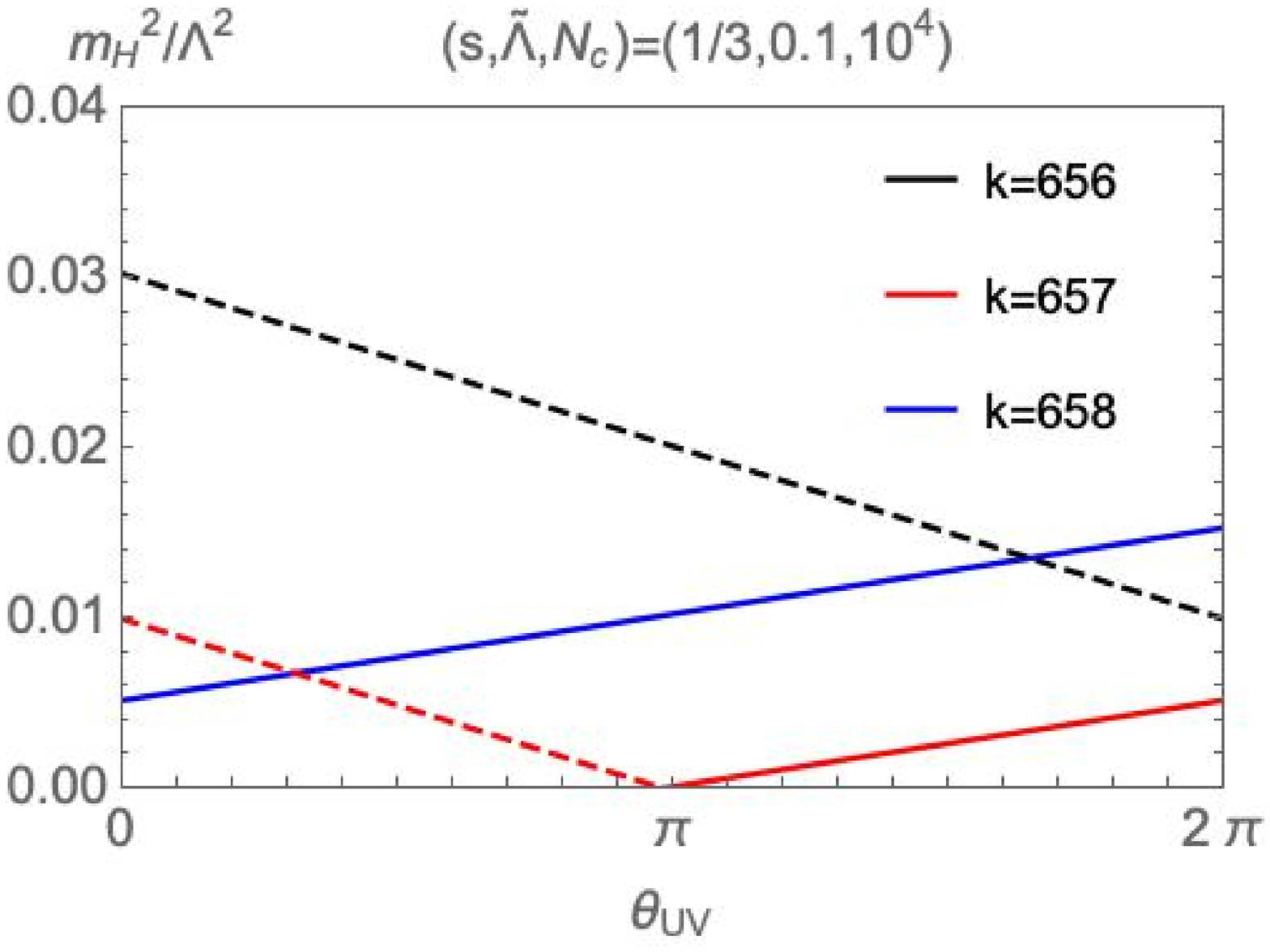}
 \end{center}
   \caption{
    Plot of the Higgs mass squared $m_H^2$ \protect\eqref{nth2} in units of $\Lambda^2$ vs.~$\theta_{UV}$ for a model with bulk functions \protect\eqref{Num1} and brane functions \protect\eqref{nfull1}.
The model parameters are \protect\eqref{nfull0}, \protect\eqref{nfull9}, $s=1/3, \, \tilde{\Lambda}=0.1, \, N_c=10^4$.
We plot the lines with $k=656, \,657, \,658$, corresponding to the branches realizing the small Higgs mass.
Note that, for $0\leq\theta_{UV}\lesssim\pi$, the symmetric ($X_H>0$) and broken ($X_H<0$) phase correspond to $k\geq658$ and $k\leq657$, respectively. For $\pi\lesssim\theta_{UV}<2\pi$, the symmetric and broken phases correspond to $k\geq657$ and $k\leq656$.
In the figure, the symmetric and broken phases are represented by the solid and dashed lines, respectively.
}
  \label{fig24-2}
 \end{figure}

\subsection{Brane potential choice $1$: No explicit axion dependence} \label{num1}
Here we consider the following choice for the brane functions in \eqref{A3}:
\be
W_B=
\frac{\Lambda^4}{M_p^3} \left[ -1 -{\f \over s} + \left(\f \over s\right)^2 \right]\sp
\label{nfull1}\ee
$$
X_H=
\frac{\Lambda^2}{M_p} \left[ 1 +{\f \over s_X} - \left(\f \over s_X\right)^2 \right]\sp
S_H=
M_p \left[ 1 + \left(\f \over s_H \right)^2 \right],
$$
where $M_p$ is the bulk Planck mass and $s, s_X$ and $s_H$ are dimensionless constants. We also introduced the parameter $\Lambda$ which has the interpretation of UV cutoff for the brane theory.
Note that \eqref{nfull1} implies that $\hat{W}_B^{eff}$ is proportional to $\Lambda^4/M_p^3$, as one can observe from \eqref{FE8}, \eqref{match4-2} and \eqref{match5}.
Therefore, it will be convenient to define the quantity
\be
\tilde{\Lambda}^4\equiv \Lambda^4/M_p^3,
\label{nfull1-2}\ee
where $\tilde{\Lambda}$ can be considered as a dimensionless quantity since we set $\ell=1$.
Besides the choice of parameters given in \eqref{nfull0}, here we also set
\be
s_X={2\over3}, \quad s_H=1,
\label{nfull9}\ee
and vary $s$ and $\tilde{\Lambda}$.

Here the brane functions do not depend explicitly on $a$ and the junction condition \eqref{nfull6} reduces to
\be
\left.
-{2\over3}\left(W_{IR} \hat{W}_B^{eff} - {1\over2} \left(\hat{W}_B^{eff}\right)^2\right)
+\left(  S_{IR} {\p \hat{W}_B^{eff}\over \p \f} - {1\over2} \left({\p \hat{W}_B^{eff}\over \p \f}\right)^2\right)
\right|_{\f=\f_0}=0.
\label{nfull7}\ee

In figure \ref{fig18} we plot solutions for $W, S, $ and $T$ for the parameter choices \eqref{nfull10}, \eqref{nfull9}, $s=1, \,\tilde{\Lambda}=1$.\footnote{ $\tilde\Lambda=1$ corresponds to a brane cutoff scale $\Lambda$ that is much smaller than  the bulk Planck scale.}
The three solutions plotted correspond to the three choices $D_{IR}=0.01, 1$ and $5$ for the axion-related integration constant $D_{IR}$. The vertical lines indicate the brane position $\f_0$ for the three different choices of $D_{IR}$. There are two solutions to \eqref{nfull7} satisfying \eqref{nfull8}. To be specific, in figure \ref{fig18} we only display the solution with the smaller value of $\f_0$. Note that the functions $W$ and $S$ are discontinuous at $\f_0$ because of the junction conditions \eqref{match1-2} and \eqref{match2-2}. The function $T$ is continuous, consistent with \eqref{match3-2}, as here the brane functions are independent of $a$, \eqref{nfull1}. However, $T$ is not smooth. The first derivative of $T$ is not continuous at $\f_0$, which follows from \eqref{a8-2} and the discontinuity of $W$ and $S$.
We also checked that the numerical result is consistent with the UV and IR asymptotic expansions (\ref{w-}, \ref{s-}, \ref{t-}, \ref{sub8-2}, \ref{sub9}, \ref{sub10}).

The main observation is that even though the brane functions in \eqref{nfull1} do not depend explicitly on the axion $a$, the equilibrium brane position $\f_0$ is affected by axion backreaction. Here, we controlled the strength of axion backreaction by adjusting the integration constant $D_{IR}$ at the IR end of the flow. A shift in $D_{IR}$ resulted in a (small but nevertheless non-vanishing) shift in $\f_0$.

In the following, we shall  also examine in more detail how the brane equilibrium position $\f_0$ and the brane functions $\hat{W}_B^{eff}, \, X_H$ and $S_H$ evaluated at $\f_0$ are affected by axion backreaction. However, rather than controlling $D_{IR}$, the axion integration constant in the IR, it will be more convenient to dial the value of $a_{\star}$, the axion integration constant in the UV, as this has a physical interpretation as the axion source in the dual field theory.\footnote{In practice, when solving numerically, we implement boundary conditions in the IR and hence we need to specify $D_{IR}$. Then, for a given solution, we read off the corresponding values of $a_{\star}$ and $\f_0$. Scanning over all values of $D_{IR}$ we can then determine $a_{\star}$, $\f_0$ as functions of $D_{IR}$, i.e.~$a_{\star}(D_{IR})$, $\f_0(D_{IR})$. Inverting $a_{\star}(D_{IR})$ then allows us to obtain $\f_0(a_{\star})$ from $\f_0(D_{IR})$. In this way we can also determine $\hat{W}_B^{eff}(\f_0), \, X_H(\f_0)$ and $S_H(\f_0)$ as functions of $a_{\star}$.}

As  examples of the types of solutions, we show  the results one obtains for  a few representative (but in no way special) values of the remaining unfixed  parameters, namely we take
\be\label{choice}
s = {1\over 3}, \;1 \qquad \tilde{\Lambda}=0.1, \; 1
\ee
Notice that choosing $\tilde{\Lambda}$ (defined in  equation (\ref{nfull1-2}))   of order one or smaller in AdS units means that we are restricting  the UV cut-off $\Lambda$ to be much  smaller than the Planck scale. Indeed, reinstating the AdS length in equation (\ref{nfull1-2}), we find that
\be
{\Lambda^4 \ell \over M^3_p} \sim O(1) \quad \Rightarrow \quad \Lambda^4 \sim {M_p^3 \over \ell} \ll M_p^4 ,
\ee
where the last inequality comes from the requirement that the bulk geometry is classical, $\ell \gg M_p^{-1}$.

In figures \ref{fig22} and \ref{fig23}, we plot the values of $\f_0, \,\hat{W}_B^{eff}, \, X_H$ and $S_H$ as functions of $a_{\star}$ for the bulk parameters \eqref{nfull10}, brane parameters \eqref{nfull9} and the combination of parameters chosen in (\ref{choice}).
We make the following observations.
\begin{itemize}
\item The range of $a_{\star}$ is typically bounded with an upper limit $a_\star^{\textrm{max}}$ whose precise value depends on the model parameters. This property of axionic RG flows was already observed in absence of the brane in \cite{Hamada}, but it also persist when a brane is included. In figures \ref{fig22}, \ref{fig23} and all following plots of functions of $a_{\star}$, we display the functions over their complete domain of support $0 < a_\star < a_\star^{\textrm{max}}$.
\item For the parameter choice $s=1/3, \, \tilde{\Lambda}=1$ there exist solutions to the junction condition \eqref{nfull7}, however, these do not satisfy the overshooting constraint \eqref{re10} in Appendix \ref{constraint} (i.e.~the solutions misses the fixed point in the UV).  Thus, as stated there, we should discard these solutions and this is why we refrain from plotting the corresponding numerical results in figure \ref{fig23}.
\item The brane cosmological constant $\hat{W}_B^{eff}$ is generically of the same magnitude as $\tilde{\Lambda}^4$, as can be seen in the top right panels of figures \ref{fig22} and \ref{fig23}. Nevertheless, the brane worldvolume is flat by construction, i.e.~the solutions exhibit self-tuning of the cosmological constant as advertised.
\item In addition to realising this self-tuning mechanism for the cosmological constant, the second objective of this work is to seek for solutions with low Higgs mass (and vev). As follows from the discussion in sec.~\ref{sec:junc}, a small  Higgs mass can be attained if $X_H$ is small. This is the case in the vicinity of $X_H \approx 0$ and thus we are particularly interested in solutions where $X_H$ as a function of $a_{\star}$ changes sign. For the parameter choices considered here, a sign change in $X_H$ exists, but only on the branch of solutions with  $s=1/3$, $\tilde{\Lambda}=0.1$, see fig.~\ref{fig22}. We shall  study the solutions on this branch in more detail next.
\end{itemize}
So far we were considering $\f_0, \, \hat{W}_B^{eff}, \, X_H$ and $S_H$ as a function of the UV parameter $a_{\star}$ in figures \ref{fig22} and \ref{fig23}. A related UV parameter is $\theta_{UV}$, the theta angle of the dual field theory supported on the UV boundary. Note that the identification between $a_{\star}$ and $\theta_{UV}$, recorded in \eqref{i2}, is many-to-one, i.e.~one fixed value of $\theta_{UV}$ corresponds to many different discrete values of $a_{\star}$. Following the notation in \eqref{i2} we can label the various vacua associated with a single value for $\theta_{UV}$ by the integer $k$. As all these different vacua for a given $\theta_{UV}$ have different values of $a_\star$, all these different vacua will generically possess different values for $\f_0, \, \hat{W}_B^{eff}, \, X_H$ and $S_H$. For large values of $N_c$ the various vacua for fixed $\theta_{UV}$ are `dense' in $a_{\star}$-space, as follows straightforwardly from \eqref{i2}. To illustrate this we consider the branch of solutions with $s=1/3$, $\tilde{\Lambda}=0.1$ in fig.~\ref{fig22}, but using \eqref{i2} we plot the brane functions $\f_0, \, \hat{W}_B^{eff}, \, X_H$ and $S_H$ as functions of $\theta_{UV}$. This is shown in fig.~\ref{fig24} where for better visibility, we only plot results for $k=0, 400, 500, 680$.\footnote{As the overall range of $a_\star$ is bounded, there is a finite number of saddle points associated with a fixed value of $\theta_{UV}$. Here, with the choice $c=1$, $N_c=10^4$ and the observed value $a_\star^{\textrm{max}} \sim 0.4$ we find that the total number of fixed-$\theta_{UV}$ saddle points is $\sim 680$, see e.g.~eq.~\eqref{nfull10}.}

Then, as long as $X_H$ as a function of $a_\star$ changes sign, there will typically exist a finite (but possibly large) number of fixed-$\theta_{UV}$ vacua with $X_H \approx 0$ and hence a small Higgs mass. Here we find that   this is the case for the branches of solutions with $k$ in the vicinity of $k \sim 657$. In figure \ref{fig24-2}, we plot of ratio of Higgs mass $m_H^2$ defined in \eqref{nth2} and the scale $\Lambda$ appearing in \eqref{nfull1} for the branches with $k= 656, 657, 658$.\footnote{Note that, for $0\leq\theta_{UV}\lesssim\pi$, the symmetric and broken phases for the Higgs correspond to $k\geq658$ and $k\leq657$, respectively. For $\pi\lesssim\theta_{UV}<2\pi$, the symmetric and broken phases correspond to $k\geq657$ and $k\leq656$. In figure \ref{fig24-2}, the symmetric and broken phases are represented by the solid and dashed lines, respectively.} Notice that the ratio $m_H^2/\Lambda^2$ is independent of $\Lambda$ and $M_p$, and we do not need to specify the values of these parameters. The scale $\Lambda$ sets  both size of the cosmological constant and the naive Higgs mass parameter $X_H$ on the brane and can be understood as the UV cutoff scale of the brane theory. We hence refer to the Higgs mass as small if
$$m_H^2\ll  \Lambda^2 \;.$$

In figure \ref{fig24-2} we observe that $m_H^2/\Lambda^2 \sim {\cal O}(10^{-2})$ for a generic value of $\theta_{UV}$ on these branches, but the precise value is also a consequence of our choice $N_c=10^4$. Branches with smaller values can be obtained if $N_c$ is chosen larger.

To summarise, the example considered here is a brane-world model that realizes self-tuning of the cosmological constant, but also possesses a (potentially) large number of saddle points, some  with a small Higgs mass. Therefore it can be seen as a proof of principle that a simultaneous self-tuning of the cosmological constant and the EW breaking scale is possible. The crucial condition for achieving this is that $X_H$ as a function of $a_\star$ changes sign for some value of $a_\star$. However, for the choice of brane functions considered here, i.e.~\eqref{nfull1}, a sign change in $X_H(a_\star)$ does not occur for  generic choice of the model parameters $s, \tilde{\Lambda}$. In the following section we shall hence consider a different choice of brane functions to see whether this short-coming of the example considered here can be overcome.

 \begin{figure}[t]
 \begin{center}
  \includegraphics[width=.45\textwidth]{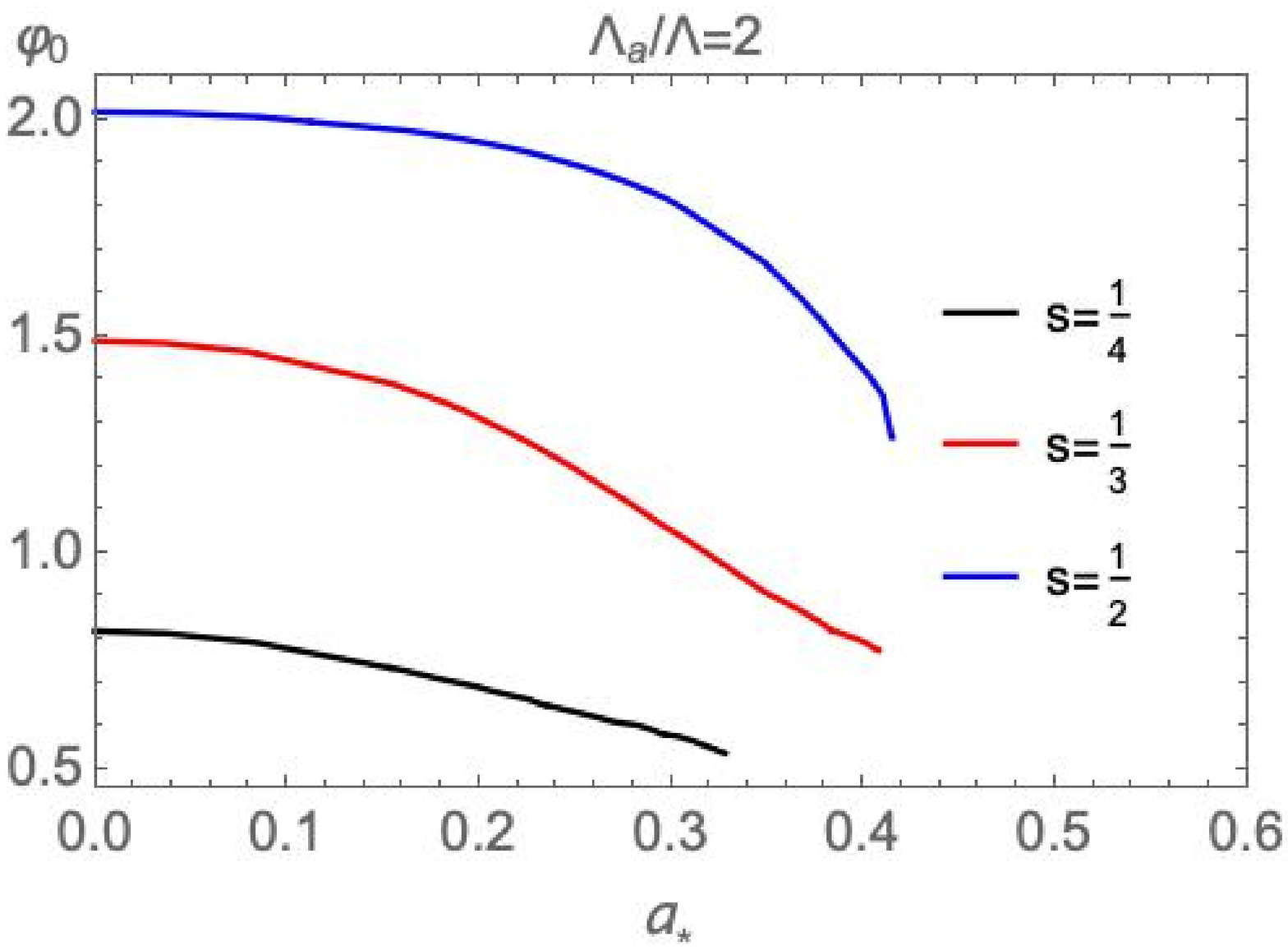}
   \includegraphics[width=.45\textwidth]{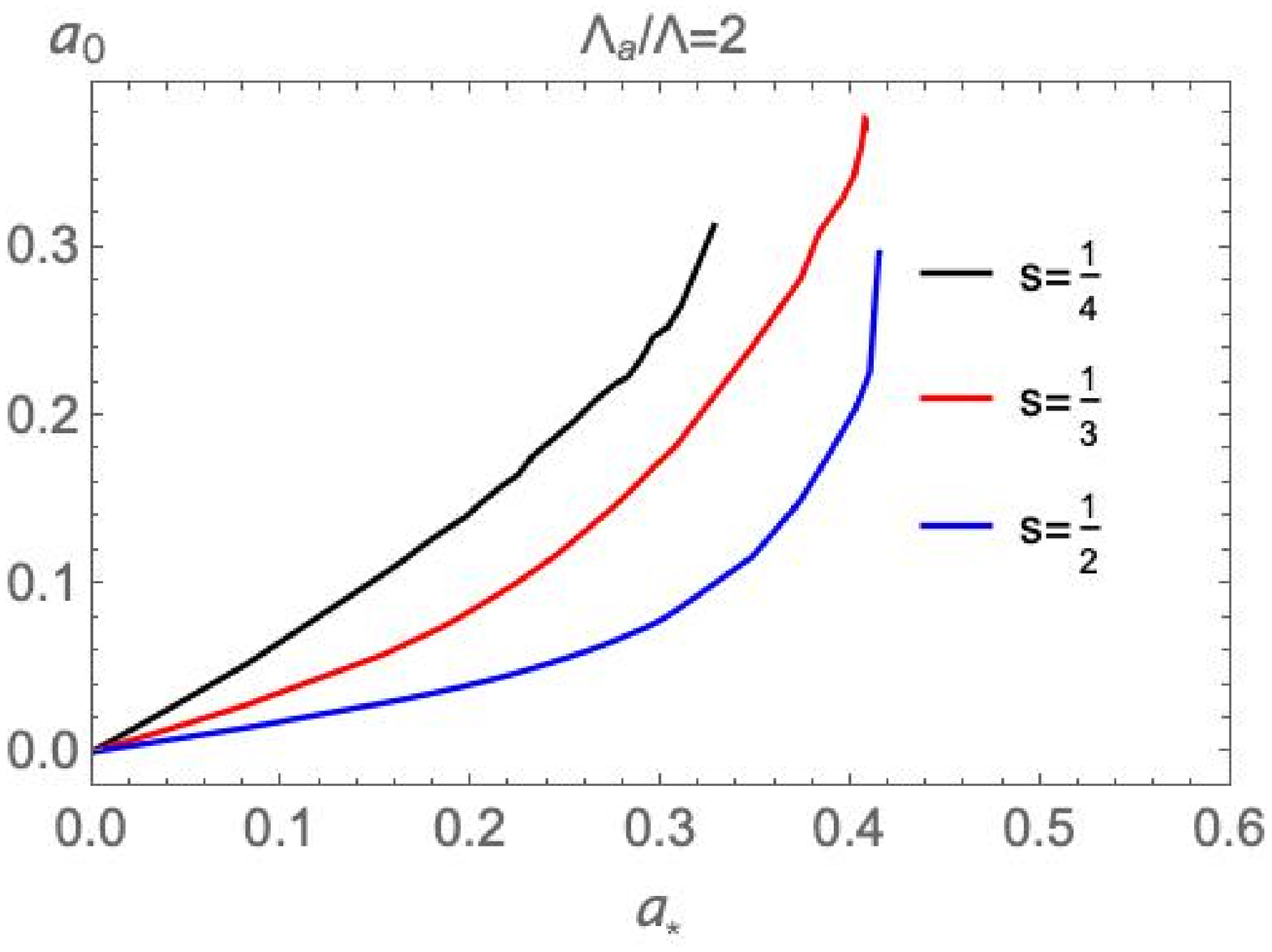}
 \end{center}
 \begin{center}
  \includegraphics[width=.45\textwidth]{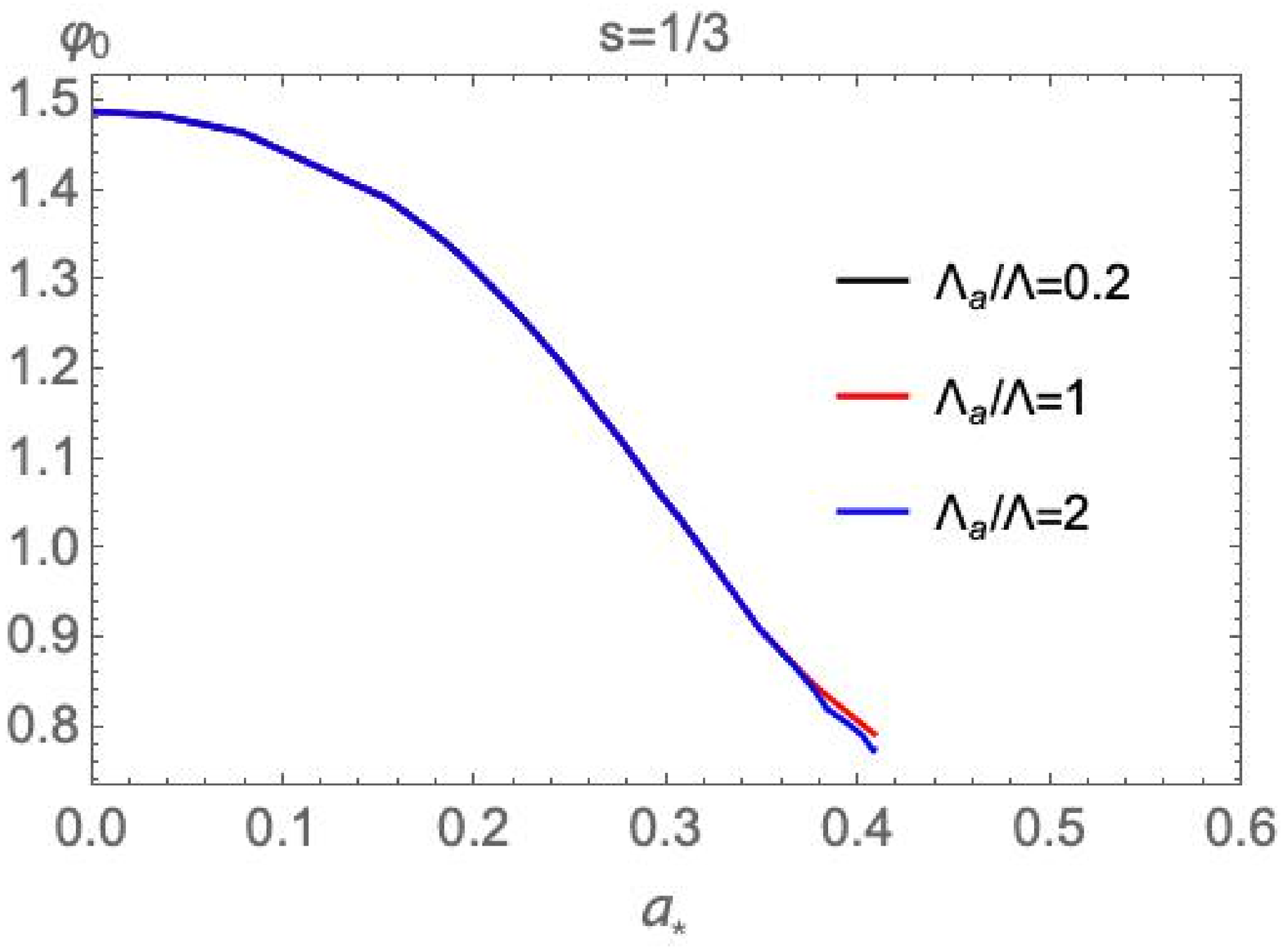}
   \includegraphics[width=.45\textwidth]{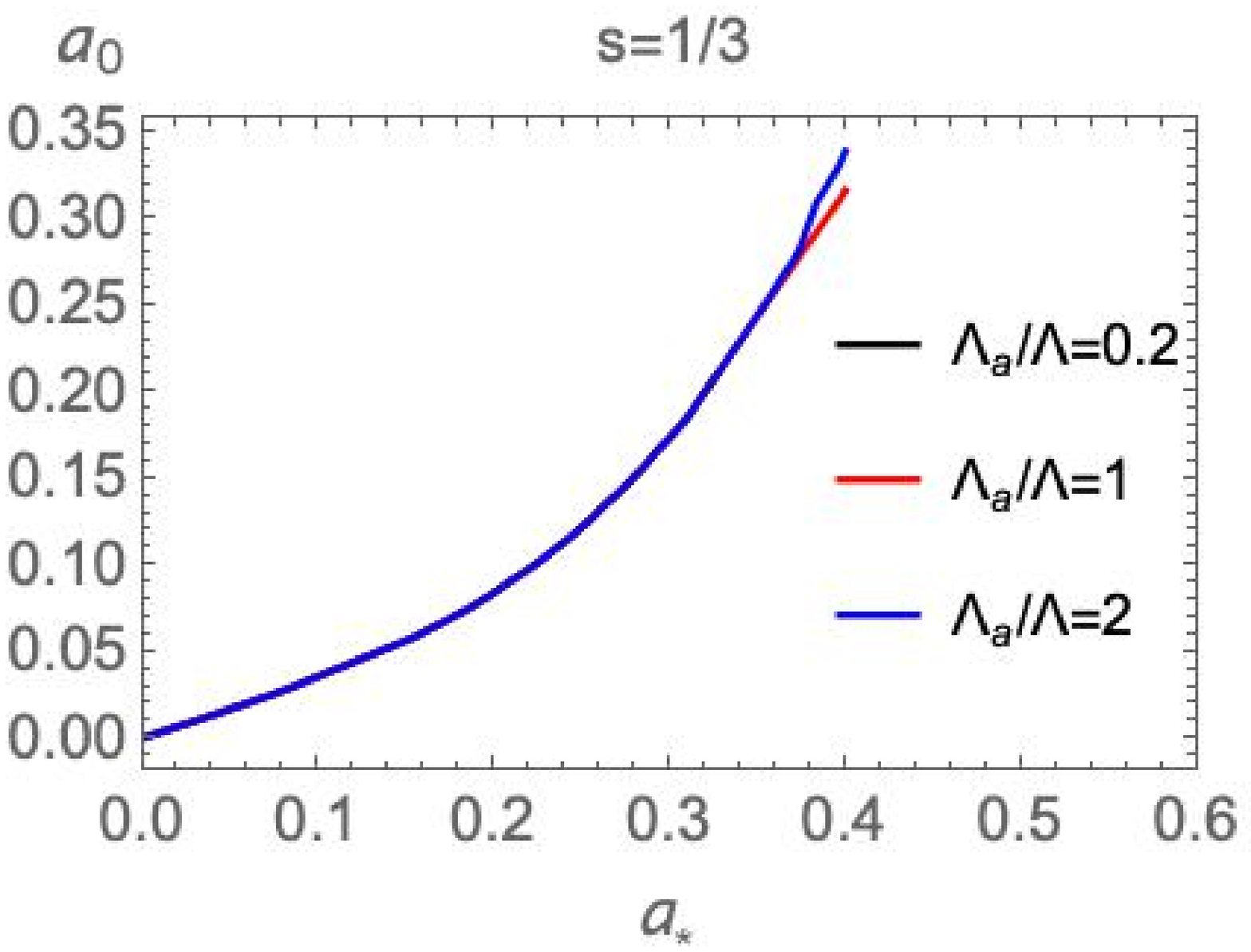}
 \end{center}
   \caption{
    Plots of $\f_0$ and $a_0$ vs.~$a_{\star}$ for a model with bulk functions \protect\eqref{Num1} and brane functions \protect\eqref{nth1}.
   The bulk parameters are \protect\eqref{nfull0}, and the brane parameters are \protect\eqref{nth3}, ${\rm sign}(Q_{IR})=-1$, $s=1/4, \, 1/3, \, 1/2$ and $\Lambda_a/\Lambda=0.2, \, 1, \, 2$.
   \textbf{Top row, left:} Plot of $\f_0$ for $s=1/4, \, 1/3, \, 1/2$ and $\Lambda_a/\Lambda=2$.
   \textbf{Top row, right:} Plot of $a_0$ for $s=1/4, \, 1/3, \, 1/2$ and $\Lambda_a/\Lambda=2$.
   \textbf{Bottom row, left:} Plot of $\f_0$ for $s=1/3$ and $\Lambda_a/\Lambda=0.2, \, 1, \, 2$. The various plots are near-indistinguishable.
   \textbf{Bottom row, right:} Plot of $a_0$ for $s=1/3$ and $\Lambda_a/\Lambda=0.2, \, 1, \, 2$. The various plots are near-indistinguishable.
}
  \label{fig33}
 \end{figure}

 \begin{figure}[t]
 \begin{center}
  \includegraphics[width=.45\textwidth]{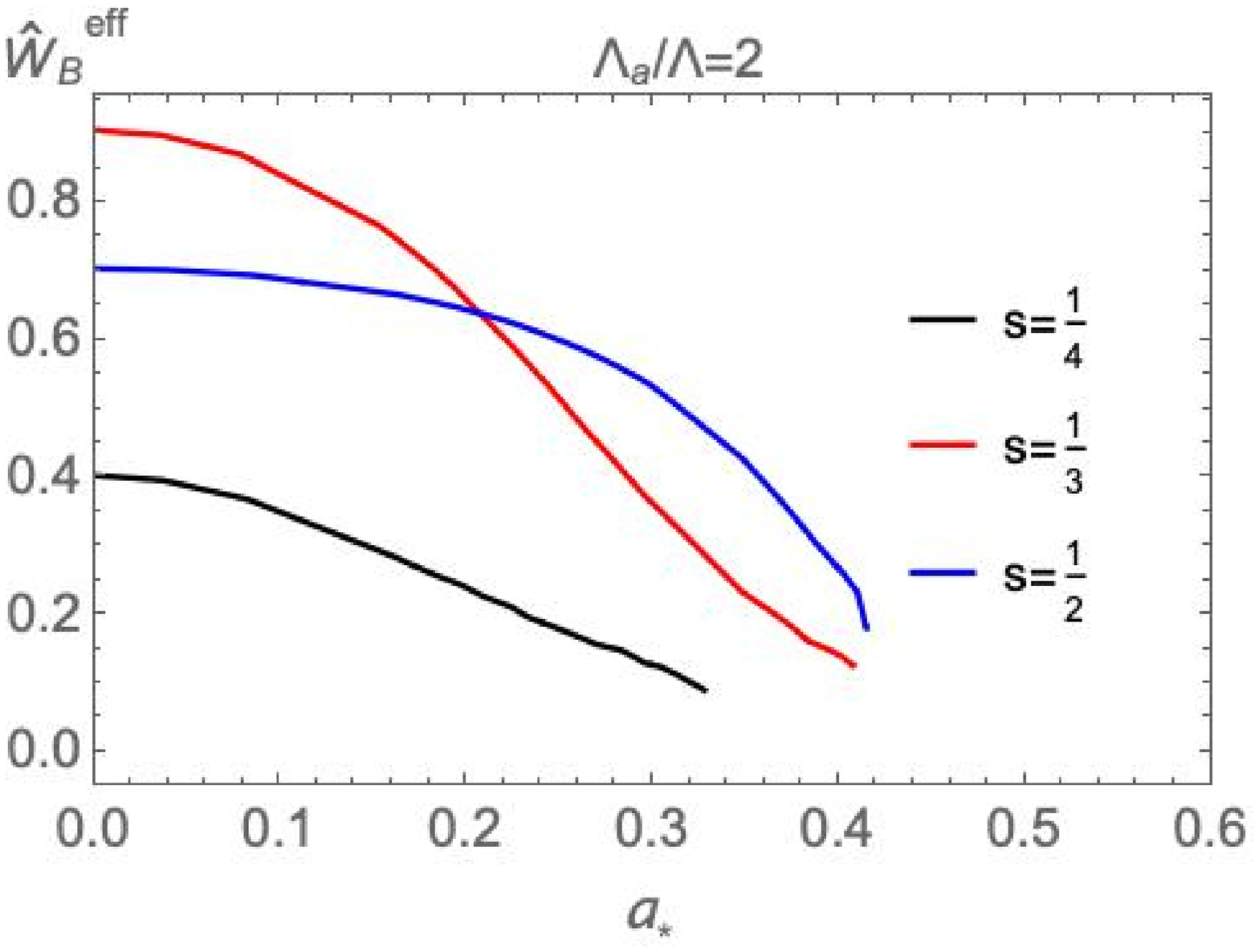}
   \includegraphics[width=.45\textwidth]{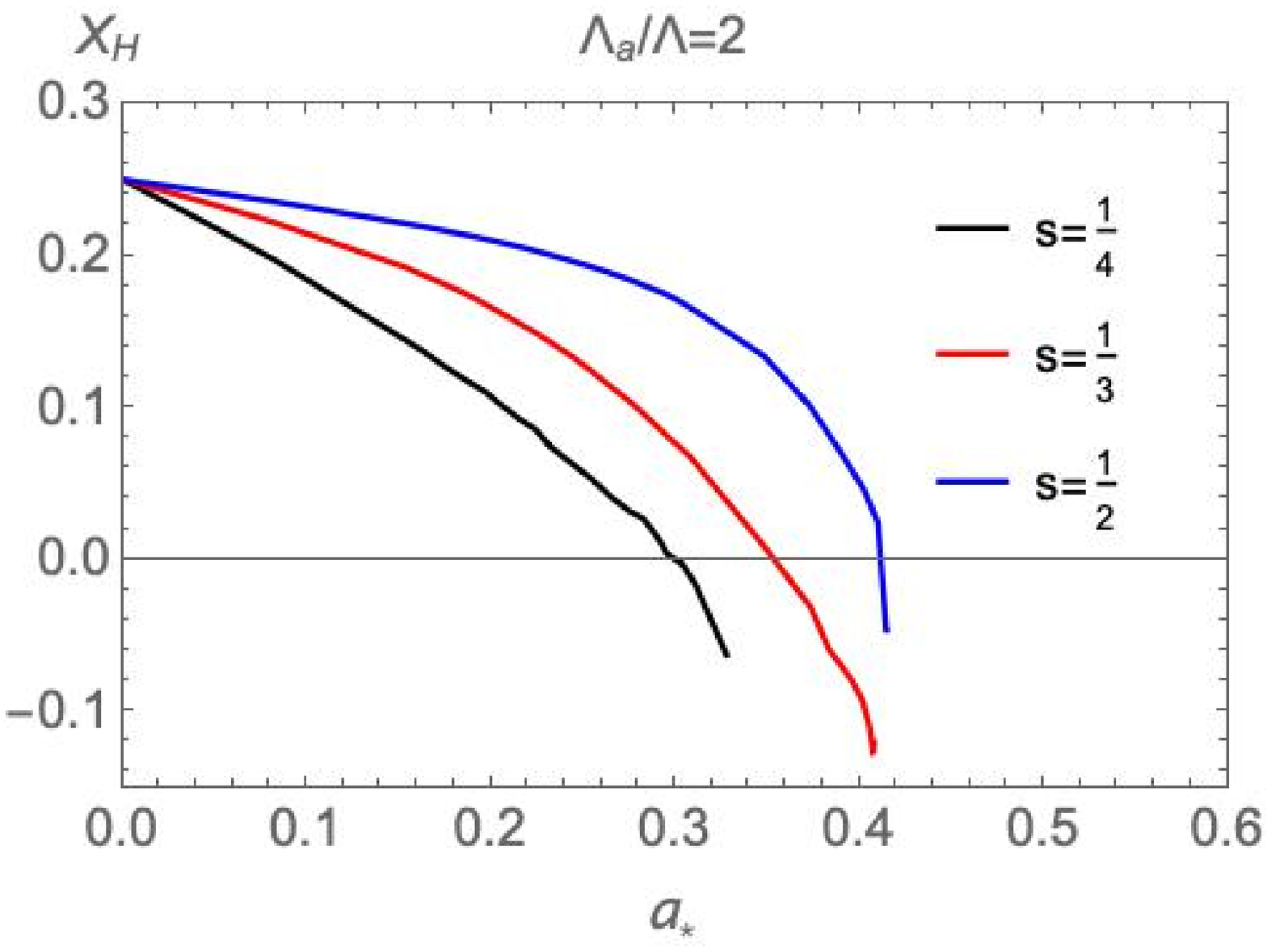}
 \end{center}
 \begin{center}
  \includegraphics[width=.45\textwidth]{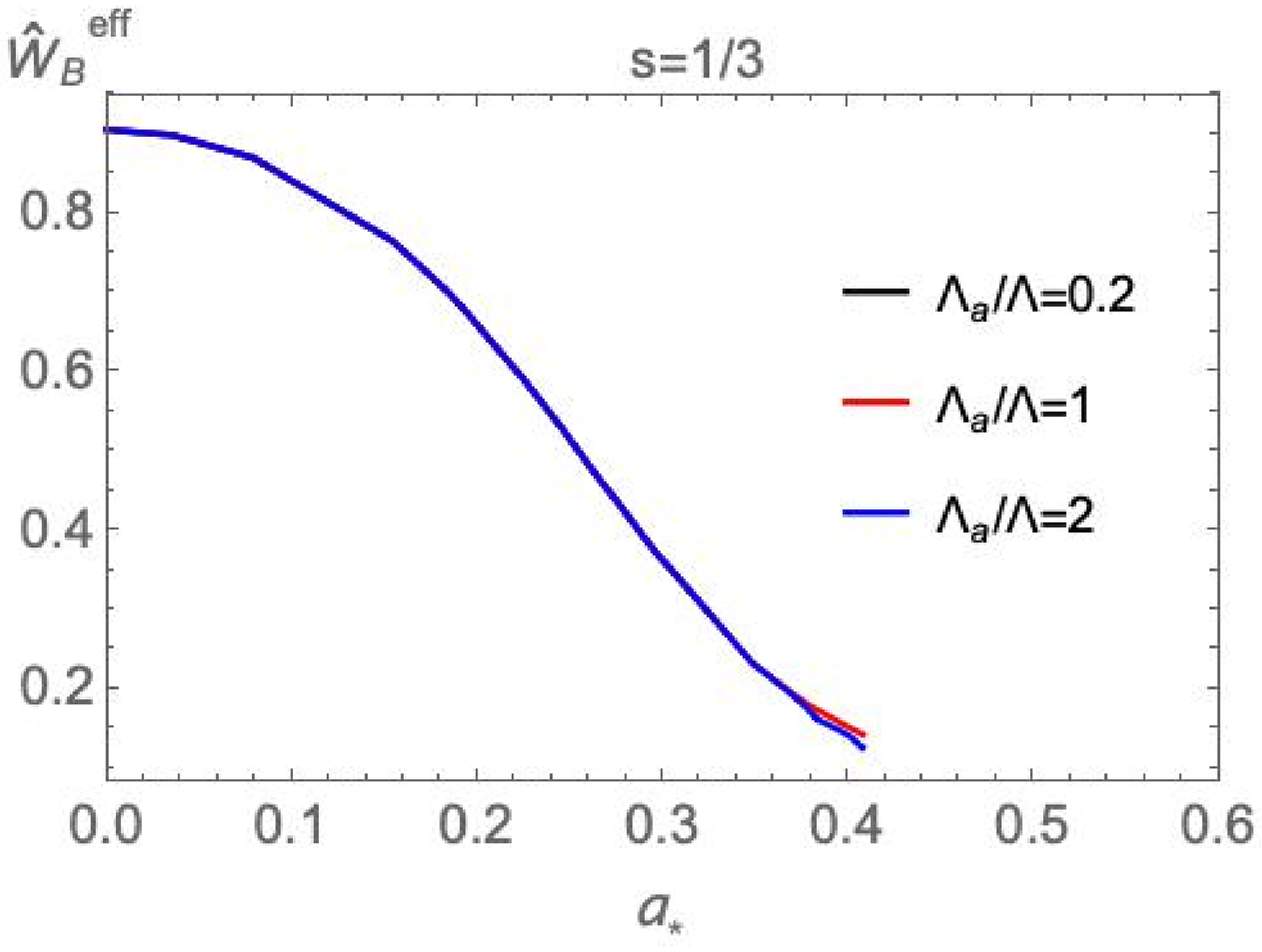}
   \includegraphics[width=.45\textwidth]{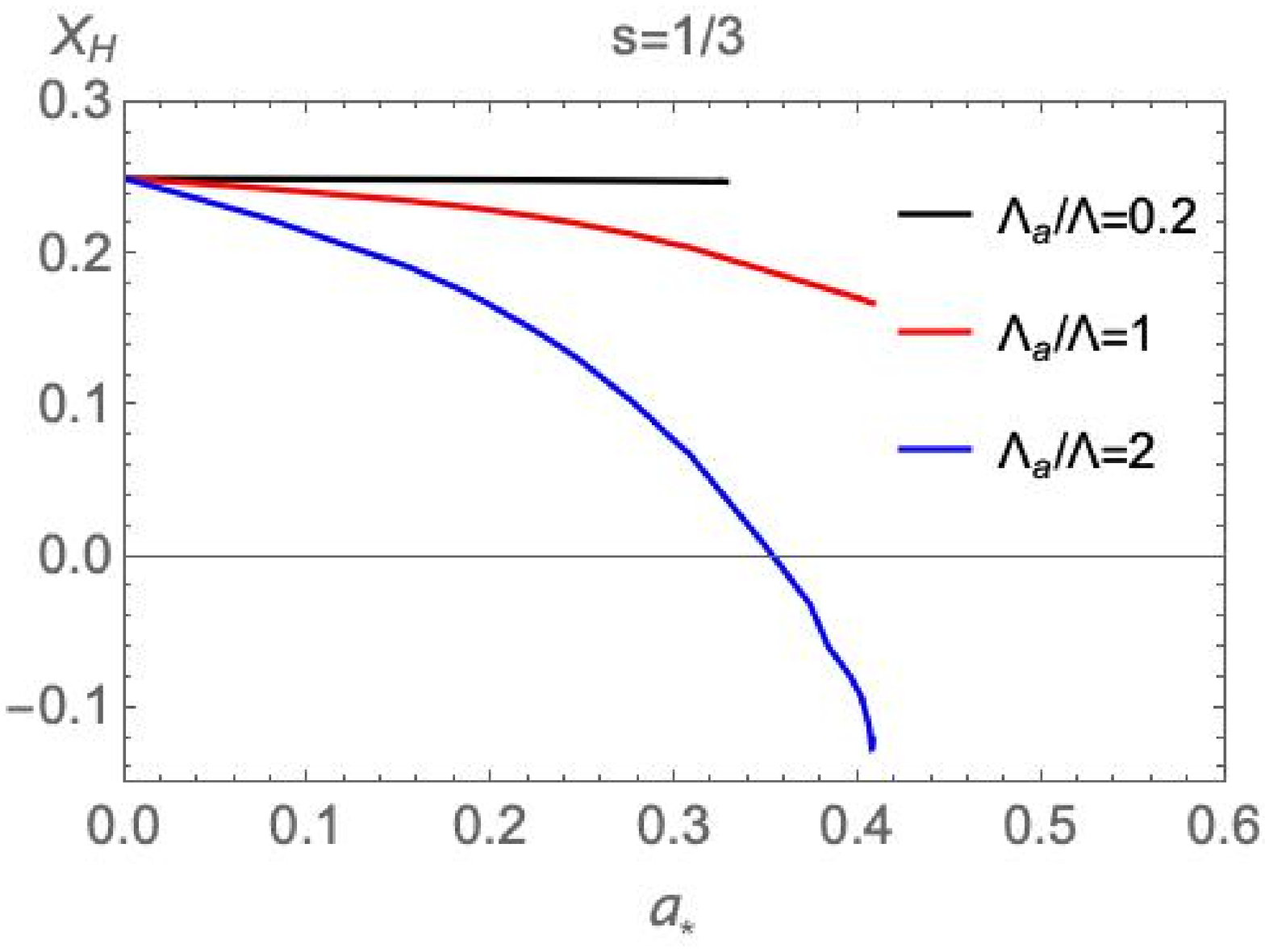}
 \end{center}
   \caption{
    Plots of $\hat{W}_B^{eff}$ (\textbf{left}) and $X_H$ (\textbf{right}) at the brane position vs.~$a_{\star}$ for a model with bulk functions \protect\eqref{Num1} and brane functions \protect\eqref{nth1}.
 The bulk parameters are \protect\eqref{nfull0}, and the brane parameters are \protect\eqref{nth3}, ${\rm sign}(Q_{IR})=-1$, $s=1/4, \, 1/3, \, 1/2$ and $\Lambda_a/\Lambda=0.2, \, 1, \, 2$.
   For top row panels, the brane position equation \protect\eqref{nfull6} with \protect\eqref{nfull8} have at most one solution for each $s$. For $s=1/2$, the solution exists only for $D_{IR}\lesssim1.5$.
   \textbf{Top row, left:} Plot of $\hat{W}_B^{eff}$ for $s=1/4, \, 1/3, \, 1/2$ and $\Lambda_a/\Lambda=2$.
   \textbf{Top row, right:} Plot of $X_H$ for $s=1/4, \, 1/3, \, 1/2$ and $\Lambda_a/\Lambda=2$.
   \textbf{Bottom row, left:} Plot of $\hat{W}_B^{eff}$ for $s=1/3$ and $\Lambda_a/\Lambda=0.2, \, 1, \, 2$. The various plots are near-indistinguishable.
   \textbf{Bottom row, right:} Plot of $X_H$ for $s=1/3$ and $\Lambda_a/\Lambda=0.2, \, 1, \, 2$.
}
  \label{fig34}
 \end{figure}

 \begin{figure}[t]
 \begin{center}
  \includegraphics[width=.45\textwidth]{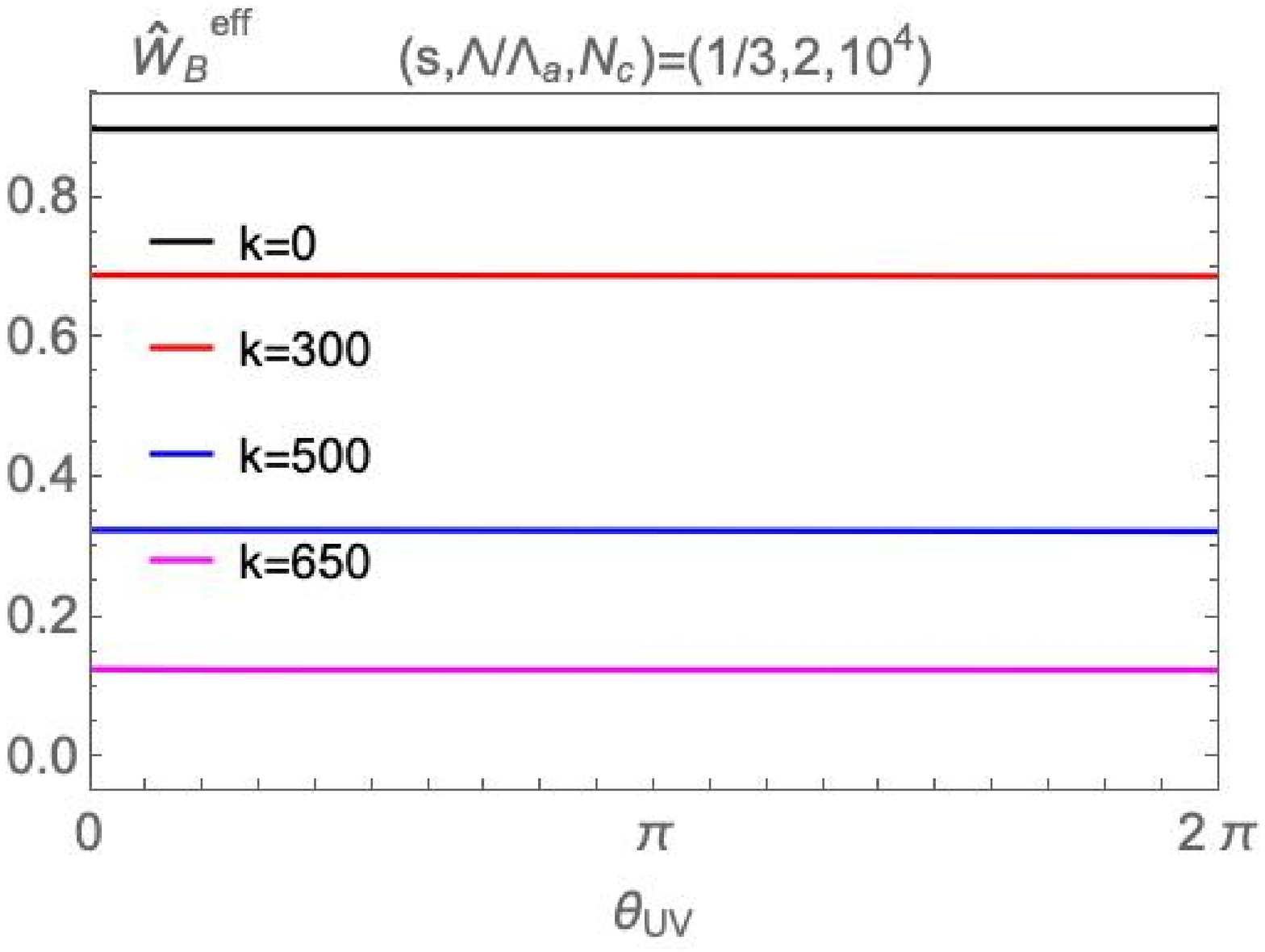}
   \includegraphics[width=.45\textwidth]{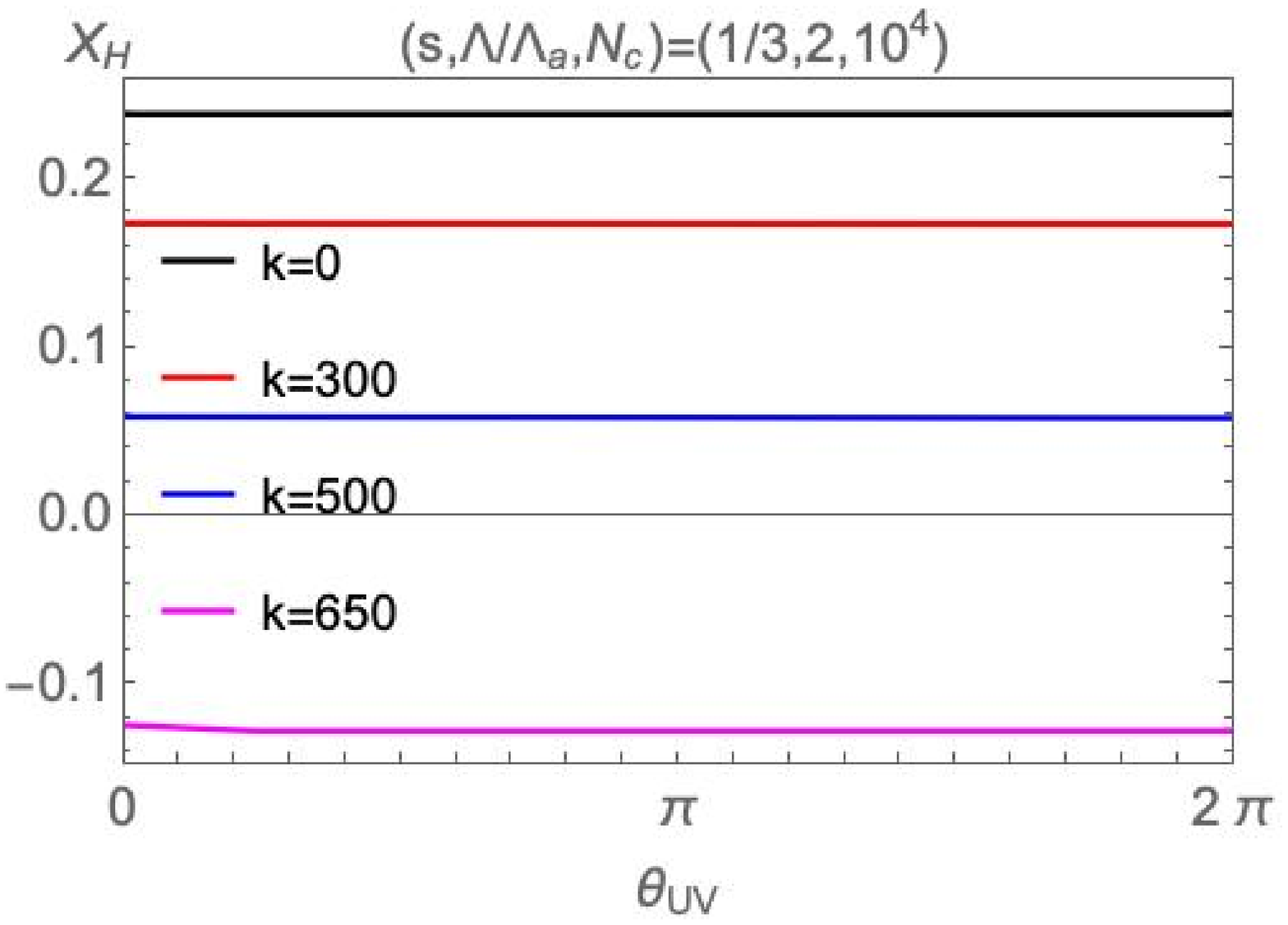}
 \end{center}
   \caption{
    Plots of $\hat{W}_B^{eff}$ (\textbf{left}) and $X_H$ (\textbf{right}) at the brane position vs.~$\theta_{UV}$ for a model with bulk functions \protect\eqref{Num1} and brane functions \protect\eqref{nth1}.
   The model parameters are \protect\eqref{nfull0}, \protect\eqref{nth3}, $s=1/3, \, \Lambda_a/\Lambda=2, \, N_c=10^4$. Here the branches for $k=0,\,300,\,500,\,650$ are shown.
}
  \label{fig34-2}
 \end{figure}

 \begin{figure}[t]
 \begin{center}
   \includegraphics[width=.5\textwidth]{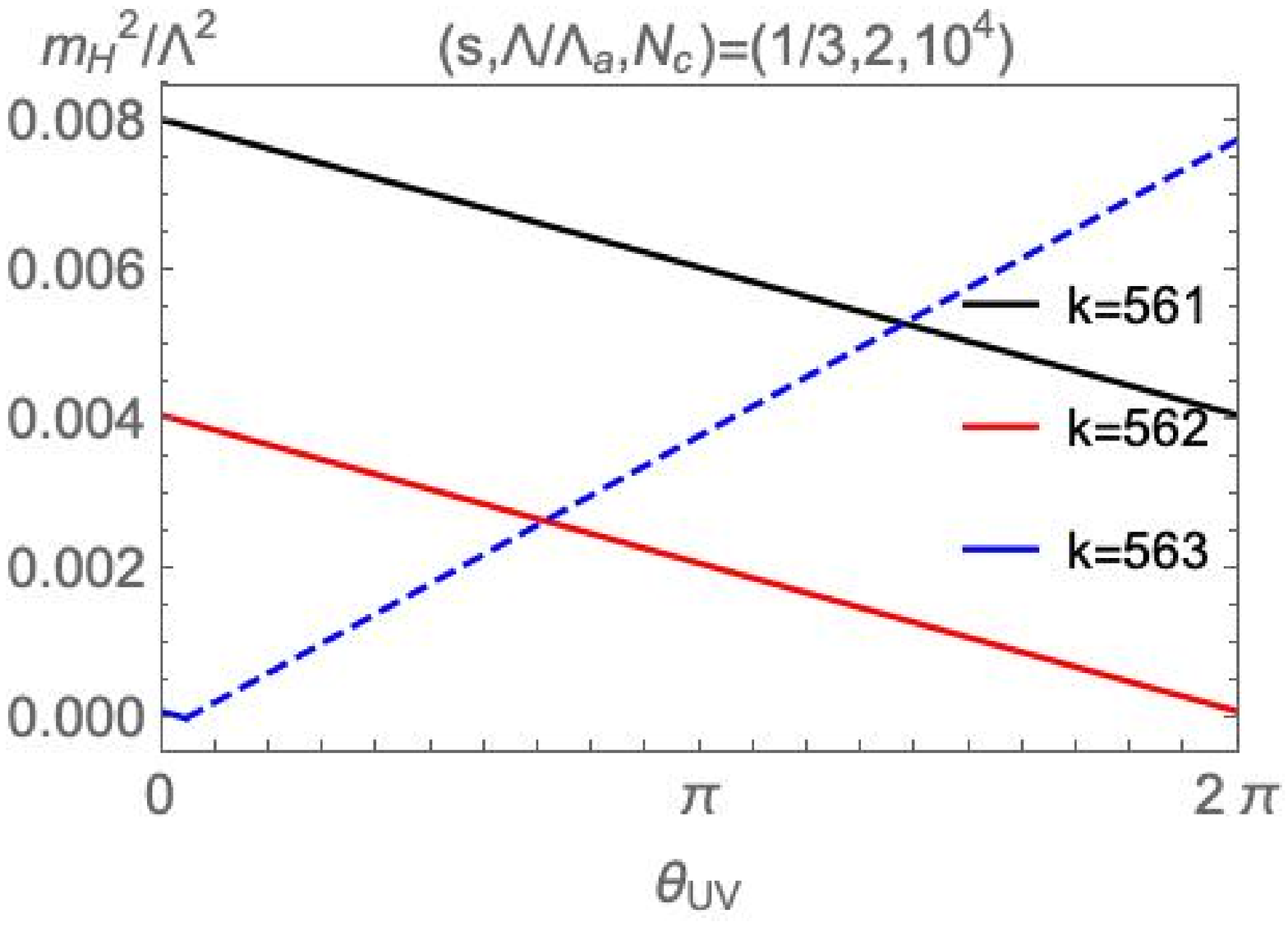}
 \end{center}
   \caption{
    Plot of the Higgs mass squared $m_H^2$ \protect\eqref{nth2} in units of $\Lambda^2$ vs.~$\theta_{UV}$ for a model with bulk functions \protect\eqref{Num1} and brane functions \protect\eqref{nth1}.
The model parameters are \protect\eqref{nfull0}, \protect\eqref{nth3}, $s=1/3, \, \Lambda_a/\Lambda=2, \, N_c=10^4$.
We plot the lines with $k=561, \,562, \,563$, corresponding to the branches realizing the small Higgs mass.
Note that, for $0\leq\theta_{UV}\lesssim 0.05\pi$, the symmetric and broken phase correspond to $k\leq563$ and $k\geq564$, respectively. For $0.05\pi\lesssim\theta_{UV}<2\pi$, the symmetric and broken phases correspond to $k\leq562$ and $k\geq563$. In the figure, the symmetric and broken phases are represented by the solid and dashed lines, respectively.
}
  \label{fig34-3}
 \end{figure}

 \begin{figure}[t]
 \begin{center}
  \includegraphics[width=.45\textwidth]{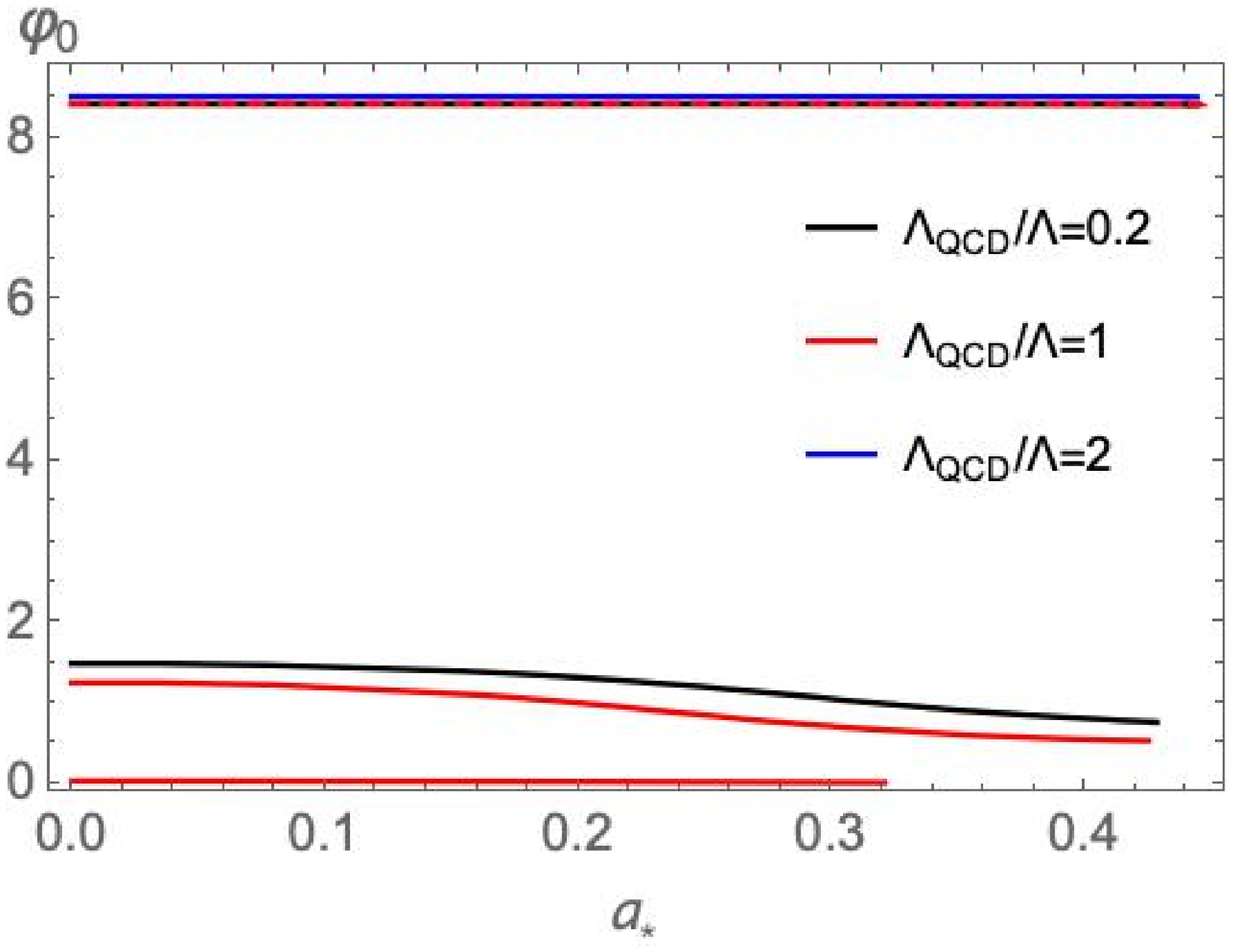}
   \includegraphics[width=.45\textwidth]{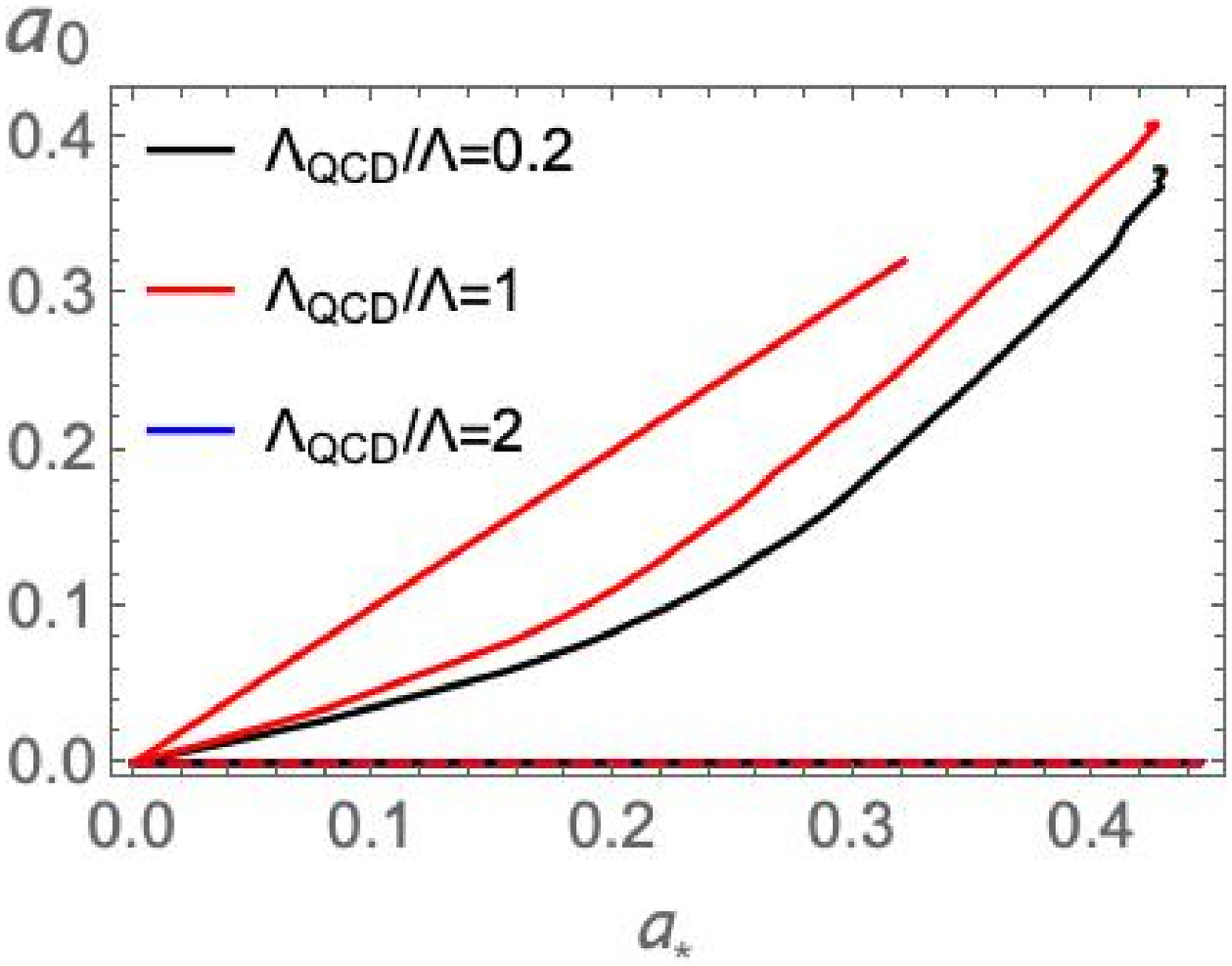}
 \end{center}
   \caption{
    Plots of $\f_0$ (\textbf{left}) and $a_0$ (\textbf{right}), i.e.~the values of $\f$ and $a$ at the brane position, vs.~$a_\star$ for a model with bulk functions \protect\eqref{Num1} and brane functions \protect\eqref{nf1}.
   The bulk parameters are \protect\eqref{nfull0}, and the brane parameters are \protect\eqref{nf2}, and $\Lambda_{QCD}/\Lambda=0.2, \, 1, \, 2$. There are two, three and one solutions of the brane position equation \protect\eqref{nfull6} with \protect\eqref{nfull8} for $\Lambda_{QCD}/\Lambda=0.2, \, 1, \, 2$, respectively. In the left (right) panel, the three lines for $\f_0\sim8.5$ ($a_0\sim0$) almost overlap one another.
}
  \label{fig38}
 \end{figure}

 \begin{figure}[t]
 \begin{center}
  \includegraphics[width=.45\textwidth]{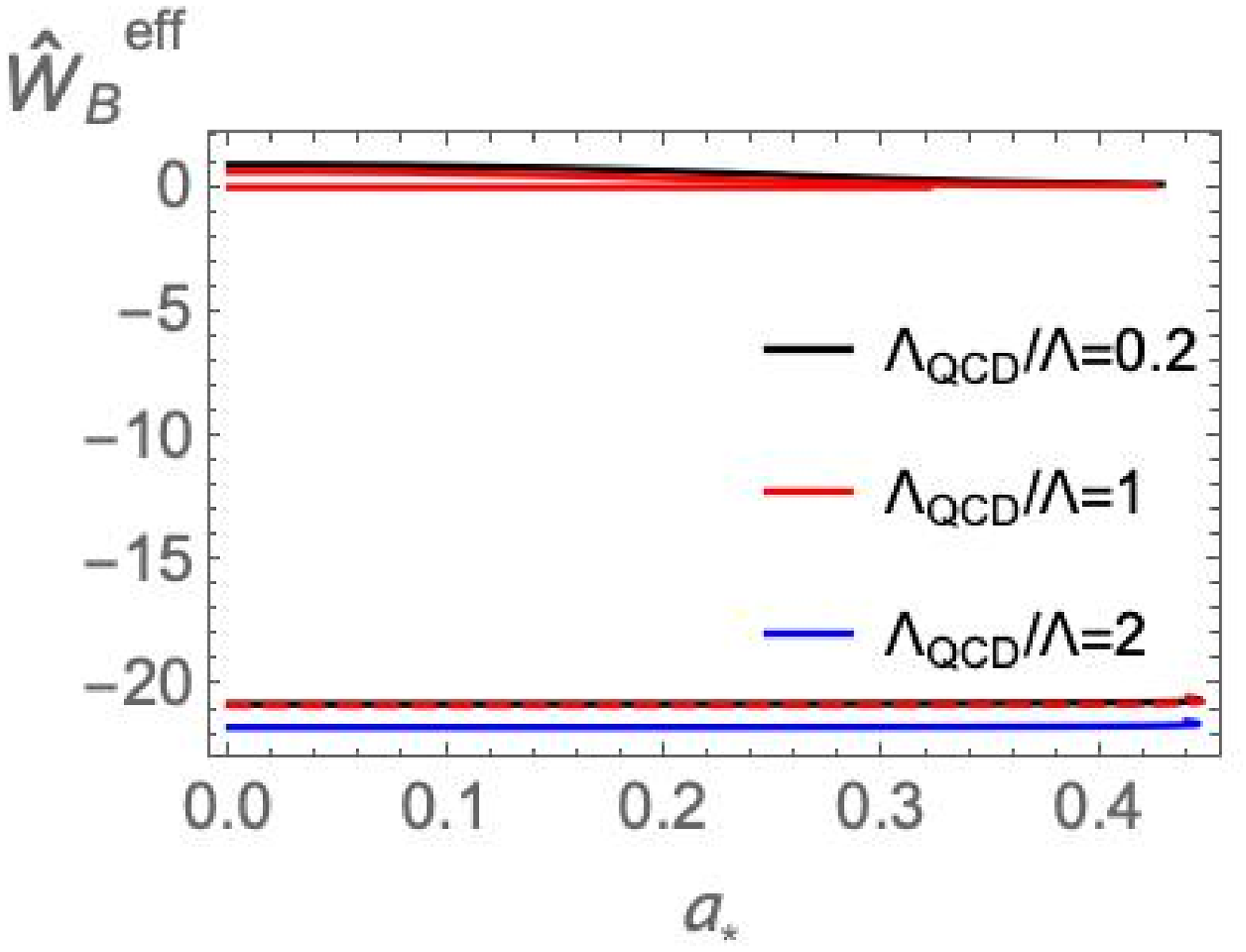}
   \includegraphics[width=.45\textwidth]{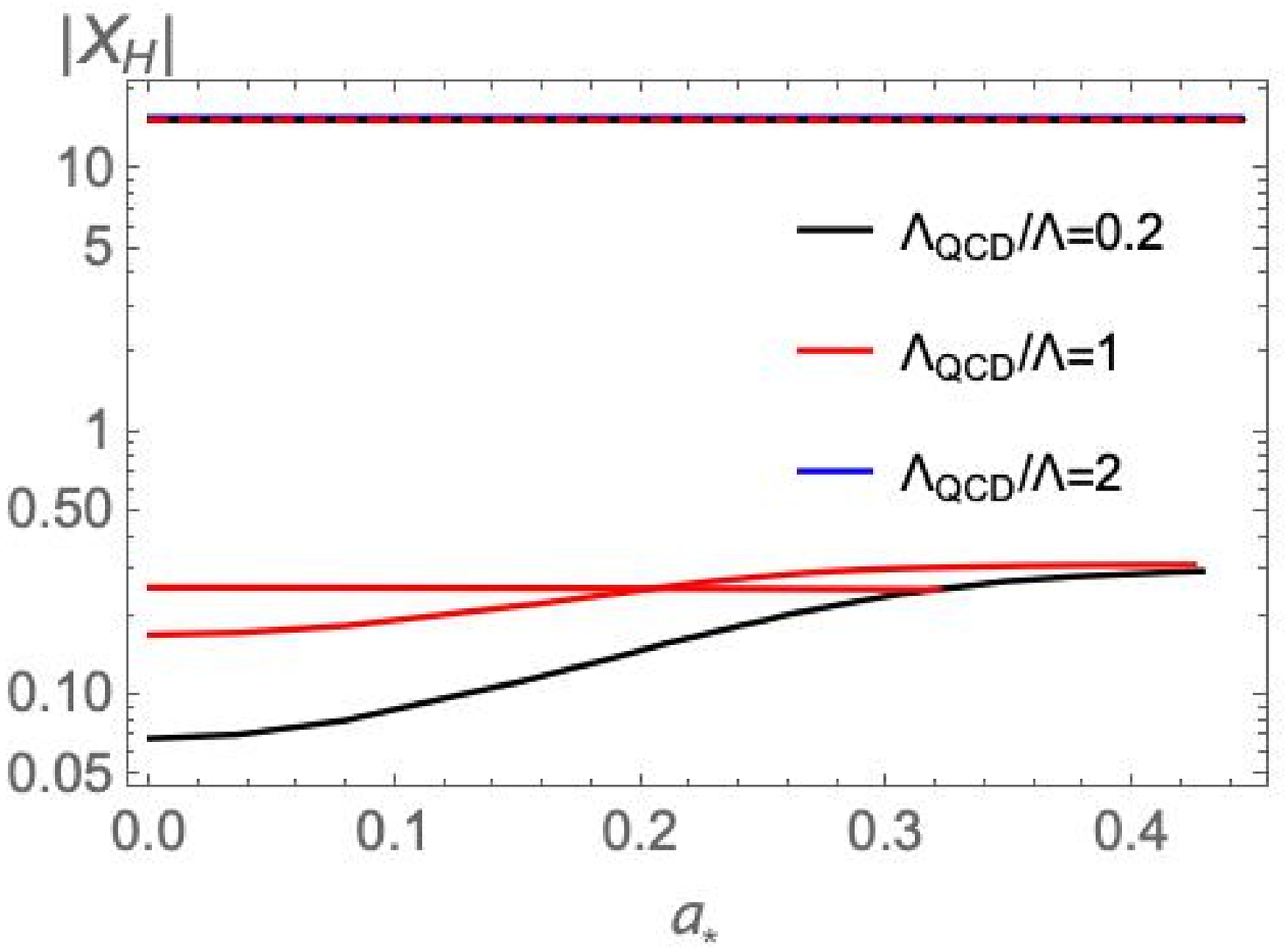}
 \end{center}
   \caption{
    Plots of $\hat{W}_B^{eff}$ (\textbf{left}) and $|X_H|$ (\textbf{right}) at the brane position vs.~$a_\star$ for a model with bulk functions \protect\eqref{Num1} and brane functions \protect\eqref{nf1}.
   The bulk parameters are \protect\eqref{nfull0}, and the brane parameters are \protect\eqref{nf2}, and $\Lambda_{QCD}/\Lambda=0.2, 1, 2$. There are two, three and one solutions of the brane position equation \protect\eqref{nfull6} with \protect\eqref{nfull8} for $\Lambda_{QCD}/\Lambda=0.2, \, 1, \, 2$, respectively.
   In the left panel, three lines $\hat{W}_B^{eff}\sim0$ almost overlap one another while two lines $\hat{W}_B^{eff}\sim-21$ also overlap one another. In the right panel, three lines at $|X_H|\sim 15$ almost overlap one another.
}
  \label{fig36}
 \end{figure}

 \begin{figure}[t]
 \begin{center}
  \includegraphics[width=.45\textwidth]{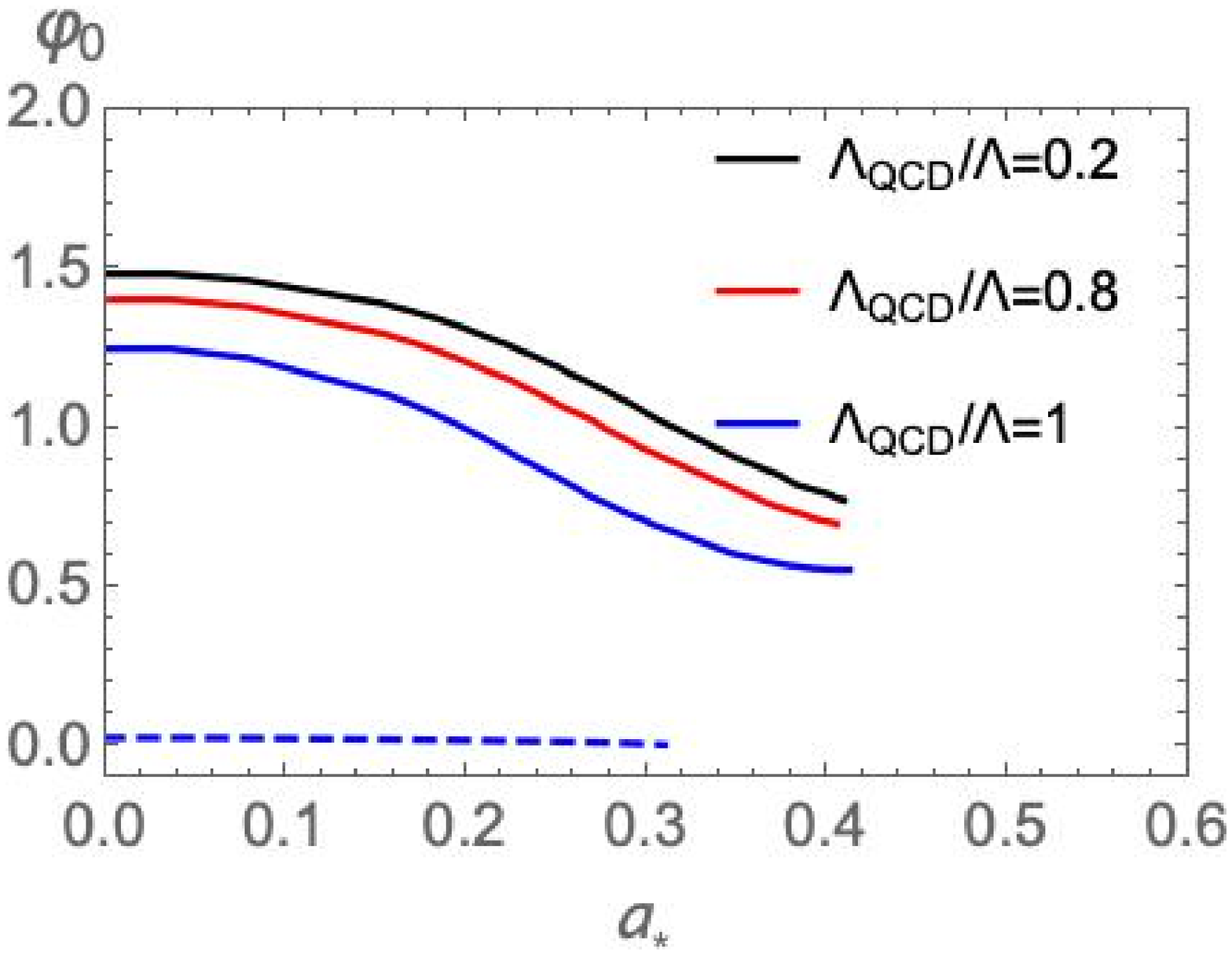}
   \includegraphics[width=.45\textwidth]{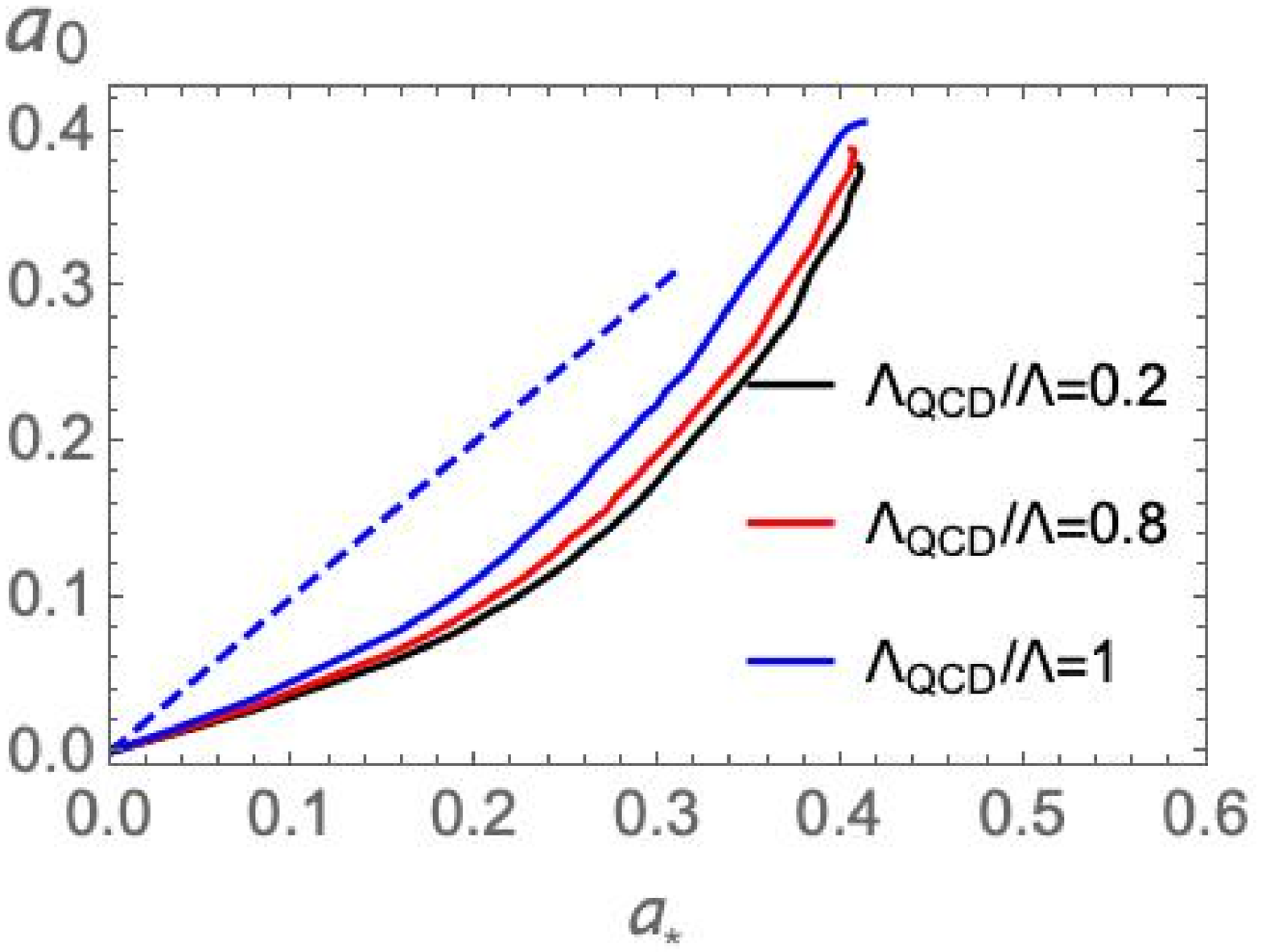}
 \end{center}
   \caption{
    Plots of $\f_0$ (\textbf{left}) and $a_0$ (\textbf{right}), i.e.~the values of $\f$ and $a$ at the brane position, vs.~$a_\star$ for a model with bulk functions \protect\eqref{Num1} and brane functions \protect\eqref{nfive1}.
   The bulk parameters are \protect\eqref{nfull0}, and the brane parameters are \protect\eqref{nfive2} and $\Lambda_{QCD}/\Lambda=0.2, \, 0.8, \, 1$. For $\Lambda_{QCD}/\Lambda=0.2$, there are two solutions of \protect\eqref{nfull6} with \protect\eqref{nfull8}, which we denote by the solid and dashed lines.
}
  \label{fig39}
 \end{figure}

 \begin{figure}[t]
 \begin{center}
  \includegraphics[width=.45\textwidth]{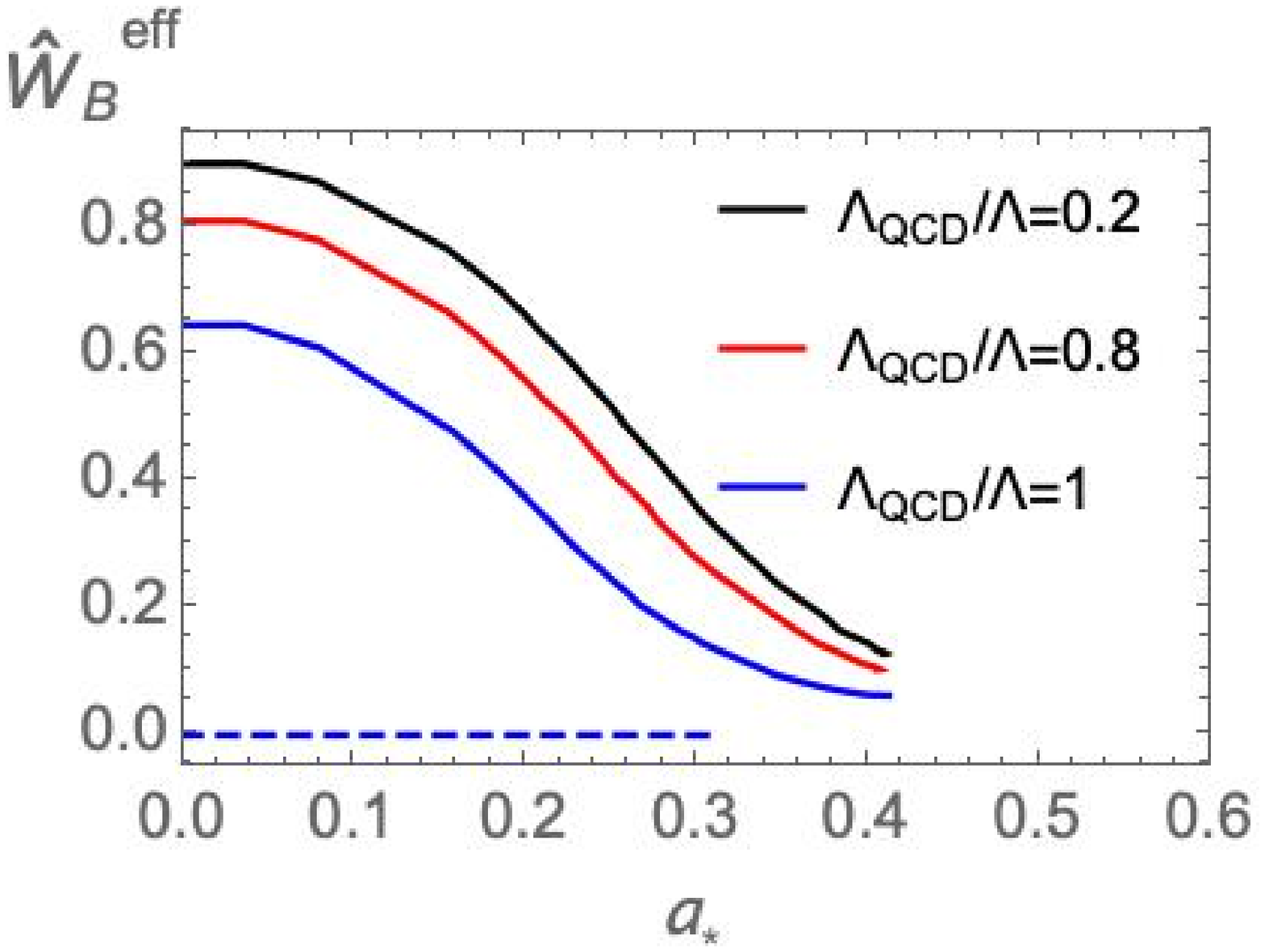}
   \includegraphics[width=.45\textwidth]{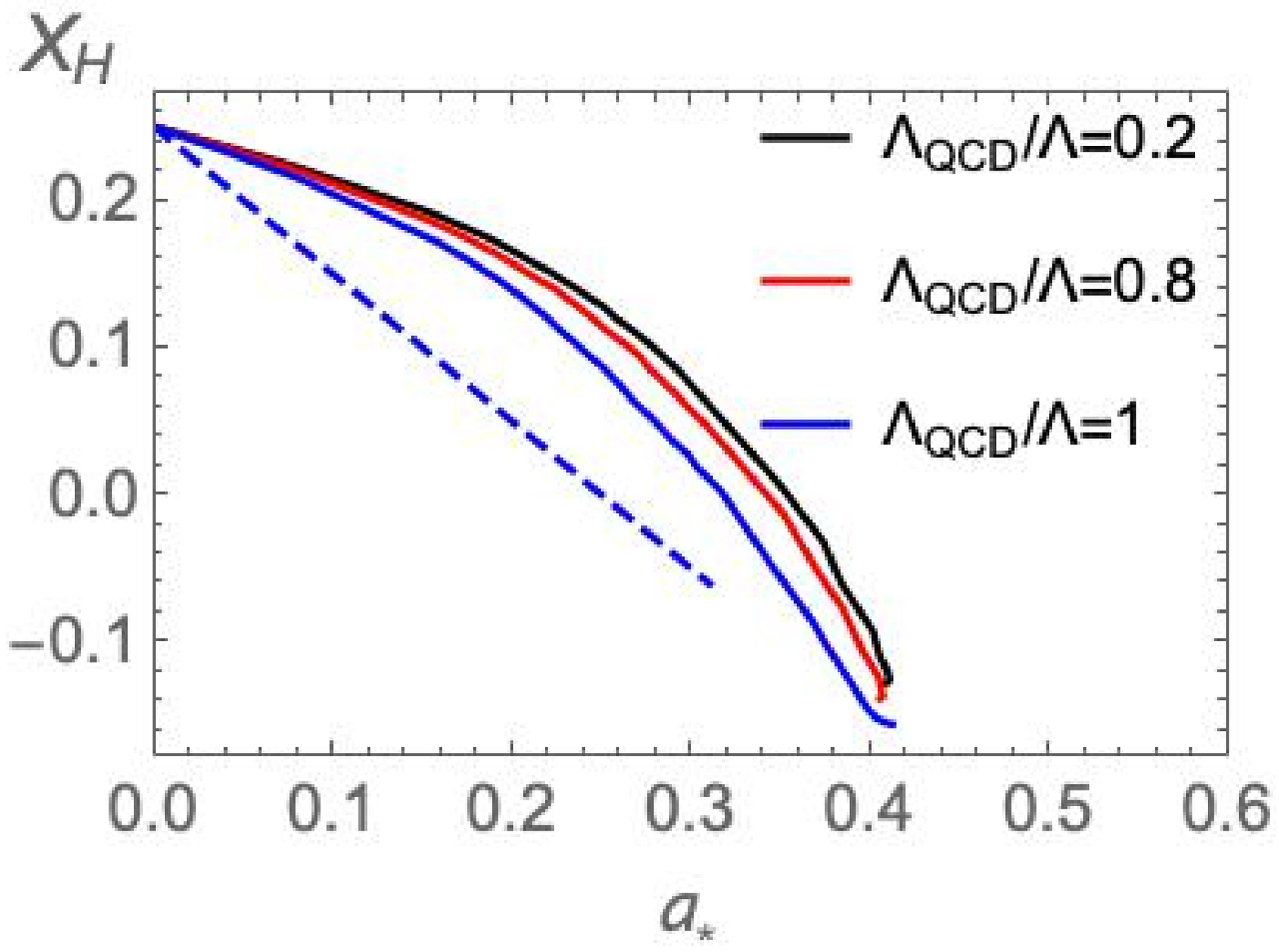}
 \end{center}
   \caption{
    Plots of $\hat{W}_B^{eff}$ (\textbf{left}) and $X_H$ (\textbf{right}) at the brane position vs.~$a_\star$ for a model with bulk functions \protect\eqref{Num1} and brane functions \protect\eqref{nfive1}.
   The bulk parameters are \protect\eqref{nfull0}, and the brane parameters are \protect\eqref{nfive2} and parameters.
   For $\Lambda_{QCD}/\Lambda=0.2$, there are two solutions of \protect\eqref{nfull6} with \protect\eqref{nfull8}, which we denote by the solid and dashed lines.
}
  \label{fig40}
 \end{figure}

 \begin{figure}[t]
 \begin{center}
  \includegraphics[width=.45\textwidth]{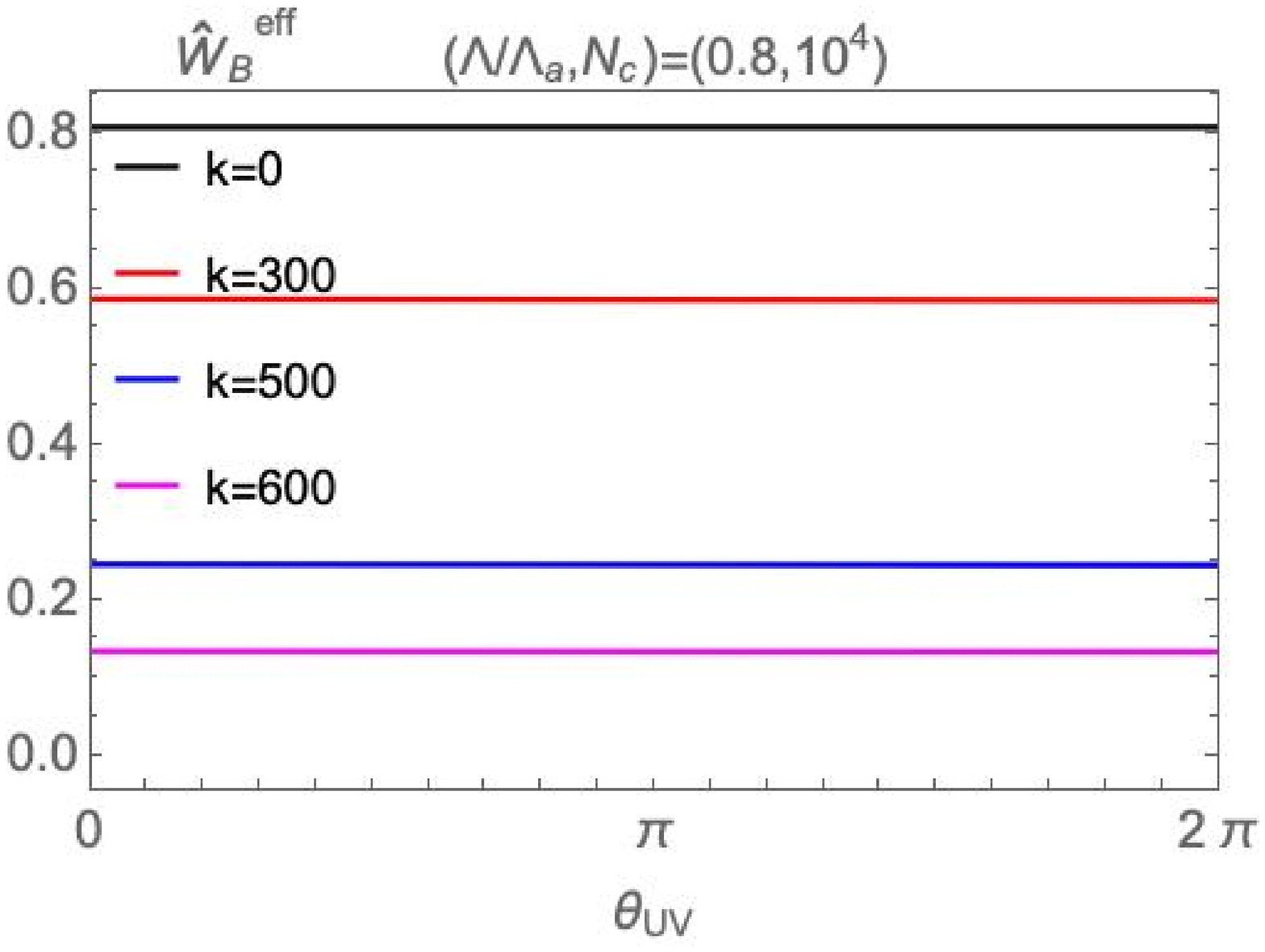}
   \includegraphics[width=.45\textwidth]{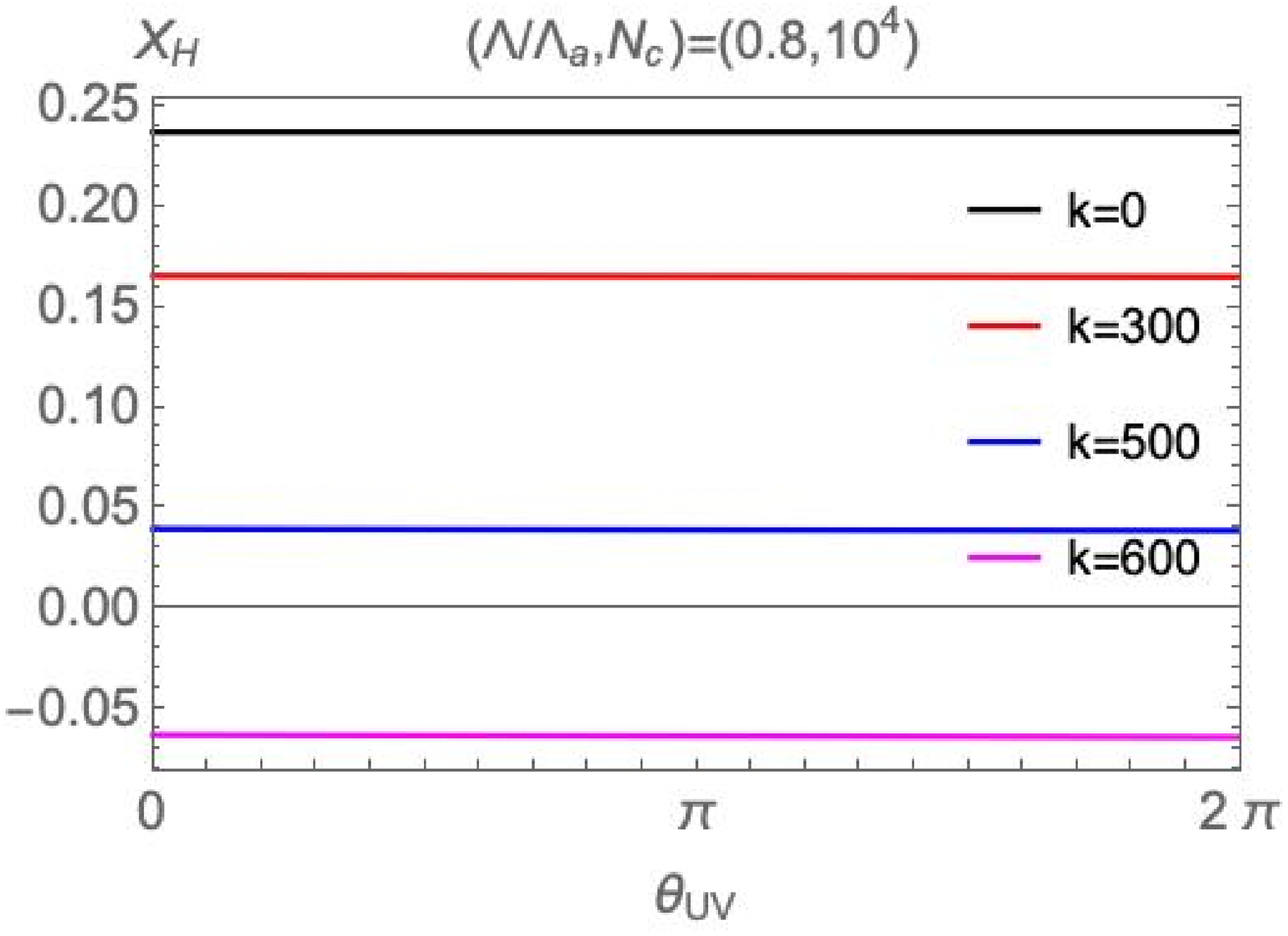}
 \end{center}
   \caption{
    Plots of $\hat{W}_B^{eff}$ and $X_H$ at the brane position vs.~$\theta_{UV}$ for a model with bulk functions \protect\eqref{Num1} and brane functions \protect\eqref{nfive1}.
   The bulk parameters are \protect\eqref{nfull0}, and the brane parameters are \protect\eqref{nfive2}, $\Lambda_{QCD}/\Lambda=0.8$, and $N_c=10^4$.
}
  \label{fig41}
 \end{figure}

  \begin{figure}[t]
 \begin{center}
  \includegraphics[width=.5\textwidth]{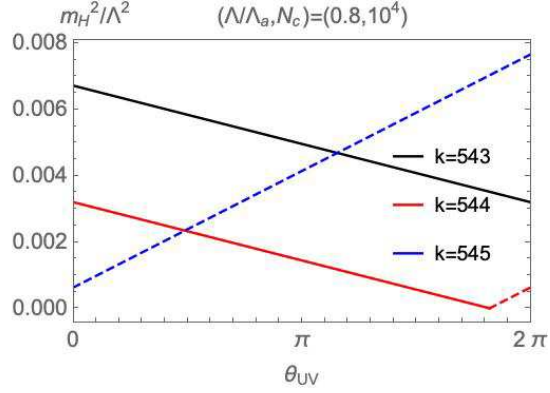}
 \end{center}
   \caption{
    Plot of the Higgs mass squared $m_H^2$ \protect\eqref{nth2} at the brane position vs.~$\theta_{UV}$ for a model with bulk functions \protect\eqref{Num1} and brane functions \protect\eqref{nfive1}.
The model parameters are \protect\eqref{nfull0}, \protect\eqref{nfive2}, $\Lambda_{QCD}/\Lambda=0.8$, and $N_c=10^4$.
We plot the lines with $k=543,\, 544, \, 545$, corresponding to the branches realizing the small Higgs mass.
In the figure, the symmetric and broken phases are represented by the solid and dashed lines, respectively.
}
  \label{fig41-2}
 \end{figure}

\subsection{Brane potential choice $2$: linear axion dependence of Higgs mass parameter} \label{num2}
Here we make the following choice for the brane potentials in \eqref{A5}:
\be
W_B=
\frac{\Lambda^4}{M_p^3} \left[ -1 -{\f \over s} + \left(\f \over s\right)^2 \right]
=\tilde{\Lambda}^4 \left[ -1 -{\f \over s} + \left(\f \over s\right)^2 \right] \sp
\label{nth1}\ee
$$
X_H = \frac{\Lambda^2}{M_p} \left(1 - \frac{\Lambda_a^2}{\Lambda^2} a \right)
= M_p^{1/2} \tilde{\Lambda}^2 \left(1 - \frac{\Lambda_a^2}{\Lambda^2} a \right)\sp
S_H = M_p,
$$
where $\Lambda_a$ is a constant.
The bulk potential and axion kinetic function are still given by \eqref{Num1} with parameters \eqref{nfull0}. The main difference with respect to the scenario examined in section \ref{num1} is the explicit axion dependence of the Higgs mass function $X_H$. Here we consider a linear dependence on the axion as in the original relaxion scenario \cite{relaxion} or as realised in the string-inspired setting of \cite{Ibanez:2015fcv}.

We further choose
\be
\tilde{\Lambda}=0.5,
\label{nth3}\ee
and vary $s$ and $\Lambda_a/\Lambda$. We take the following as representative values:
\be\label{choice2}
s = {1\over 4},\; {1\over 3},\;  {1\over 2}, \qquad {\Lambda_a \over \Lambda} = 0.2, \; 1, \; 2.
\ee

We again derive the brane position by solving \eqref{nfull6} and \eqref{nfull8}.
As in section \ref{num1} we can plot the various brane quantities as functions of $a_\star$. In fig.~\ref{fig33} we hence display $\f_0, \, a_0$ vs.~$a_\star$, while in fig.~\ref{fig34} we show the brane cosmological constant $\hat{W}_B^{eff}$ and the Higgs mass parameter $X_H$ as functions of $a_\star$. The top row panels in figures \ref{fig33}, \ref{fig34} correspond to $\Lambda_a/\Lambda=2$ with $s=1/4, \, 1/3, \, 1/2$. There is only one branch of solutions to the junction conditions for every parameter choice here. The bottom row panels in figures \ref{fig33}, \ref{fig34} correspond to $\Lambda_a/\Lambda=0.2, \, 1, \, 2$ with $s=1/3$.

We are mainly interested in the possibility of solutions with a small Higgs mass. Recall that such solutions generically exist if $X_H$ as a function of $a_\star$ exhibits a sign change (see the discussion in section \ref{num1}).
Here we find that this is the case for all solutions we considered with $\Lambda_a/\Lambda = 2$. Overall, we observe that for the brane functions considered here it is much easier to find solutions where $X_H(a_\star)$ changes sign compared to the choice for the brane functions in sec.~\ref{num1}. That is, an explicit axion-dependence in $X_H$ as in \eqref{nth1} is advantageous for finding solutions with a  small Higgs mass.

Using \eqref{i2} we can again write a function of $a_\star$ as a multi-branched function of $\theta_{UV}$ as we have done in sec.~\ref{num1}. To be specific, we pick the example with model parameters $s=1/3,  \, \Lambda_a/ \Lambda=2$ which exhibits a sign change in $X_H(a_\star)$.  In figure \ref{fig34-2} we then plot $\hat{W}_B^{eff}$ and $X_H$ as functions of $\theta_{UV}$ for this model, further choosing $c=1$ and $N_c=10^4$. For better visibility we only show the branches with $k=0,\,300,\,500,\,650$. Then, for this example, we can find that the branches with the lowest value of the Higgs mass can be found for $k\sim562$. In figure \ref{fig34-3} we plot of ratio of Higgs mass squared $m_H^2$ defined in \eqref{nth2} and the scale $\Lambda$ for the branches with $k= 561, 562, 563$. In the figure, solid and dashed lines correspond to solutions with intact and broken EW symmetry, respectively. For the branches displayed one finds $m_H^2/\Lambda^2= {\cal O}(10^{-3})$, i.e.~the solutions exhibit a hierarchy between $|m_H|$ and the UV cutoff scale on the brane $\Lambda$. The numerical separation between $|m_H|$ and $\Lambda$ can be further increased by choosing a larger value for $N_c$.

\subsection{Brane potential choice $3$: axion cosine potential} \label{num3}
Once more, the bulk potential and axion kinetic function are given by \eqref{Num1} with parameter choice \eqref{nfull0}.
For the brane potentials \eqref{A5} we now take
\begin{align}
\nonumber W_B &= \frac{1}{M_p^3}\left\{
\Lambda^4 \left[ -1 -{\f \over s} + \left(\f \over s\right)^2 \right] + \Lambda_{QCD}^4 \cos(a) \right\} \\
&= \tilde{\Lambda^4} \left[ -1 -{\f \over s} + \left(\f \over s\right)^2 + \frac{\Lambda_{QCD}^4}{\Lambda^4} \cos(a)\right] ,
\label{nf1}\end{align}
$$
X_H= \frac{\Lambda^2}{M_p} \left[ 1 +{\f \over s_X} - \left(\f \over s_X\right)^2 \right]
= M_p^{1/2} \tilde{\Lambda}^2 \left[ 1 +{\f \over s_X} - \left(\f \over s_X\right)^2 \right] \sp
S_H= M_p,
$$
where $\Lambda_{QCD}$ is a dimension one parameter. That is, here we revert to an axion-independent Higgs mass parameter $X_H$ as in section \ref{num1}, but now the brane potential $W_B$ is given a periodic dependence on $a$ as observed in the instanton-generated potential axion potential in QCD.
We take
\be
s=1/3, \quad s_X=1, \quad \tilde{\Lambda}=0.5,
\label{nf2}\ee
and vary $\Lambda_{QCD}/\Lambda$. As representative values, we take
\be\label{choice3}
{\Lambda_{QCD} \over \Lambda} = 0.2,\;  1, \;  2.
\ee


The dilaton and axion field value at the brane position as functions of $a_\star$ are plotted in figure \ref{fig38} with the parameters \eqref{nfull0}, \eqref{nf2}, and $\Lambda_{QCD}/\Lambda=0.2, \, 1, \, 2$.
For the same parameter set, the brane cosmological constant $\hat{W}_B^{eff}$ and the absolute value of Higgs mass parameter $|X_H|$ are shown in figure \ref{fig36} as functions of $a_\star$.
There are two, three and one solutions of the brane position equation \eqref{nfull6} with \eqref{nfull8} for $\Lambda_{QCD}/\Lambda=0.2, \, 1, \, 2$, respectively.
The brane cosmological constant $\hat{W}_B^{eff}$ is generically of the order $ \sim \tilde{\Lambda}^4$. For the parameter values chosen here $X_H$ does not change sign as a function of $a_\star$, nor does it closely approach zero anywhere.
For the (generic) parameter choices considered here, we are not able to find solutions with a small Higgs mass. Therefore, we conclude that the sinusoidal axion dependence of the brane potential $W_B$ is generically not helpful for the existence of saddle points with a small Higgs mass.

\subsection{Brane potential choice $4$: $\Lambda_{QCD}+\Lambda_a$} \label{num4}
In the final numerical example we include both a periodic axion dependence in $W_B$ as in section \ref{num3} while at the same time allowing for a linear dependence of $X_H$ on $a$ as in section \ref{num2}. Hence here the brane potentials are given by
\begin{align}
\nonumber W_B &= \frac{1}{M_p^3}\left\{
\Lambda^4 \left[ -1 -{\f \over s} + \left(\f \over s\right)^2 \right] + \Lambda_{QCD}^4 \cos(a)\right\} \\
&= \tilde{\Lambda}^4 \left\{
\left[ -1 -{\f \over s} + \left(\f \over s\right)^2 \right] + \frac{\Lambda_{QCD}^4}{\Lambda^4} \cos(a)\right\},
\label{nfive1}\end{align}
$$
X_H= \frac{\Lambda^2}{M_p} \left(1 - \frac{\Lambda_a^2}{\Lambda^2} a \right),
\quad
S_H = M_p.
$$
with
\be
s=1/3, \quad \tilde{\Lambda}=0.5, \quad \Lambda_a/\Lambda=2.
\label{nfive2}\ee
As in the previous subsection, we vary $\Lambda_{QCD}/\Lambda$ and we show results for the representative values:
\be\label{choice4}
{\Lambda_{QCD} \over \Lambda} = 0.2,\;  0.8, \;  1.
\ee

In figure \ref{fig39} we plot the dilaton and axion at the brane locus, $\f_0, \, a_0$, as functions of $a_\star$ for $\Lambda_{QCD}/\Lambda =0.2, 0.8, 1$. In figure \ref{fig40} we plot the corresponding values for $\hat{W}_B^{eff}$ and $X_H$ at the brane locus as functions of $a_\star$. Note that modifying $\Lambda_{QCD} / \Lambda$ does not affect the solutions much, as $\f_0, a_0$ in fig.~\ref{fig40} or $\hat{W}_B^{eff}, X_H$ in fig.~\ref{fig41} do not differ significantly for different values of $\Lambda_{QCD} / \Lambda$. Hence the effect of the $\cos(a)$-term in $W_B$ on the overall solution is fairly mild. However, from fig.~\ref{fig40} we observe that for all parameter choices considered the Higgs mass parameter as a function of $a_\star$ exhibit a sign change. This can be traced back to the linear axion-dependence of $X_H$ as in section \ref{num2}, which facilitates the appearance of a sign change in $X_H(a_\star)$. As observed previously, this sign change will guarantee the existence of saddle points with a small Higgs mass, which therefore arise generically in the model considered here.

Focussing on the solution with $\Lambda_{QCD}/\Lambda =0.8$ we once more rewrite $\hat{W}_B^{eff}$ and $X_H$ as multi-branched functions of $\theta_{UV}$ with the help of \eqref{i2}. Here we choose $c=1$ and $N_c=10^4$. The corresponding plots can be seen in fig.~\ref{fig41} where for visibility we only display the branches for $k=0,300,500,600$. For this solution the minimal value for the Higgs mass squared is observed on the branches with $k \sim 544$. In fig.~\ref{fig41-2} we plot of ratio of Higgs mass squared $m_H^2$ defined in \eqref{nth2} and the scale $\Lambda$ defined in \eqref{nfive2} for the branches with $k= 543, 544, 545$. In the figure, solid and dashed lines correspond to solutions with intact and broken EW symmetry, respectively. For the branches displayed one finds $m_H^2/\Lambda^2= {\cal O}(10^{-3})$ for generic $\theta_{UV}$, i.e.~the solutions exhibit a hierarchy between $|m_H|$ and the UV cutoff scale on the brane $\Lambda$. The numerical separation between $|m_H|$ and $\Lambda$ can be further increased by choosing a larger value for $N_c$.

\subsection{Summary of the section}

Our goal in this section has been to numerically check the existence of vacua which realize the self-tuning of the cosmological constant as well as a large electroweak hierarchy. A small Higgs mass is obtained if the Higgs mass parameter $X_H$ as a function of $a_\star$ exhibits at least one zero. If this happens, we expect that, around this value, we will have electroweak symmetry breaking with a small Higgs mass. Then, for any value of $\theta_{UV}$, as long as there exist branches that satisfy
\be
c \frac{\theta_{UV}+2\pi k}{N_c} \approx a_{\star,0}
\ee
these correspond to vacua with a small Higgs mass.
We performed a numerical analysis for static solutions of our brane-bulk system for four types of the brane potentials (\ref{nfull1}, \ref{nth1}, \ref{nf1}, \ref{nfive1}).
Here we summarize the main observations from this section.

\begin{itemize}

\item In section \ref{num1}, we used brane potentials \eqref{nfull1} which only depend on the bulk scalar $\f$, but not the axion $a$. In this case the effect of axion backreaction on brane quantities only enters via a shift in the value of the brane position $\f_0$.
We observed that $\f_0$ typically exhibits ${\cal O}(1)$ shifts when we take into account the axion backreaction (Top row, left panel in figure \ref{fig22}).
Correspondingly, the effective Higgs potential on the brane changes.
In particular, we find that (for certain choices of model parameters) there exist solutions where the Higgs mass parameter $X_H$ crosses zero as a function of the axion source. In this case, we find that there exist saddle points which allow for a  small Higgs mass (figure \ref{fig24-2}).

\item In section \ref{num2}, we used a brane potential \eqref{nth1}, where the Higgs mass function $X_H$ depends linearly on the axion $a$.
If this linear coupling is large enough, by increasing the value of the axion source, the Higgs mass parameter $X_H$ generically changes sign as a function of the axion source (figure \ref{fig34}). The model is reminiscent of the relaxion scenario whose mechanism will be compared to our setup in section \ref{gauge_hierarchy}.

\item In section \ref{num3}, we used brane potentials where the brane cosmological constant depends on the axion as $\sim \cos(a)$, with a coefficient of the order of SM scales, see \eqref{nf1}. This is the standard QCD-instanton-generated potential. The effect of $\cos(a)$ on the dynamics is mild. It only slightly changes the brane position $\f_0$. For the parameter choices given in section \ref{num3}, we do not obtain a sign change of $X_H$. All branches parameterized by $k$ have the same sign for $X_H$.

\item In section \ref{num4}, we used the brane potentials \eqref{nfive1}. Here the Higgs mass function depends linearly on the axion $a$, and the brane cosmological constant is proportional to $\cos(a)$. This is a combination of the ans\"atze in \ref{num2} and \ref{num3}. As in section \ref{num2}, we can easily obtain solutions with a ``small''  Higgs mass (figure \ref{fig41-2}). The effect of $\cos(a)$ on the brane cosmological constant and Higgs mass is not large even if we take $\Lambda_{QCD}$ to be of the order of the cutoff $\Lambda$ (figure \ref{fig40}).

\end{itemize}

\section{The gauge hierarchy problem and outlook} \label{gauge_hierarchy}

A formulation of solutions in brane-world models in terms of holographic RG flows has been shown to be advantageous for realizing self-tuning of the cosmological constant \cite{self-tuning}. Here we observe that generalizing to axionic RG flow solutions leads to further promising applications in brane-world phenomenology. As discussed in section \ref{num}, thanks to the relation \eqref{i2}, adding a non-trivial axion provides an avenue for obtaining a large number of inequivalent vacua (parametrized by a different periodicity branch of the $\theta$-angle and therefore a different parameter $a_\star$) over which to scan,\footnote{Different mechanisms to scan the Higgs mass are proposed in \cite{Herraez:2016dxn,Giudice:2019iwl,Kaloper:2019xfj,Lee:2019efp}.} with possibly different physical properties such as the Higgs mass and vev. This may help finding a stabilized vacuum with a naturally small value of the Higgs mass, as in the relaxion scenario \cite{relaxion}.

Indeed, in figures \ref{fig24}, \ref{fig34-2}, and \ref{fig41}, we found that different values of the Higgs mass are realized in different axionic saddle points labelled by $k$.
Especially, when the Higgs mass squared parameter $X_H$ crosses zero as a function of the axion source $a_\star$, the Higgs expectation value is much smaller than the cutoff scale of the brane physics $\Lambda$. This can be obtained in one of the axionic vacua parametrized by $k$, for sufficiently large $N_c$, as we observed in figures \ref{fig24-2}, \ref{fig34-3}, and \ref{fig41-2}.

A first question is whether the setup can accommodate a large hierarch of scales, like many orders of magnitude as the naive version of the cosmological constant problem suggests.\footnote{The actual hierarchy scale may be smaller as running bulk fields may also contribute to this.}
This has two sub-questions, the first addressing the  cosmological constant self-tuning mechanism and the second the existence of small Higgs mass  vacua.
These two questions are currently under study.

If the answer to the previous two questions is in the affirmative, the next question to ask is how the vacuum realizing the light Higgs mass is selected in our world.
If the vacuum with a small Higgs mass minimizes the free energy \eqref{f13-main}, then the system evolves to this state after a sufficiently long time.

On the other hand, if the vacuum with a small Higgs mass does not minimize the free energy \eqref{f13-main}, this state could be realized as a metastable vacuum.  In the absence of the brane, it is well known that the minimum free energy occurs for minimal values of $k=0,1$. However,  the brane contributes importantly to the free energy and the minimization problem becomes complex, especially as it is affected by the scalar-dependent functions on the brane.

As was done already for the self-tuning setup, the relevant dynamics  is the bulk motion of the brane that will generate the associated cosmology. This was studied in the absence of the axion in \cite{Kehagias:1999vr,relaxion,Amariti:2019vfv} in the probe approximation,  which is solvable. What was found is that the setup corresponds to a brane moving in a radial bulk potential whose minimum (or minima) are at the positions which correspond to a stabilised flat  brane, where the vacuum energy is  cosmologically invisible. Once the brane starts in a different bulk position it will move generating a non-trivial brane cosmology. This motion is affected, beyond the initial velocity and potential, by the presence of matter densities on the brane and  brane-bulk energy exchange, \cite{bb1,bb2}.

In our case, we have two effects  which can happen in tandem, and which can change the position of the brane:
the first is a semiclassical tunneling that interpolates between different $k$-bulk solutions;
the second is classical brane motion in a single bulk solution which will also be affected by the axion.

For the second effect, we expect a similar behavior to the one mentioned above,  but now the brane motions will also be affected by the axion.
The solutions we found for different values of the integer $k$ (the oblique vacua of the dual QFT) will correspond to the minima of the effective potential felt by the brane.
It is important to find how the system may evolve to the metastable vacuum by studying the associated cosmology. At the same time, the lifetime of this vacuum should be long enough. An alternative possibility is to rely on anthropic arguments for the Higgs mass \cite{Agrawal:1997gf,Damour:2007uv,Donoghue:2009me,Hall:2014dfa,Meissner:2014pma}.

We finally compare our scenario to the standard relaxion scenario \cite{relaxion}.
The scalar potential of the relaxion model is given by
\be
V = \left( \Lambda^4 - \Lambda^2 g \tilde{a} + \cdots\right) + \left( \Lambda^2 - g \tilde{a} + \cdots\right) |H|^2 + \Lambda_{QCD}^4 \cos\left({\tilde{a}\over f_a}\right),
\label{g3}\ee
where $\Lambda$ is the cutoff scale, $\tilde{a}$ is the relaxion, $f_a$ is the relaxion decay constant, and $g$ is the shift-symmetry-breaking small parameter which has mass dimension one. In order to obtain the vacua, a certain balance between the $ga$ term and $\cos(a)$ needs to be imposed,
\be
\Lambda^2 g \sim {\Lambda_{QCD}^4 \over f_a} \, ,
\label{g4}\ee
which indicates that the parameter $g$ needs to be hierarchically small.

On the other hand, in our brane potential \eqref{nth1}, such an extreme fine-tuning of couplings is not required, because the existence of the multiple axionic vacua emerges naturally from holography \eqref{i2}.
From the 4d dual field theory viewpoint, the brane scalar potential can be written as
\be
V =  \left(\Lambda^2 - \Lambda_a^2 a \right) \frac{|\tilde{H}|^2}{T_H}  + \ldots
\sim\left( \Lambda^2 - {\Lambda_a^2 \over N_c} (\theta_{UV}+2\pi k) \right) \frac{|\tilde{H}|^2}{T_H} + \ldots,
\label{g2}\ee
where the canonical Higgs field is
\be
\tilde{H} \equiv \frac{M_P}{\sqrt{T_H(\f_0, a_0)}} H \, . \label{g5}
\ee
and $\ldots$ in eq.~\eqref{g2} stands for terms in the scalar potential other than the Higgs mass term.
In the second expression for $V$ in eq.~\eqref{g2}, we used $a\sim (\theta_{UV}+2\pi k)/N_c$ assuming that $a\sim a_\star$ and $c=1$ in \eqref{i2}.\footnote{This assumes that the equilibrium position of the brane is near the IR end point where the value of $a$ becomes very small.}
From \eqref{g2} and for large $N_c$, the coupling between the $\theta_{UV}$ and $\tilde{H}$ is suppressed. In this sense, in our scenario, the breaking of the shift symmetry of $\theta_{UV}$ is small, as in the relaxion model.

\section*{Acknowledgements}
\addcontentsline{toc}{section}{Acknowledgements}

We would like to thank Pascal Anastasopoulos, Mina Arvanitaki, Panos Betzios, Savas Dimopoulos, Matti J\"arvinen and Olga Papadoulaki for discussions.

This work was supported in part by the European Union via  the Advanced ERC grant SM-GRAV, No 669288. LW also acknowledges support from the European Research Council under the European Union's Horizon 2020 research
and innovation programme (grant agreement No 758792,
project GEODESI).

\newpage
\appendix

\begin{appendix}
\renewcommand{\theequation}{\thesection.\arabic{equation}}
\addcontentsline{toc}{section}{Appendices}
\section*{APPENDIX}

\section{Calculation of on-shell action, free energy and topological susceptibility}\label{Free}
In this Appendix, we present the calculation of the on-shell action and free energy.
The bulk and brane actions are given in (\ref{A2}, \ref{A4}, \ref{A5}).
The relevant calculation of the Einstein-dilaton-axion theory without the brane was performed in \cite{Hamada}.

First, we calculate the on-shell bulk action.
From the metric ansatz \eqref{FE7}, we obtain
\be
R=-8\ddot{A}-20\dot{A}^2={1\over2}\dot{\f}^2+{1\over2}Y\dot{a}^2+{5\over3}V.
\label{Free1}\ee
In the second equality, \eqref{a1} and \eqref{a2} are used.
Substituting \eqref{Free1} into \eqref{A2}, the bulk on-shell action is
\be
S_{\text {bulk, on-shell}}=
M_p^{3} \int d^4x \left(\int^{u_{IR}}_{u_0} du + \int^{u_0}_{u_{UV}} du \right) e^{4A} \left[R - {1\over 2} \dot{\f}^2 -{1\over 2}Y \dot{a}^2 - V \right] +S_{GHY}
\label{f1}\ee
$$
={2\over 3} M_p^{3} V_4  \left(\int^{u_{IR}}_{u_0} du + \int^{u_0}_{u_{UV}} du \right) e^{4A} V + S_{GHY}
$$
$$
=-2M_p^{3} V_4 \left\{ \left( \left[e^{4A}\dot{A}\right]_{u_{IR}}- \left[e^{4A}\dot{A}\right]_{u_0+\epsilon} \right) + \left(\left[e^{4A}\dot{A}\right]_{u_0-\epsilon} - \left[e^{4A}\dot{A}\right]_{u_{UV}}\right)\right\}+ S_{GHY}
$$
where $V_4$ is the $4$-dimensional space-time volume, \eqref{Free1} has been used in the second line, and in the third line, we used
\be
V=-3\ddot{A}-12\dot{A}^2.
\label{Free3}\ee

For the Gibbons-Hawking-York term, we obtain
\be
S_{GHY}= -8 M_p^{3} V_4 \bigg\{
\left(\left[ e^{4A} \dot{A} \right]_{u_{UV}} - \left[ e^{4A} \dot{A} \right]_{u_0-\epsilon}\right)
+\left(\left[ e^{4A} \dot{A} \right]_{u_0+\epsilon} - \left[ e^{4A} \dot{A} \right]_{u_{IR}}\right)
\bigg\}.
\label{f2}\ee

Here we are exclusively interested in solutions which have a behavior in the IR (i.e.~for $\f \rightarrow + \infty$) as described in section \ref{first_order}. The corresponding expression for $W$ and $A$ as functions of $\f$ can be read from the equations (\ref{a27}, \ref{sub8-2}, \ref{sub10}). Using these expressions and \eqref{a6}, the IR contribution to the on-shell action can be shown to give
\begin{align}
\left[e^{4A}\dot{A} \right]_{\textrm{IR}} \sim \left[e^{4A} W \right]_{\f \rightarrow + \infty} \sim  e^{- \frac{8 - 3 b^2}{6 b} \f} \, .
\end{align}
Note that if the parameter $b$ satisfies the Gubser bound \eqref{a41} the exponent in the above is negative and the IR contribution vanishes.

Combining \eqref{a6}, \eqref{f1} and \eqref{f2} we arrive at
\be
S_{\text {bulk, on-shell}}=-6M_p^{3}V_4
\left\{ \left[ e^{4A} \dot{A} \right]_{u_{UV}} - \left[{1\over 6} e^{4A_0} W\right]^{IR}_{UV}\right\}
\label{f3}\ee
$$
=
-6 M_p^{3}V_4 \left[ e^{4A} \dot{A} \right]_{u_{UV}} + M_p^{3}V_4 \, e^{4A_0} W_B^{eff}(\f_0, a_0),
$$

Next, we calculate the on-shell brane action starting from (\ref{A3}, \ref{A4}, \ref{A5}).
As for $S_g$, the only nonzero term is the brane cosmological constant term $W_B$ by using \eqref{FE7}.
Similarly, we can observe that the first and last terms in \eqref{A5} vanish on-shell.
Therefore, we obtain
\be
S_{\text{brane, on-shell}}=
M_p^{3} V_4 \, e^{4A_0} \bigg[ -W_B(\f_0, a_0) - X(\f_0) |H|^2 - S_H(\f_0) |H|^4\bigg]
\label{f4}\ee
$$
=
- M_p^{3} V_4 \, e^{4A_0} W_B^{eff}(\f_0, a_0),
$$
where $W_B^{eff}$ is defined in \eqref{match5}.

In total, the on-shell action is
\be
S_{\text{on-shell}}=S_{\text {bulk, on-shell}}+S_{\text{brane, on-shell}}
=-6 M_p^{3}V_4 \left[ e^{4A} \dot{A} \right]_{u_{UV}}
=M_p^{3}V_4 \left[ e^{4A} W\right]_{UV},
\label{f6}\ee
where \eqref{a6} is used in the last equality.
This is the same form as the case without the brane \cite{Hamada}.
As we can observe from (\ref{a27}, \ref{w-}, \ref{t-}), the on-shell action as written in \eqref{f6} is divergent and requires renormalization.
The procedure of renormalization is same as for the case without the brane.
The divergences can be removed by adding a counterterm $S_{ct}$ to the on-shell action (see e.g. \cite{Papa2}), with $S_{ct}$ given by
\be
S_{ct} = -M_p^{3} \Big[ \int d^4x \sqrt{|\g|}\, W_{ct}(\f) \Big]_{\substack{u=\ell \log \e \\ \f=\f(\ell \log \e)}} = - (M_p \ell)^{3} V_4 \, |\f_-|^{\frac{4}{\Delta_-}} \Big[ \Lambda^{4} \, \ell \, W_{ct}(\f_\e) \Big] \, ,
\label{Free13}\ee
where
\be
\Lambda\equiv \left.e^{A(u)}\over\ell \, |\f_-|^{1/\Delta_-}\right|_{u=\ell \log\epsilon} \, .
\label{Free8}\ee
The function $W_{ct}$ is defined as the solution of equation \eqref{a9} with $T=0$, i.e.
\be
{1\over3}W_{ct}^2-{1\over2}(W_{ct}')^2+V=0.
\label{Free17}\ee
One can show that the renormalized on-shell action can be written as
\be
S_{\text{on-shell}}^{\text{ren}}=
M_p^{3} V_4 \, \ell^{3} |\f_-|^{4\over\Delta_-} (C_{UV}(q_{UV})-C_{UV, ct}).
\label{f7}\ee
As in the case without the brane, $C_{UV, ct}$ is a free parameter corresponding to the choice of the renormalization scheme, and $C_{UV}(q_{UV})$ depends on $q_{UV}$ (or $a_\star$) through the IR regularity condition.
The relation between $q_{UV}$ and $a_\star$ is given through \eqref{Q19} and \eqref{Q20}.
The free energy is given by $-S_{\text{on-shell}}^{\text{ren}}$:
\be
F_k \equiv - S_{\rm on-shell}^{\textrm{ren}}=- \left(M_p \ell\right)^{3}  V_4 \, |\f_-|^{4\over\Delta_-} \, \big( C_{UV}(q_{UV, k})-C_{UV,ct} \big) \, ,
\label{f12}\ee
$$
=- \left(M_p \ell\right)^{3}  V_4 \, |\f_-|^{4\over\Delta_-} \, \left[ C_{UV} \left({\theta_{UV}+2\pi k\over N_c}\right)-C_{UV, ct} \right] \, .
$$
where we have written the $k$-dependence of $q_{UV}$ explicitly for clarity.
Note that this exhibits several features familiar from QCD. The parameter $\f_-$ corresponds to the mass scale of the theory and is the analogue of $\Lambda_{\textrm{QCD}}$.\footnote{In the case at hand $\f_-$ is a dimensionful coupling, but one can also modify the setup so that the operator deforming the UV theory is marginally relevant like the QCD coupling. This can be achieved  by setting the mass term to zero the UV expansion of the potential, in which case the running is driven by the cubic or higher terms \cite{exotic}. Alternatively one can realize the UV as a runaway AdS solution, as in the Improved Holographic QCD models \cite{iQCD}. In either case, the scale $\Lambda_{\textrm{QCD}}$ is dynamically generated.}  Further, like in QCD, there is another dimensionless coupling which here is given by $\theta_{UV}$. As $C_{UV}$ is a dimensionless parameter, it only depends on the dimensionless coupling $\theta_{UV}$ (through $q_{UV}$).
Then, we can recognize in \eqref{f12} the structure of the free energy familiar from large $N_c$ QCD \cite{witten1,VW,witten}, i.e.
\be
F_k \sim \Lambda_{\textrm{QCD}}^4 V\left({\theta_{UV}+2\pi k\over N_c}\right).
\label{F21}\ee
This is a general feature of holographic QCD-like theories \cite{glueball}.

The physical free energy is the minimization over $k$ of the free energies $F_k$ for fixed $\theta_{UV}$.
\be
F\left(\f_-,\theta_{UV}\right)={\rm Min_ {k\in\mathbb{Z}}}~F_k\left(\f_-,\theta_{UV}\right).
\label{f13}\ee
The topological susceptibility becomes
\be
\chi\equiv {1\over V_4}{\pa^2 F\over \pa \theta_{UV}^2}=
\left. - {\rm Min}_k |\f_-|^{4\over\Delta_-}{(M_p \ell)^{3}\over N_c^2}{\partial ^2 C_{UV} (a_{\star,k})\over \partial {a_{\star,k}}^2}\right|_{a_{\star,k} = {\theta_{UV}+2\pi k\over N_c}}
\label{f9}\ee

\subsection{Small axion backreaction approximation}\label{FreeProbe}

At small $q_{UV}$, the free energy can be written as
\begin{align}
\nonumber F_k &=F^{(q0)}+ q_{UV, k} F^{(q1)}+\mathcal{O}(q_{UV, k}^2) \\
&= - \left(M_p \ell\right)^{3}  V_4 \, |\f_-|^{4\over\Delta_-} \, \Big[ \big(C_{UV}^{(q0)} - C_{UV,ct} \big) + C_{UV}^{(q1)} q_{UV, k} + \mathcal{O} ( q_{UV, k}^2) \Big] \, ,
\label{f10}\end{align}
where the expression of $C_{UV}^{(q1)}$ can be found in \eqref{Q49}.

Next, from \eqref{Q50}, we observe that
\be
q_{UV, k} = \frac{1}{f^2} a_{\star,k}^2
\ee
at leading order for small axion backreaction. Note that $q_{UV, k}$ is positive due to equations \eqref{a27} and \eqref{t-}, and the constant $f$ is defined by the last line of \eqref{Q50}.
Then, the free energy becomes
\be
F_k =F^{(q0)} - {V_4 \, |\f_-|^{4\over\Delta_-}} \, \left(M_p \ell\right)^{3} {C_{UV}^{(q1)}\over f^2 } a_{\star,k}^2 + {\cal O} \big( \left(a_{\star,k}\right)^4 \big) \, .
\label{f11}\ee
Note that the subsubleading term is ${\cal O} \big( a_{\star,k}^4 \big)$ because we assume CP invariance.

Finally, using the relation \eqref{i2} between $a_\star$ and the theta-parameter $\theta_{UV}$ one obtains
\be
F_k =F^{(q0)} - {V_4 \, |\f_-|^{4\over\Delta_-}} \, {\left(M_p \ell\right)^{3} \over N_c^2} {C_{UV}^{(q1)}\over f^2 } \left(\theta_{UV}+2\pi k \right)^2 + \mathcal{O} \big( N_c^{-2} \big) \, .
\label{f14}\ee
From the definition \eqref{f9} of the topological susceptibility, we obtain
\be
\chi
= -2
{|\f_-|^{4\over\Delta_-}} \,
{\left(M_p \ell\right)^{3}\over N_c^2} {C_{UV}^{(q1)}\over f^2 }
+\mathcal{O}(N_c^{-2}) \, ,
\label{f15}\ee
at leading order for small axion backreaction. The leading order of $\chi$ is ${\cal O}(N_c^0)$ because we identify $\left(M_p \ell\right)^{3} \sim N_c^2$ in holography.

\section{Small axion backreaction approximation} \label{small_axion_detail}
First, we derive the relation between $Q_{IR}$ and $q_{UV}$.
From \eqref{Q18} and \eqref{Q7}, $Q_{IR}$ is
\be
Q_{IR}=
{\rm sign}(Q_{UV}) {\sqrt{q_{UV}}\over \ell} \,\left(\ell |\f_-|^{1\over\Delta_-}\right)^{4}
+ Y_0 e^{4A_0}{\partial \hat{W}_B^{eff} \over\partial a}(\f_0, a_0)
\label{Q21}\ee
$$
={\rm sign}(Q_{UV}) {\sqrt{q_{UV}}\over \ell} \,\left(\ell |\f_-|^{1\over\Delta_-}\right)^{4}
- Y_0 e^{4A_0}{\partial^2 \hat{W}_B^{eff} \over\partial a^2}(\f_0, 0) Q_{IR} \int^{u_{IR}}_{u_0} {du \over Y e^{4A}}
+{\cal O}(q_{UV}),
$$
from which we obtain
\be
\ell Q_{IR}=
 \frac{{\rm sign}(Q_{UV}) \left(\ell |\f_-|^{1\over\Delta_-}\right)^4}{1 +  Y_0 e^{4A_0} {\partial^2 \hat{W}_B^{eff} \over\partial a^2} \int^{u_{IR}}_{u_0} {du \over Y e^{4A}}} \sqrt{q_{UV}}
 + {\cal O} (q_{UV})
=\frac{\ell Q_{UV} }{1 +  Y_0 e^{4A_0} {\partial^2 \hat{W}_B^{eff} \over\partial a^2} \int^{u_{IR}}_{u_0} {du \over Y e^{4A}}}
+ {\cal O} (q_{UV}) ,
\label{aQ52}\ee
where the argument of ${\partial^2 \hat{W}_B^{eff} \over\partial a^2}$ is suppressed.
In the following, we do not write the argument of the functions to avoid clutter.
All the functions are evaluated at the brane position in the trivial axion solution.

Next, we shall calculate the perturbation of the bulk equations of motion \eqref{a8}, \eqref{a8-2} and \eqref{a9} to derive the corrections to $W, S$ and $T$.
At the linear order in $q_{UV}$, the bulk equations are
\be
S^{(q1)}=W'^{(q1)}-{T^{(q1)}\over YS^{(q0)}},
\label{Q34}\ee
\be
{T'^{(q1)} \over T^{(q1)}}={4\over3}{W^{(q0)}\over S^{(q0)}},
\label{Q35}\ee
\be
S^{(q0)}W'^{(q1)}={T^{(q1)}\over 2Y}+{2\over3}W^{(q0)}W^{(q1)}.
\label{Q36}\ee
The general solution of \eqref{Q34}, \eqref{Q35}, \eqref{Q36} is
\be
\ell W^{(q1)}=
\lim_{\f(u_{UV})\to0} \left[e^{{2\over3}\int^\f_{\f(u_{UV})}d\f'{W^{(q0)}\over S^{(q0)}}}
\left(
F_1 + {F_2 \over 2} \int^\f_{\f(u_{UV})}d\f' {e^{{2\over3}\int^{\f'}_{\f(u_{UV})}d{\f''}{W^{(q0)}\over S^{(q0)}}}\over Y \ell S^{(q0)}}\right)\right],
\label{Q37}\ee
\be
\ell S^{(q1)}=
\lim_{\f(u_{UV})\to0}
\bigg[
{2\over3}{W^{(q0)}\over S^{(q0)}}e^{{2\over3}\int^\f_{\f(u_{UV})}d\f'{W^{(q0)}\over S^{(q0)}}}
\left(
F_1 + {F_2 \over 2} \int^\f_{\f(u_{UV})}d\f' {e^{{2\over3}\int^{\f'}_{\f(u_{UV})}d{\f''}{W^{(q0)}\over S^{(q0)}}}\over Y \ell S^{(q0)}}\right)
\label{Q38}\ee
$$
-{F_2\over 2Y \ell S^{(q0)}} e^{{4\over3}\int^{\f}_{\f(u_{UV})}d\f'\,{W^{(q0)}\over S^{(q0)}}}\bigg],
$$
\be
\ell^2 T^{(q1)}= \lim_{\f(u_{UV})\to0} F_2 \, e^{{4\over3}\int^{\f}_{\f(u_{UV})}d\f'\,{W^{(q0)}\over S^{(q0)}}},
\label{Q39}\ee
where $F_1, F_2$ are integration constants. There are four integration constants $F_1^{UV}, F_1^{IR}, F_2^{UV}$ and $F_2^{IR}$ corresponding to the UV and IR regions.
The IR regularity condition (Gubser's bound) imposes
\be
F_1^{IR} + {F_2^{IR} \over 2} \int^\infty_{\f(u_{UV})}d\f' {e^{{2\over3}\int^{\f'}_{\f(u_{UV})}d{\f''}{W^{(q0)}\over S^{(q0)}}}\over Y \ell S^{(q0)}}=0.
\label{Q42}\ee
From the UV limit of \eqref{Q37} and \eqref{Q39}, we obtain
\be
F_1^{UV}= C_{UV}^{(q1)} |\f(u_{UV})|^{4\over\Delta_-},
\label{Q41}\ee
\be
F_2^{UV}= |\f(u_{UV})|^{8\over\Delta_-},
\label{Q40}\ee
where \eqref{w-} are used.

In order to deduce the value of the integration constants, we need to use the junction conditions. For $q_{UV}=0$, the junction conditions \eqref{match1-2} and \eqref{match2-2} become
\be
\left[W^{(q0)}\right]^{IR}_{UV}
=\hat{W}_B^{eff},
\label{Q29-2}\ee
\be
\left[S^{(q0)}\right]^{IR}_{UV}=\left[W'^{(q0)}\right]^{IR}_{UV}
={\p \hat{W}_B^{eff} \over \p \f}.
\label{Q30}\ee
We notice that, to have a consistent solution without axion, we must have
\be
{\partial \hat{W}_B^{eff} \over\partial a}
=0,
\label{Q20}\ee
from \eqref{match3-2}.
Moreover, we assume CP invariance:
\be
{\partial^2 \hat{W}_B^{eff} \over\partial a \,\partial \f}
=0,
\label{Q29}\ee
which shall be used in \eqref{Q9}.

At linear order in $q_{UV}$, by using \eqref{Q29} and \eqref{Q30}, \eqref{match1-2} and \eqref{match2-2} become
\be
\left[W^{(q1)}\right]^{IR}_{UV} + {\partial \hat{W}_B^{eff} \over \partial \f} \f_0^{(q1)}
={\partial^2 \hat{W}_B^{eff} \over \partial a^2} g^2 + {\partial \hat{W}_B^{eff} \over \partial \f} \f_0^{(q1)},
\label{Q8}\ee
\be
\left[S^{(q1)}\right]^{IR}_{UV} + \left[{\partial^2 W^{(q0)} \over \partial \f^2}\right]^{IR}_{UV} \f_0^{(q1)}
={\partial^3 \hat{W}_B^{eff} \over \partial \f \partial a^2} g^2 + {\partial^2 \hat{W}_B^{eff} \over \partial \f^2} \f_0^{(q1)},
\label{Q9}\ee
where $g$ is defined in \eqref{Q32},
while the last junction condition \eqref{match3-2} is
\be
\left[{\rm sign}(Q)\sqrt{{T^{(q1)}}}\right]^{IR}_{UV}=
 {\partial \hat{W}_B^{eff} \over \partial a^2}  Y_0 g.
\label{Q12}\ee
The first condition \eqref{Q8} fixes $W^{UV(q1)}(\f_0)$ in terms of $W^{IR(q1)}(\f_0)$:
\be
W^{(q1)}_{UV}(\f_0)=
W^{(q1)}_{IR}(\f_0) - {\partial^2 \hat{W}_B^{eff} \over \partial a^2} g^2 .
\label{Q10}\ee
The second junction condition determines the perturbation of the brane position $\f_0^{(q1)}$,
\be
\f_0^{(q1)}=
\frac{\left[S^{(q1)}\right]^{IR}_{UV} - {\partial^3 \hat{W}_B^{eff} \over \partial \f \partial a^2} g^2} {{\partial^2 \hat{W}_B^{eff} \over \partial \f^2} - \left[{\partial^2 W^{(q0)} \over \partial \f^2}\right]^{IR}_{UV}}.
\label{Q11}\ee

Now we can calculate the integration constants, $(F_1^{UV}, F_1^{IR}, F_2^{UV}, F_2^{IR})$, using the junction conditions.
By using \eqref{Q32} and \eqref{Q39}, the condition \eqref{Q12} leads to
\be
\lim_{\f(u_{UV})\to0} F_2^{IR} \left( 1+ {\partial \hat{W}_B^{eff} \over \partial a^2}  Y_0 \int^{\infty}_{\f_0^{(q0)}} {e^{{2\over3}\int^{\f}_{\f(u_{UV})}d\f'\,{W^{(q0)}\over S^{(q0)}}} \over Y \ell S^{(q0)}} d\f \right)^2=
\lim_{\f(u_{UV})\to0} F_2^{UV} e^{{4\over3}\int^{\f_0^{(q0)}}_{\f(u_{UV})}d\f'\,{W^{(q0)}\over S^{(q0)}}}
\label{Q43}\ee
$$
=
\lim_{\f(u_{UV})\to0} |\f(u_{UV})|^{8\over\Delta_-} e^{{4\over3}\int^{\f_0^{(q0)}}_{\f(u_{UV})}d\f'\,{W^{(q0)}\over S^{(q0)}}},
$$
where \eqref{Q40} is used in the last equality.

From \eqref{Q10}, we obtain
\be
\lim_{\f(u_{UV})\to0} e^{{2\over3}\int^{\f_0^{(q0)}}_{\f(u_{UV})}d\f'{W^{(q0)}\over S^{(q0)}}}
\left(
\left[F_1\right]^{IR}_{UV} + {\left[F_2\right]^{IR}_{UV} \over 2} \int^{\f_0^{(q0)}}_{\f(u_{UV})}d\f' {e^{{2\over3}\int^{\f'}_{\f(u_{UV})}d{\f''}{W^{(q0)}\over S^{(q0)}}}\over Y \ell S^{(q0)}}\right)
\label{Q44}\ee
$$
=
\lim_{\f(u_{UV})\to0} {\partial^2 \hat{W}_B^{eff} \over \partial a^2} F_2^{IR}
\left( \int^{\f_{IR}}_{\f_0^{(q0)}} {e^{{2\over3}\int^{\f}_{\f(u_{UV})}d\f'\,{W^{(q0)}\over S^{(q0)}}} \over Y \ell S^{(q0)}} d\f \right)^2.
$$

By combining \eqref{Q44} with \eqref{Q42}, \eqref{Q40}, \eqref{Q43}, The integration constants are given by
\be
{F_1^{UV} \over |\f(u_{UV})|^{8\over\Delta_-} }=
-{1\over2}  \int^{\f_0^{(q0)}}_{\f(u_{UV})}d\f' {e^{{2\over3}\int^{\f'}_{\f(u_{UV})}d{\f''}{W^{(q0)}\over S^{(q0)}}}\over Y \ell S^{(q0)}},
\label{Q45}\ee
$$
-\frac{
\left(\int^\infty_{\f_0^{(q0)}} {e^{{2\over3}\int^{\f}_{\f(u_{UV})}d\f'\,{W^{(q0)}\over S^{(q0)}}} \over Y \ell S^{(q0)}} d\f\right)
\left(
{1\over2} e^{{4\over3}\int^{\f_0^{(q0)}}_{\f(u_{UV})}d\f'\,{W^{(q0)}\over S^{(q0)}}}
+{\partial^2 \hat{W}_B^{eff} \over \partial a^2} \int^\infty_{\f_0^{(q0)}} {e^{{2\over3}\int^{\f}_{\f(u_{UV})}d\f'\,{W^{(q0)}\over S^{(q0)}}} \over Y \ell S^{(q0)}} d\f
\right)
}
{\left( 1+ {\partial \hat{W}_B^{eff} \over \partial a^2}  Y_0 \int^{\infty}_{\f_0^{(q0)}} {e^{{2\over3}\int^{\f}_{\f(u_{UV})}d\f'\,{W^{(q0)}\over S^{(q0)}}} \over Y \ell S^{(q0)}} d\f \right)^2},
$$
\be
{F_1^{IR}\over |\f(u_{UV})|^{8\over\Delta_-}}=
- {1\over2} \frac{e^{{4\over3}\int^{\f_0^{(q0)}}_{\f(u_{UV})}d\f'\,{W^{(q0)}\over S^{(q0)}}}
\int^\infty_{\f(u_{UV})}d\f' {e^{{2\over3}\int^{\f'}_{\f(u_{UV})}d{\f''}{W^{(q0)}\over S^{(q0)}}}\over Y \ell S^{(q0)}}}
{\left( 1+ {\partial \hat{W}_B^{eff} \over \partial a^2}  Y_0 \int^{\infty}_{\f_0^{(q0)}} {e^{{2\over3}\int^{\f}_{\f(u_{UV})}d\f'\,{W^{(q0)}\over S^{(q0)}}} \over Y \ell S^{(q0)}} d\f \right)^2},
\label{Q46}\ee
\be
{F_2^{UV}\over |\f(u_{UV})|^{8\over\Delta_-}}= 1,
\label{Q47}\ee
\be
{F_2^{IR} \over |\f(u_{UV})|^{8\over\Delta_-}}=
\frac{ e^{{4\over3}\int^{\f_0^{(q0)}}_{\f(u_{UV})}d\f'\,{W^{(q0)}\over S^{(q0)}}}}
{\left( 1+ {\partial \hat{W}_B^{eff} \over \partial a^2}  Y_0 \int^{\infty}_{\f_0^{(q0)}} {e^{{2\over3}\int^{\f}_{\f(u_{UV})}d\f'\,{W^{(q0)}\over S^{(q0)}}} \over Y \ell S^{(q0)}} d\f \right)^2}.
\label{Q48}\ee

From \eqref{Q41} and \eqref{Q45}, we obtain
\be
C_{UV}^{(q1)}=
-{1\over2}  \int^{\f_0^{(q0)}}_{\f(u_{UV})}d\f' {e^{{2\over3}\int^{\f'}_{\f(u_{UV})}d{\f''}{W^{(q0)}\over S^{(q0)}}}\over Y \ell S^{(q0)}}-
\label{Q49}\ee
$$
-\frac{
\left(\int^\infty_{\f_0^{(q0)}} {e^{{2\over3}\int^{\f}_{\f(u_{UV})}d\f'\,{W^{(q0)}\over S^{(q0)}}} \over Y \ell S^{(q0)}} d\f\right)
\left(
{1\over2} e^{{4\over3}\int^{\f_0^{(q0)}}_{\f(u_{UV})}d\f'\,{W^{(q0)}\over S^{(q0)}}}
+{\partial^2 \hat{W}_B^{eff} \over \partial a^2} \int^\infty_{\f_0^{(q0)}} {e^{{2\over3}\int^{\f}_{\f(u_{UV})}d\f'\,{W^{(q0)}\over S^{(q0)}}} \over Y \ell S^{(q0)}} d\f
\right)
}
{\left( 1+ {\partial \hat{W}_B^{eff} \over \partial a^2}  Y_0 \int^{\infty}_{\f_0^{(q0)}} {e^{{2\over3}\int^{\f}_{\f(u_{UV})}d\f'\,{W^{(q0)}\over S^{(q0)}}} \over Y \ell S^{(q0)}} d\f \right)^2}.
$$

By using \eqref{Q19}, \eqref{Q39}, \eqref{Q47}, \eqref{Q48}, the axion source is, to leading order,
\be
\frac{a_\star}{\sqrt{q_{UV}}}=
- {\rm sign}(Q_{UV})\int^{\f_0^{(q0)}}_{\f(u_{UV})} d\f{\sqrt{T^{(q1)}}  \over S^{(q0)} \,Y }
- {\rm sign}(Q_{IR}) \int^{\infty}_{\f_0^{(q0)}} d\f{\sqrt{T^{(q1)}} \over S^{(q0)} \,Y }
\label{Q50}\ee
$$
=
- {\rm sign}(Q_{UV}) {\sqrt{F_2^{UV}}}
\int^{\f_0^{(q0)}}_{\f(u_{UV})}d\f {e^{{2\over3}\int^{\f}_{\f(u_{UV})}d{\f'}{W^{(q0)}\over S^{(q0)}}}\over Y \ell S^{(q0)}}
- {\rm sign}(Q_{IR}) {\sqrt{F_2^{IR}}}
\int_{\f_0^{(q0)}}^{\infty}d\f {e^{{2\over3}\int^{\f}_{\f(u_{UV})}d{\f'}{W^{(q0)}\over S^{(q0)}}}\over Y \ell S^{(q0)}}
$$
$$
=-\lim_{\f(u_{UV})\to0} |\f(u_{UV})|^{4\over\Delta_-}
\bigg[
{\rm sign}(Q_{UV}) \int^{\f_0^{(q0)}}_{\f(u_{UV})}d\f {e^{{2\over3}\int^{\f}_{\f(u_{UV})}d{\f'}{W^{(q0)}\over S^{(q0)}}}\over Y \ell S^{(q0)}}
$$
$$
+{\rm sign}(Q_{IR})
\frac{ e^{{2\over3}\int^{\f_0^{(q0)}}_{\f(u_{UV})}d\f'\,{W^{(q0)}\over S^{(q0)}}}}
{\left| 1+ {\partial \hat{W}_B^{eff} \over \partial a^2}  Y_0 \int^{\infty}_{\f_0^{(q0)}} {e^{{2\over3}\int^{\f}_{\f(u_{UV})}d\f'\,{W^{(q0)}\over S^{(q0)}}} \over Y \ell S^{(q0)}} d\f \right|}
\int_{\f_0^{(q0)}}^{\infty}d\f {e^{{2\over3}\int^{\f}_{\f(u_{UV})}d{\f'}{W^{(q0)}\over S^{(q0)}}}\over Y \ell S^{(q0)}}
\bigg]
$$
Here ${\rm sign}(Q_{UV})$ is a free parameter, and ${\rm sign}(Q_{IR})$ is determined by \eqref{Q52}.

Finally, we provide the dilaton and axion at the brane position (\eqref{Q11} and \eqref{Q32}) in terms of the unperturbed quantities:
\be
\f_0^{(q1)}=
\frac{1}{{\partial^2 \hat{W}_B^{eff} \over \partial \f^2} - \left[{\partial^2 W^{(q0)} \over \partial \f^2}\right]^{IR}_{UV}}
\Bigg\{
{2\over3}e^{{2\over3}\int^{\f_0^{(q0)}}_{\f(u_{UV})}d\f'{W^{(q0)}\over S^{(q0)}}}
\Bigg(
\left[{W^{(q0)}F_1\over S^{(q0)}}\right]^{IR}_{UV}+
\label{Q51}\ee
$$
+ \left[{W^{(q0)}F_2\over 2S^{(q0)}}\right]^{IR}_{UV}  \int^{\f_0^{(q0)}}_{\f(u_{UV})}d\f' {e^{{2\over3}\int^{\f'}_{\f(u_{UV})}d{\f''}{W^{(q0)}\over S^{(q0)}}}\over Y \ell S^{(q0)}}
\Bigg)
-{1\over 2Y_0} \left[{F_2\over  \ell S^{(q0)}}\right]^{IR}_{UV} e^{{4\over3}\int^{\f_0^{(q0)}}_{\f(u_{UV})}d\f'\,{W^{(q0)}\over S^{(q0)}}} -
$$
$$
- {\partial^3 \hat{W}_B^{eff} \over \partial \f \partial a^2} F_2^{IR}
\left( \int^{\f_{IR}}_{\f_0^{(q0)}} {e^{{2\over3}\int^{\f}_{\f(u_{UV})}d\f'\,{W^{(q0)}\over S^{(q0)}}} \over Y \ell S^{(q0)}} d\f \right)^2
\Bigg\},
$$
\be
a_0=
- {\rm sign}(Q_{IR}) {\sqrt{F_2^{IR}}}
\int_{\f_0^{(q0)}}^{\infty}d\f {e^{{2\over3}\int^{\f}_{\f(u_{UV})}d{\f'}{W^{(q0)}\over S^{(q0)}}}\over Y \ell S^{(q0)}}
+ {\cal O}(q_{UV}),
\label{Q57}\ee
where $(F_1^{UV}, F_1^{IR}, F_2^{UV}, F_2^{IR})$ are given in \eqref{Q45}, \eqref{Q46}, \eqref{Q47}, \eqref{Q48} in terms of the unperturbed quantities, and \eqref{Q32} and \eqref{Q38} are used in \eqref{Q51}, while \eqref{Q39} is used in \eqref{Q57}.

The calculations of the on-shell axion, free energy and topological susceptibility in the small axion backreaction are given in Appendix \ref{FreeProbe}, where the known large-$N_c$ result \cite{witten1,VW,witten} is reproduced.

\section{Brane equilibrium position in the near-UV or near-IR regions: analytic results} \label{brane_UV_IR}

\subsection{Brane equilibrium position in the near-UV region}
Suppose that $\f_0$ is close to the $\f(u_{UV})=0$, and we can use the UV asymptotic expressions at the locus of the brane position.
By using the junction conditions \eqref{match1-2}, \eqref{match2-2}, \eqref{match3-2} and the UV asymptotic expressions for the scalar functions \eqref{w-}, \eqref{s-}, \eqref{t-}, we obtain
\be
{1\over\ell} \left(  C_{IR} - C_{UV} \right) |\f_0|^{4\over\Delta_-} = \hat{W}_B^{eff} (\f_0, a_0),
\label{nearUV1}\ee
\be
{1\over\ell} {4\over\Delta_-} (C_{IR}-C_{UV}) |\f_0|^{{4\over\Delta_-}-1} = {\p \hat{W}_B^{eff} \over \p \f} (\f_0, a_0),
\label{nearUV2}\ee
\be
{1\over\ell} \bigg( {\rm sign}(Q_{IR}) \sqrt{q_{IR}} - {\rm sign}(Q_{UV}) \sqrt{q_{UV}} \bigg) |\f_0|^{4\over\Delta_-} = Y_0 {\p \hat{W}_B^{eff} \over \p a} (\f_0,a_0),
\label{nearUV3}\ee
at leading order in $\f_0$.
Here $C_{IR}$ and $q_{IR}$ are the integration constant appearing in the UV expansions of the scalar functions $(W_{IR}, S_{IR}, T_{IR})$.

By solving \eqref{nearUV1}, \eqref{nearUV2} and \eqref{nearUV3}, one can express the IR integration constants $C_{IR}$ and $q_{IR}$ and the brane position $\f_0$ in terms of the UV integration constants.
If $\hat{W}_B^{eff} (\f_0, a_0)\neq0$, from the first two equations \eqref{nearUV1} and \eqref{nearUV2}, one can see that $\f_0$ is determined by solving
\be
\f_0 {\p \log \hat{W}_B^{eff} \over \p \f}(\f_0, a_0)= {4\over \Delta_-},
\label{nearUV4}\ee
where $a_0$ is related to the function of $\f_0$ through
\be
a_0= - {\rm sign}(Q_{IR}) \int^{\f(u_{IR})}_{\f_0} {\sqrt{T}\over Y S}.
\label{nearUV5}\ee
By using the solution for $\f_0$ we can determine $C_{IR}$ and $q_{IR}$ as
\be
C_{IR} = C_{UV} + |\f_0|^{-{4\over\Delta_-}}{\ell \,\hat{W}_B^{eff} (\f_0,a_0) } ,
\label{nearUV6}\ee
\be
 {\rm sign}(Q_{IR}) \sqrt{q_{IR}} =
 {\rm sign}(Q_{UV}) \sqrt{q_{UV}} +  |\f_0|^{-{4\over\Delta_-}} \ell \, Y_0 {\p \hat{W}_B^{eff} \over \p a} (\f_0,a_0).
\label{nearUV7}\ee
On the other hand, if $\hat{W}_B^{eff}(\f_0, a_0)=0$, then $C_{IR}=C_{UV}$ but \eqref{nearUV7} is still valid. The brane position is determined by the condition ${\p \hat{W}_B^{eff} \over \p a} (\f_0, a_0)=0$.

\subsection{Brane equilibrium position in the near-IR region}
Suppose that the brane is in the region where the IR asymptotic expansions \eqref{sub8-2}, \eqref{sub9}, \eqref{sub10} can be used.
Again we can obtain relations among the various integration constants analytically in this case. The junction conditions \eqref{match1-2}, \eqref{match2-2}, \eqref{match3-2} lead to
\be
-{1\over2}{D_{IR}-D_{UV}\over {b\over2}+\g-{4\over3b}} e^{\left({8\over3b}-{b\over2}-\g\right)\f_0}
-E_{UV} \,e^{{4\over3b}\f_0} = \ell \, \hat{W}_B^{eff}(\f_0, a_0),
\label{nearIR1}\ee
\be
-{D_{IR}-D_{UV}\over2}{{b\over2}+\g\over {b\over2}+\g-{4\over3b}} e^{\left({8\over3b}-{b\over2}-\g\right)\f_0}
-{4E_{UV}\over3b} \,e^{{4\over3b}\f_0}
=\ell {\p \hat{W}_B^{eff} \over \p \f} (\f_0,a_0),
\label{nearIR2}\ee
\be
\sqrt{{b W_\infty \over 2Y_\infty} } \bigg({\rm sign}(Q_{IR}) \sqrt{D_{IR}} - {\rm sign}(Q_{UV}) \sqrt{D_{UV}}\bigg)e^{\left({4\over3b}-\g\right) \f_0}
=\ell {\p \hat{W}_B^{eff} \over \p a} (\f_0, a_0),
\label{nearIR3}\ee
where $D_{UV}$ and $E_{UV}$ are the integration constant appearing in the IR expansion of the scalar functions $(W_{UV}, S_{UV}, T_{UV})$ on the UV side of the brane:
\be
\ell W_{UV}=
W_\infty \,e^{{b\over2}\f}
-{D_{UV}\over2}{1\over {b\over2}+\g-{4\over3b}} e^{\left({8\over3b}-{b\over2}-\g\right)\f}
+E_{UV} \,e^{{4\over3b}\f}+\cdots,
\label{asub16}\ee
\be
\ell S_{UV}=
{b\over2}W_\infty\, e^{{b\over2}\varphi}
-{D_{UV}\over2}{{b\over2}+\g\over {b\over2}+\g-{4\over3b}} e^{\left({8\over3b}-{b\over2}-\g\right)\f}
+{4E_{UV}\over3b} \,e^{{4\over3b}\f}+\cdots,
\label{asub17}\ee
\be
\ell^2 T_{UV}=
{b\over2}D_{UV} W_\infty Y_\infty \,e^{{8\over3b}\f}+\cdots.
\label{asub18}\ee
Note that the integration constant $E$ has to be set to zero in $(W_{IR}, S_{IR}, T_{IR})$ due to Gubser's bound.

From \eqref{sub11} we observe
\be
{8\over3b}-{b\over2}-\g \leq {4\over3b}.
\label{nearIR7}\ee
Therefore, the left hand side of \eqref{nearIR1} and \eqref{nearIR2} is generally dominated by the $E_{UV}$ term, and we neglect $D_{IR}-D_{UV}$ term in the following.
Combining \eqref{nearIR1} with \eqref{nearIR2}, we obtain
\be
{\p \log \hat{W}_B^{eff} \over \p \f}(\f_0, a_0)={4\over3b},
\label{nearIR4}\ee
for $\hat{W}_B^{eff}(\f_0, a_0)\neq0$. The position $\f_0$ can be deduced from this equation.
The integration constants are determined by
\be
E_{UV}=-\ell {\p \hat{W}_B^{eff} \over \p \f} (\f_0,a_0) e^{-{4\over3b}\f_0},
\label{nearIR5}\ee
\be
 {\rm sign}(Q_{IR}) \sqrt{D_{IR}}= {\rm sign}(Q_{UV}) \sqrt{D_{UV}}  + \sqrt{2 Y_\infty \over b W_\infty Y_\infty} \ell {\p \hat{W}_B^{eff} \over \p a} (\f_0, a_0) e^{-\left({4\over3b}-\g\right)\f_0}.
\label{nearIR6}\ee
In the case of $\hat{W}_B^{eff}(\f_0,a_0)=0$, we get $E_{UV}=0$, and $\f_0$ is determined by solving ${\p \hat{W}_B^{eff} \over \p \f} (\f_0, a_0)=0$. Then, $D_{UV}$ is fixed by \eqref{nearIR6}.

\section{Overshooting constraint} \label{constraint}

One can show that there exists an additional constraint on the brane potential $W_B$, if the solution for $W$ on the UV side of the brane is to exhibit the expected UV asymptotics recorded in \eqref{w-}.
For simplicity, we construct the argument in the case without a bulk axion and generalise to case with bulk axion later. In the absence of a bulk axion the equation of motion for $W_{UV}$ can be written as
\be
W_{UV}' = \sqrt{{2\over3}W_{UV}^2+2V}.
\label{re1}\ee
which is \eqref{a9} with $S_{UV}=W_{UV}'$ as follows from \eqref{a8} in absence of the axion.
The function $W_{UV}$ is then given by the solution to \eqref{re1} subject to the boundary condition
\be
\lim_{\epsilon\to0}W_{UV}(\f_0-\epsilon)\equiv W_0 \, ,
\label{re9}\ee
where we assume $W_0 > 6$.
To construct the argument, it will be useful to introduce an auxiliary function $\tilde{W}$, which is defined as the solution to
\be
\tilde{W}' = \sqrt{{2\over3}\tilde{W}^2-24}\sp
\tilde{W}(\f_0)=W_0.
\label{re2}\ee
By definition, the function $\tilde{W}$ is monotonic in $\f$.
Equation \eqref{re2} can be solved analytically to find
\be
{\tilde{W} (\f) \over 3} =  \tilde{C} e^{\sqrt{2\over 3}(\f-\f_0)} + \tilde{C}^{-1} e^{-\sqrt{2\over 3}(\f-\f_0)},
\label{re3}\ee
with
\be
\tilde{C} = {W_0\over 6} + \sqrt{\left({W_0\over 6}\right)^2-1}.
\label{re3-2}\ee
The solution \eqref{re3} is valid only for
\be
\f \geq \f_{\rm min} \, , \quad \textrm{with} \quad
\f_{\rm min} = \f_0 - \sqrt{3\over 2} \log \tilde{C} \, ,
\label{re3-3}\ee
as for $\f=\f_{\rm min}$ one finds
\be
\tilde{W}(\f_{\rm min})=6,
\label{re3-4}\ee
and the square root in \eqref{re2} vanishes.

From \eqref{Num1}, we have that $V>-12$. Then, as a consequence of the definition of $\tilde{W}$ in \eqref{re2} it follows that
\be
\tilde{W}(\f) < W_{UV}(\f) \quad \textrm{for} \quad \f<\f_0 \, .
\label{re4}\ee
For $W_{UV}$ to exhibit the desired UV behaviour of a RG flow solution, we require $W_{UV}(0)=6$. Then, consistency of the condition $W_{UV}(0)=6$ with the property \eqref{re4} and the monotonicity of $\tilde{W}$ require
\be
\f_{\rm min} > 0,
\quad \Rightarrow
\quad e^{\sqrt{4\over3}\f_0} > \tilde{C}.
\label{re6}\ee
This provides a non-trivial constraint on the brane potential through the dependence of $\tilde{C}$ on $W_0$.

  \begin{figure}[t]
 \begin{center}
  \includegraphics[width=.5\textwidth]{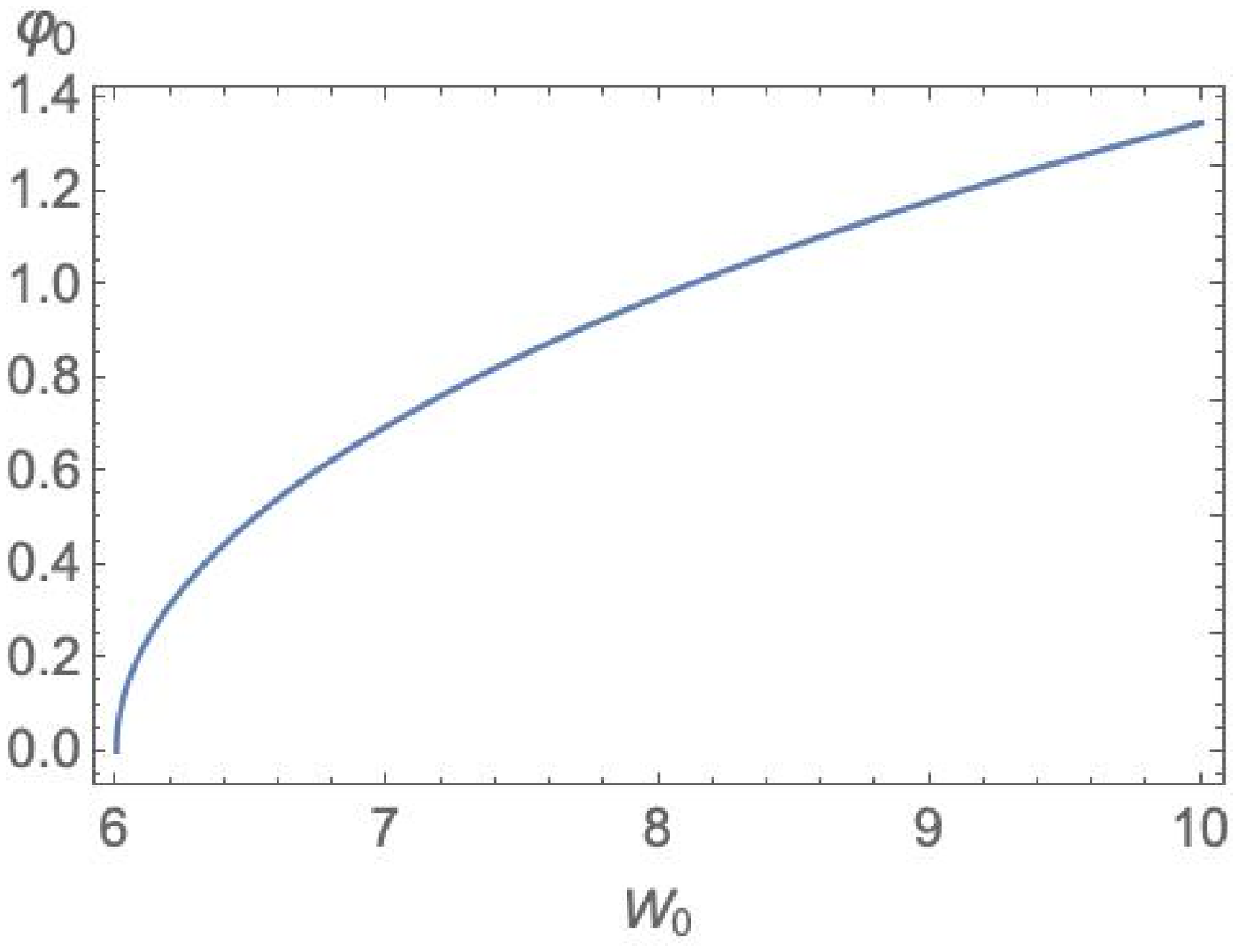}
 \end{center}
 \begin{center}
  \includegraphics[width=.49\textwidth]{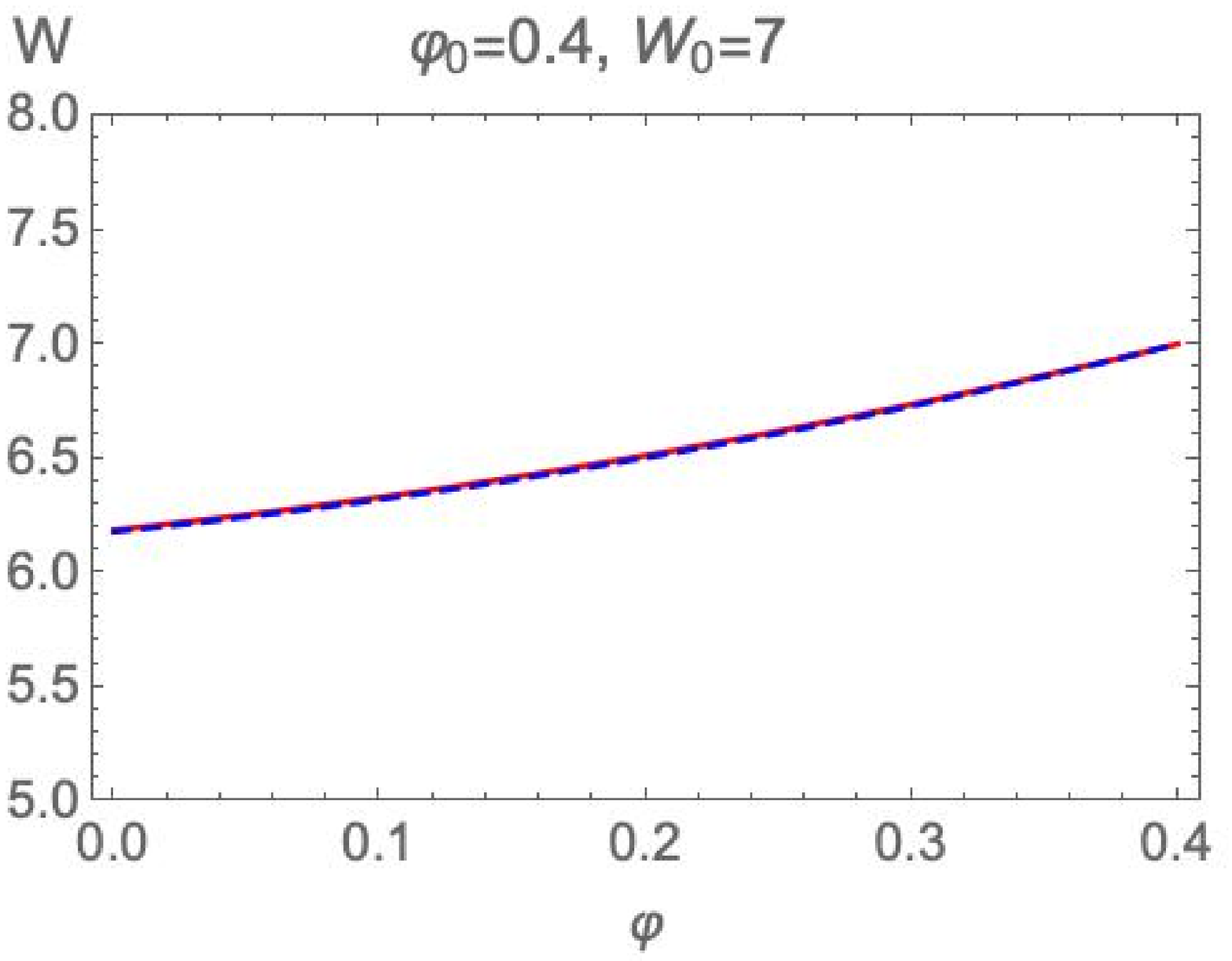}
  \includegraphics[width=.49\textwidth]{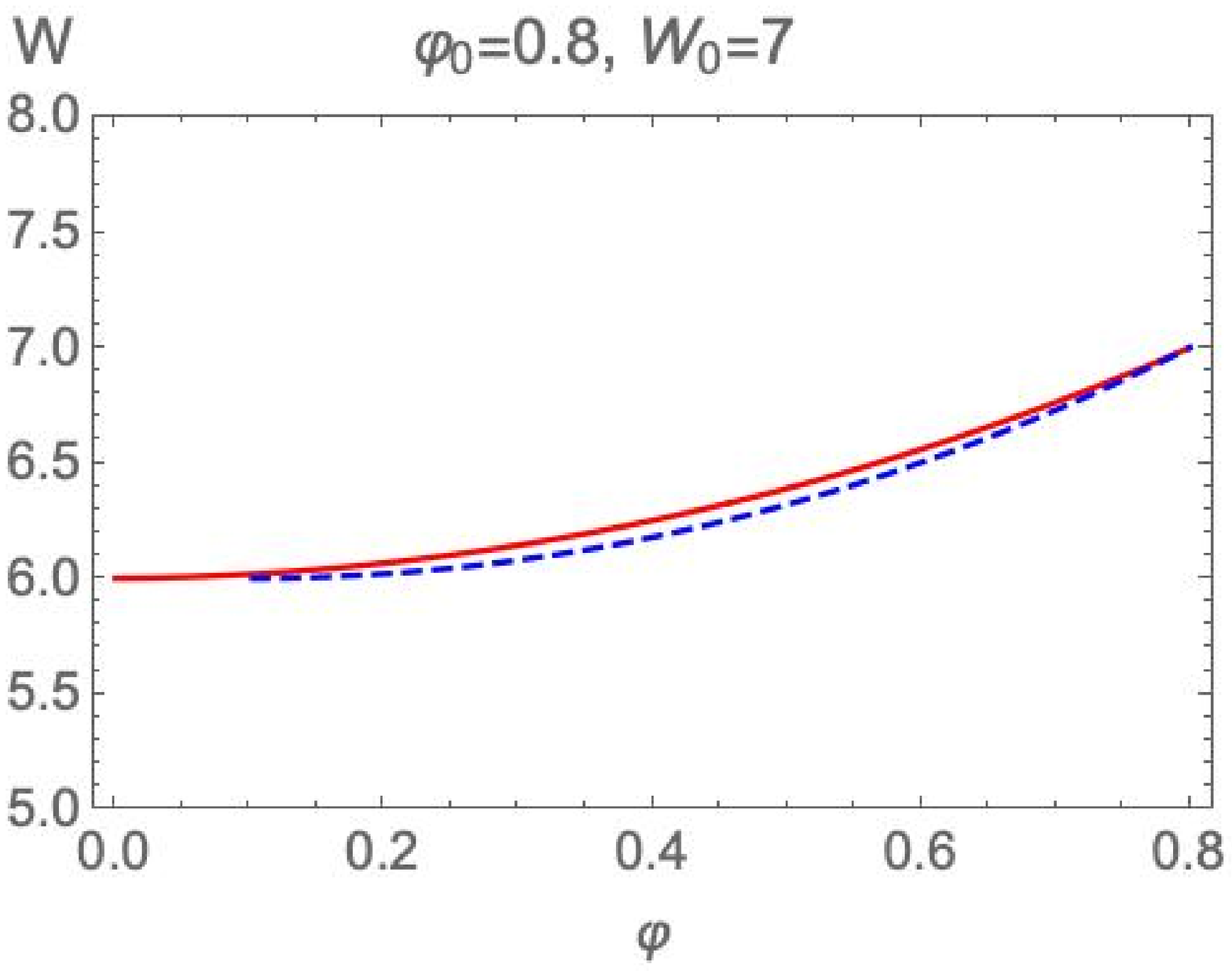}
 \end{center}
   \caption{
   \textbf{Top row:} Lower bound on $\f_0$ as a function of $W_0$ as arising from the constraint \protect\eqref{re6}.
   \textbf{Bottom row, left:} Plot of $W_{UV}$ and $\tilde{W}$ vs.~$\f$ by solving \protect\eqref{re1} and \protect\eqref{re2} for the parameter choice $\f_0=0.4$ and $W_0=7$. The solid and dashed lines correspond to $W_{UV}$ and $\tilde{W}$, respectively. For this parameter choice $W_{UV}(0)=6$ is not realized.
   \textbf{Bottom row, right:} Plot of $W_{UV}$ and $\tilde{W}$ vs.~$\f$ by solving \protect\eqref{re1} and \protect\eqref{re2} for the parameter choice $\f_0=0.8$ and $W_0=7$. The solid and dashed lines correspond to $W_{UV}$ and $\tilde{W}$, respectively. For this parameter we find $W_{UV}(0)=6$.
}
  \label{fig42}
 \end{figure}

We numerically check this constraint in fig.~\ref{fig42}.
The condition \eqref{re6} can be interpreted as a lower bound on $\f_0$. In the top row panel we plot this bound on $\f_0$ as a function of $W_0$. In the bottom row panels, we numerically solve \eqref{re1} and \eqref{re2} for $W_{UV}$ and $\tilde{W}$, respectively. The left panel corresponds to a parameter choice which does not satisfy \eqref{re6}. We observe that $W_{UV}(0)=6$ is not realized in this case. On the other hand, the parameter choice in the right panel satisfies \eqref{re6}, and $W_{UV}(0)=6$ can be attained.

Inserting for $\tilde{C}$ with \eqref{re3-2} and substituting for $W_0= W_{UV}(\f_0)$ using the junction condition \eqref{match1-2}, the condition \eqref{re6} can be written as
\be
\left. e^{\sqrt{2\over3}\f} \left(
{W_{IR}-\hat{W}_B^{eff}\over6} + \sqrt{\left({W_{IR}-\hat{W}_B^{eff}\over6}\right)^2-1}
\right)^{-1}
\right|_{\f=\f_0}>1
\label{re10}\ee
where $\f_0$ is determined by \eqref{nfull6}.
We expect a similar condition to also exist in the presence of nontrivial axion backreaction, but an analytical statement along the lines of the argument above is more difficult in this case and is left for future work. Instead, in the numerical examples including axion backreaction considered here, we check explicitly whether the condition $\lim_{\f\to0} W_{UV}=6$ is satisfied, at least within our numerical precision. If a solution does not satisfy this condition it is discarded. Hence, all numerical examples shown in sec.~\ref{num} exhibit $\lim_{\f\to0} W_{UV}=6$ as expected for a RG flow solution.

\section{Positivity constraints}
\label{app:positivityanalysis}

In this appendix we give a preliminary assessment of the question: can the mixing of the Higgs with the bulk modes lead to the emergence of ghost-like or tachyonic instabilities? As we shall see, the detailed answer will depend on the specific features of the model. Nevertheless, some general conclusions can be drawn which show that, broadly speaking,  this is not an issue that would make the theory inconsistent.

In what follows we analyze this question at the level of linear perturbations around a self-tuning vacuum. The answer about stability  is read off from the action at the  quadratic level in the fluctuations: the absence of ghosts requires positivity of  kinetic terms, whereas the absence of  tachyons implies constraints on the mass matrix.

The model contains tensor modes (which come from the traceless and transverse part of the bulk graviton) and scalar modes. The latter are a mixture of the trace part of the metric fluctuations, of the bulk dilaton and axion  fluctuations, of the brane-localized Higgs, plus the fluctuations in the brane position around equilibrium. Tensor modes are healthy (i.e.~non-ghost like and with a positive definite mass)  if both the brane and bulk Einstein-Hilbert term have the correct sign, which we assume to be always true in the models under considerations. In the absence of brane-localized fields, the conditions for stability of the Einstein-dilaton theory were analyzed in \cite{self-tuning}, and the addition  of the axion does not change the pictures qualitatively (as long as its bulk and brane kinetic terms are healthy). However, new and qualitatively different constraints may arise from the mixing of the brane-localized Higgs with the bulk KK modes, which is a new feature in this model with respect to \cite{self-tuning}.

Since the Higgs on the brane does not affect tensor fluctuations at linear order,  we restrict the analysis to  scalar perturbations. The bulk perturbations are also unaffected by the Higgs. However  the matching conditions are modified.

To simplify the discussion, we make one more simplification: we neglect the  bulk axion and we consider a model of Einstein-dilaton 5d gravity (with generic  bulk and brane-induced terms up to two derivatives) coupled to the brane-Higgs field. In other words, we set $a=0$  everywhere in the action (\ref{A2}), (\ref{A4})  and (\ref{A5}).  This does not change the picture qualitatively.  In the presence of a (backreacting) axion, one would first need to diagonalize the bulk fluctuations before  performing  the analysis, and add another set of KK modes for the corresponding new bulk field.  However, as long as the axion has healthy brane and bulk kinetic terms, this will not change the qualitative picture.

This appendix extends to the present model the techniques described in detail in Section 5 of \cite{self-tuning}, and we refer the reader to that work for more details.

\subsection{Linear Perturbations}

The relevant bulk perturbations in the scalar sector (with a convenient gauge fixing, see \cite{self-tuning} for details)  are:
\be \label{per1}
ds^2  = a(r)^2\left\{(1 +2 \phi) dr^2 + [(1+ 2\psi)\eta_{\mu\nu} + 2\de_\mu\de_\nu E] dx^\mu dx^\nu \right\}, \quad \f = \bar{\f}(r) \, ,
\ee
where $\phi, E, \psi$ are small  perturbations which depend on $(r,x^\mu)$.  We use a gauge where the dilaton is unperturbed in the bulk.

On the brane, we have two more perturbations: the fluctuation in the brane position,
\be \label{per2}
r_{brane}(x_\mu) = r_0 + \rho(x_\mu) \, ,
\ee
and the perturbation in the Higgs field, which we parametrize as\footnote{For simplicity we take a $U(1)$ Higgs. For the SM Higgs we  have 3 goldstones  instead of one}
\be \label{per3}
H =( v_0 + h(x) )e^{i\theta(x)} \, ,
\ee
where  $v_0 \geq 0$ and $h(x)$ is a  real scalar.

We follow closely the discussion in \cite{self-tuning}, section 3 and 5, and appendix D. This leads to a quadratic action for the perturbation, which is a  modified version of equation (5.9) of \cite{self-tuning}. We define the doublet:
\be
\Psi =  \left(\begin{array}{c} \psi_{UV} \\ \psi_{IR}\end{array}\right) \, ,
\ee
where $\psi_{UV}(r)$ is in principle defined for $r>r_0$ and  $\psi_{IR}(r)$ for $r<r_0$, but we can extend them over the whole range of $r$ with no consequence.

The bulk+brane action at quadratic order reads,
\bea \label{ac1}
S_{5} = &&-{M_p^3\over 2}\int d^4x \Bigg[\int dr\, \left[\de_r \Psi^\dagger {\cal B}(r) \de_r \Psi + \de_\mu \Psi^\dagger {\cal B}(r) \de^\mu \Psi \right]  \nonumber \\  &&
\nonumber \\ &&
 + \Psi^\dagger (r_0) \,\Sigma \hat{\Gamma}_1\, \Psi (r_0) \,- \,   \de_\mu\Psi^\dagger(r_0)\,  \Sigma  \hat{\Gamma}_2 \, \de^\mu \Psi(r_0) \nonumber  \\  &&
\nonumber \\ &&
 + T_H a_0^2 \left[ \de_\mu h \de^\mu h +a_0^2  m_H^2 h^2 + v_0^2 \de_\mu
\theta \de^\mu \theta \right] \nonumber \\ &&
 + a_0^4 {\de^2 \hat{W}_B \over \de \f \de H}\Big|_{\f_0, H_0} \hat{\chi}(r_0) h + 6 a_0^2 U_H(\f_0)  v_0  \,\de^\mu \hat{\psi}(r_0) \de_\mu h  \Bigg] \, .
\eea
The factors of $a_0 \equiv a(r_0)$ come from the induced metric on the brane, $\gamma_{\mu\nu} = a_0^2 \eta_{\mu\nu}$.
 The first two lines in equation (\ref{ac1})  have the same form as equation  (5.9) in \cite{self-tuning} 
The $2\times 2$ matrices ${\cal B}(r)$ and $\Sigma$ are:
 \be\label{ac2}
{\cal B}(r) = \left( \begin{array}{cc} e^{2B_{UV}}\theta(r_0-r) & 0 \\0 &e^{2B_{IR}}\theta(r-r_0)\end{array}\right) , \qquad \Sigma \equiv \left( \begin{array}{cc} -e^{2B_{UV}(r_0)} & 0 \\0 & e^{2B_{IR}(r_0)}\end{array}\right) ,
\ee
where on each side $e^{2B} = a^3(r)z^2(r)  $ and $z = a\bar{\f}'/a'$.
The matrices   $\hat{\Gamma}_1$ and $\hat{\Gamma}_2$ are given explicitly in (D.79) in \cite{self-tuning},  where  now   $W_B$ and $U_B$ have to be  replaced everywhere by   $\hat{W}_B$ and $\hat{U}_B$ from equations (\ref{FE8}-\ref{FE8-2}).

The last line in the quadratic action (\ref{ac1}) contains the mixing between the Higgs fluctuation  and the bulk fields. Notice that the $U(1)$  Goldstone mode is decoupled from the rest.  Notice also that the mixing is very simply written in terms of  $\hat{\psi}(r_0)$ and $\hat{\chi}(r_0)$,   the gauge-invariant fluctuations which couple to the dilaton charge and trace of the stress-tensor, respectively:
\be \label{ac2-ii}
\hat{\psi}(r_0) = \psi(r_0) + {a' (r_0) \over a(r_0)} \rho, \quad \hat{\chi}(r_0) =  \bar{\f}'(r_0) \rho.
\ee
 They also correspond to the ``heavy'' and ``light'' scalar fluctuations \cite{self-tuning}. Interestingly,  the first one has only a mass-mixing with $h$, the second only kinetic mixing.  In the gauge we are using, they are given by the following linear combinations  of $\psi_{UV}(r)$  and  $\psi_{IR}(r)$:
\be \label{gi3}
\hat{\chi}(r_0) = -{[\psi]\over [1/z]},  \qquad \hat{\psi} (r_0) =  {[z \psi] \over [z]},
\ee
where $[...]$ denotes the jump across the brane.

To summarize, when we compare (\ref{ac1}) with the quadratic action found in \cite{self-tuning}, the effect of the Higgs field is to change $W_B,U_B$ into $\hat{W}_B, \hat{U}_B$ and to generate the mixing terms in the last line. Notice that the mixing is  proportional to  the Higgs vev $v_0$: this is manifest in the second term on the last line in (\ref{ac1}), and the first term is explicitly:
\be \label{d2W}
{\de^2 \hat{W}_B  \over \de \f \de H} = 2 X_H'(\f) v_0 + 4  S_H'(\f) v_0^3.
\ee
Therefore, the mixing vanishes in the trivial Higgs vacuum $v_0=0$.

\subsection{KK expansion}

In order to understand the effect of the  mixing, we expand the action on
 ``Kaluza-Klein'' modes with  eigenvalue $m^2$ for the radial Hamiltonian. These modes have  form:
\be\label{ac3}
\Psi(r, x^\mu) = \Psi(r) \phi(x) ,
\ee
where the radial wave-function $\Psi$ satisfies (for $r\neq r_0$):
\be \label{ac4}
-{\cal B}^{-1} {d\over dr}\left({\cal B}(r) {d\Psi(r) \over dr}\right) = m^2 \Psi(r) ,
\ee
plus the boundary condition:
\be
\Psi'(r_0) = \Big(\hat{\Gamma}_1 + \hat{\Gamma}_2\, m^2\Big) \Psi(r_0).  \label{sl-sec2}
\ee
This radial  problem plus boundary conditions define a self-adjoint radial Hamiltonian  with orthogonal eigenstates.  If we expand a generic normalizable  function  $\Psi(r,x^\mu)$ in a complete basis of radial eigenstates, insert in (\ref{ac1}), and integrate over the bulk,   we obtain at quadratic order an effective 4d action which is a sum over decoupled 4d  KK modes, except for the last line in (\ref{ac1}), where each mode separately mixes with the brane Higgs field:
\be \label{kk}
S = M^3_p  \sum_n S^{(n)} \, ,
\ee
\bea  \label{kk2}
S^{(n)} = && -{1\over 2} \int d^4x \, {\cal N}_n \left[\de^\mu \phi_n \de_\mu \phi_n + m^2_n \phi_n^2\right]
+ a_0^2 T_H \left[\de_\mu h \de^\mu h + a_0^2 m_H^2 h^2 \right] \nonumber \\ &&
   +  a_0^2 \lambda_n  \phi_n  h + a_0 q_n \,\de^\mu \phi_n \de_\mu h
\eea
with
\be\label{ac6}
{\cal N}_n = \int_{r<r_0} dr\,  e^{2B_{UV}} \psi_{UV,n}^2 + \int_{r>r_0} dr\,  e^{2B_{IR}} \psi_{IR,n}^2 -  \Psi^\dagger_n (r_0) \,\Sigma \hat{\Gamma}_2\, \Psi_n (r_0) \, ,
\ee
and
\be \label{lambda-q}
\lambda_n  \equiv a_0^2 \left( 2 X_H'(\f_0) v_0 + 4  S_H'(\f_0) v_0^3\right) \hat{\chi}_n(r_0), \quad q_n \equiv  6 a_0 U_H(\f_0) v_0 \hat{\psi}_n(r_0) \, ,
\ee
where $\Psi_n(r)$ are the radial (doublet) eigenfunctions corresponding to the $n$-th eigenvalue   $m_n^2$. We see that the mixing terms are proportional to the values of the radial eigenfunctions taken on the brane.

We have written equation (\ref{kk})  for the case of  a discrete KK spectrum. For a continuous spectrum, the mass $m$ becomes a continuous variable and the sums are replaced by integrals. We continue to write the symbol of ``sum'' but it is understood that this may represent both cases.

The action (\ref{kk}) describes an infinite tower of  four-dimensional scalar modes with masses $m_n$, plus the Higgs field. All of the modes do not mix with each other but only with the Higgs. To check whether there are ghosts and/or tachyons we have to separately analyse the kinetic mixing and then the mass mixing.

\subsubsection{Ghosts}

To unmix the kinetic terms, it is sufficient to shift each mode by an appropriate multiple of $h$.  Define:
\be \label{gh0}
\phi_n = \tilde\phi_n - {a_0 q_n \over 2 {\cal N}_n} h \, .
\ee
This redefinition diagonalizes the kinetic action,
\bea \label{gh1}
S_{kin} = && -{M^3_p\over 2}\Bigg[ \sum_n {\cal N}_n \int d^4x \, \left(\de^\mu \tilde{\phi}_n \de_\mu \tilde{\phi}_n \right)  + \nonumber \\
 && + a_0^2 \left(T_H - \sum_n {q_n^2 \over 2 {\cal N}_n}\right)\int d^4x \, \de_\mu h \de^\mu h \Bigg] \, .
\eea
From this expression  we see that the mixing only affects the kinetic term of the Higgs, and leaves unchanged the KK kinetic terms.  Absence of ghosts requires that:
\begin{enumerate}
\item All the ${\cal N}_n >0$. If this is the case, none of the KK modes are ghosts.
A sufficient condition for this was obtained in \cite{self-tuning} (equation (5.19)). Here,  it has to be satisfied with the effective superpotentials  $\hat{W}_B$ and $\hat{U}_B$, which contain the Higgs vev $v_0$ and the new functions $X_H$ and $S_H$.
\item On top of that, the Higgs must not be a ghost, i.e. we need:
\be \label{gh2}
T_H > \Delta Q \, , \qquad  \Delta Q \equiv \sum_n {q_n^2 \over 2 {\cal N}_n} \, ,
\ee
where $q_n$ is defined in (\ref{lambda-q}) and ${\cal N}_n$  in (\ref{ac6}).
Under the assumption that point 1 holds, i.e. if no KK mode is a ghost, then $\Delta Q$  is the sum (or integral) of non-negative terms. In  section \ref{simple}, we show that the sum is finite and we will evaluate it in simple cases.
\end{enumerate}

\subsubsection{Tachyons}
Let us assume that there are no ghosts, i.e. all kinetic terms are positive definite. Then, the absence of tachyons is equivalent to the positivity of the mass eigenvalues of the KK system. We can  again disentangle the Higgs from the KK modes by redefining:
\be \label{tach1}
\phi_n = \tilde{\tilde\phi}_n- {a_0^2 \lambda_n \over 2 {\cal N}_n m_n^2 } h \, .
\ee
Notice that this new redefinition does not remove the kinetic mixing, but this doesn't matter for the discussion of the sign of the mass eigenvalues, as long as the kinetic matrix is positive definite (i.e.~there are no ghosts). If this is the case,  we only need to look at the eigenvalues of the mass terms which after the redefinition (\ref{tach1}) take the form:
\be \label{tach2}
S_{mass} =  -{M^3_p\over 2} \int d^4x \, \Bigg[ \sum_n {\cal N}_n m_n^2  \tilde{\tilde\phi}_n^2  + a_0^4 \left(T_H m_H^2 - \sum_n {\lambda_n^2 \over 2 {\cal N}_n  m_n^2}\right) h^2 \Bigg] \, .
\ee
Absence of tachyons requires:
\begin{enumerate}
\item
All KK masses-squared have to be positive. A sufficient conditions for this to happen was derived in \cite{self-tuning} (equation (5.25) there)  and the same holds here except that one has to replace $W_B$ with $\hat{W}_B$.
\item The Higgs must not be a tachyon, i.e.
\be \label{tach3}
m_H^2 >  \Delta M^2 \, , \qquad   \Delta M^2 \equiv {1\over T_H}\sum_n {\lambda_n^2 \over 2 {\cal N}_n  m_n^2} \, ,
\ee
where $\lambda_n$ is defined in (\ref{lambda-q}) and ${\cal N}_n$  in (\ref{ac6}). We shall estimate $\Delta M^2$ in the next section.
\end{enumerate}

\subsection{Simple models estimates} \label{simple}

Having obtained a general expression for the effective Higgs kinetic and mass terms which include the KK contribution,  we now proceed to assess the positivity of these terms. We do that in two simple toy-models which roughly mimic the situation of two general classes of theories: those in which the bulk spectrum is discrete, and those in which it is continuous.

\subsubsection{Discrete models}
These arise if the bulk potential  $V(\f)$ is such that as $\f \to +\infty$
\be \label{est1}
V \sim \exp b\f \, , \quad b > \sqrt{2\over 3} \quad \text{or} \quad V \sim \f^P \exp \sqrt{2\over 3}\f \, ,  \quad P>0 \, .
\ee
In this case, the KK spectrum is gapped and discrete, and the KK masses $m_n$ behave asymptotically as:
\be \label{est2}
m_n^2 \simeq \left\{ \begin{array}{ll}k^2 n^{2P} & b =\sqrt{2\over 3}\; \text{and}\; 0<P<1 \, , \\
& \\
k^2 n^{2} & b =\sqrt{2\over 3}\; \text{and}\; P \geq 1 \quad \text{or} \; b > \sqrt{2\over 3} \, , \end{array} \right.
\ee
where $k$ is the IR scale of the holographic model.

Except for the subclass with $ b =\sqrt{2/3}$ and $0<P<1$, all other cases above lead to the same KK spectrum asymptotics as a one-dimensional  compactification on a circle of radius $R = 1/k$. Therefore, in order to make estimates in these models, we shall use as a proxy the model constituted by a brane localized in a flat 5-dimensional bulk,  compactified  on a circle of radius $R$, and with an induced Einstein-Hilbert term on the brane characterized by a crossover scale $r_c$.
This model was discussed in detail in \cite{Power}. The crossover scale is given by   $r_c = M_4^2/M_p^3$, which in the full holographic model is (roughly) given by $\hat{U}(\f_0, H_0)$ \cite{self-tuning}.
This is parametrically similar to the DGP setup in bulk AdS space, with $R\leftrightarrow \ell$ as analyzed in \cite{ktt2}.

Before we continue, we  pause to assess whether this simple circle compactification can capture the qualitative features of the full holographic model. The main difference between the full holographic  setup and the  toy model is the presence, in the latter, of a normalizable zero mode for bulk fields (including the graviton), due to which the model matches GR on very large scales. In the holographic model this mode is not part of the spectrum, since the bulk volume is infinite on the UV side. However this is irrelevant for the effect we are studying, i.e.~the mixing of the Higgs field with the entire tower of  of KK modes. Another difference is the fact that the 5th dimension in the toy model is flat whereas in the holographic model it is warped. However, this does not affect the features of the KK states, except for the fact that in the toy model the IR scale $k$ is simply the inverse radius, as shown in \cite{ktt2}.

In \cite{Power} analytical expressions were obtained for  the masses $m_n$, wave-function normalizations ${\cal N}_n$, and values of the wave-function at the brane position (which we set at $r=0$)  $\psi_n(0)$ for  the tower of KK modes. We use  wave-functions which are normalized to unity in the bulk, i.e.~such that (in the real model) the first two terms in (\ref{ac6}) add up to one. Then we have:
\be \label{est3}
m_n^2 = {n^2 \over R^2}, \qquad {\cal N}_n = 1 + r_c |\psi_n(0)|^2,  \qquad \psi_n (0) = {1\over \sqrt{2\pi R}}{1 \over \sqrt{1 + {r_c^2 m_n^2 \over4}  }}.
\ee
With these expressions we can easily estimate the sums in equations (\ref{gh2}) and (\ref{tach3}). The result is controlled by  the value of the dimensionless quantity $r_c/R$. Since the fifth dimension is flat, we can choose the constant scale factor to be $a=1$. Therefore we do not have factors of $a_0$ around in our equations.

\paragraph{Higgs kinetic term correction.}
First, notice that the normalization factor ${\cal N}_n$ is given explicitly by:
\be \label{est4}
 {\cal N}_n  = 1 + {{r_c / R} \over 1 + {r_c^2 n^2 \over4 R^2}} \, .
\ee
No matter the value of $r_c/R$, the above expression is always near unity, therefore  we  make the approximation:
\be \label{est5}
{\cal N}_n \approx 1.
\ee

The correction $\Delta Q$ to the Higgs kinetic term in  (\ref{gh2}) is then given by:
\be \label{est6}
\Delta Q = (6 U_H v_0)^2  {1\over 2\pi R} \sum_{n=1}^{\infty} {1\over  1 + {r_c^2 n^2 \over4 R^2}} \, .
\ee
We can distinguish two cases:
\begin{enumerate}
\item $r_c / R \gg 1$. \\
This is the  case in which there is no five-dimensional regime for gravitational propagation \cite{Power,self-tuning}.  We can drop the ``$1$'' in the denominator of (\ref{est6}) for any $n>0$, and we can estimate the sum as
\be \label{est7}
 \sum_n  {1\over  1 + {r_c^2 n^2 \over4 R^2}} \approx  {4R^2 \over r_c^2} \sum_{n=1}^{\infty} {1\over n^2} \, .
\ee
The series converges to  a finite constant and we arrive at (dropping numerical factors of order one):
\be \label{est8}
\Delta Q \sim (U_H v_0)^2 {R \over r_c^2} \, , \qquad {r_c \over R} \gg 1 \, .
\ee
\item  $r_c /R \ll 1$. \\
In this case gravity has an intermediate five-dimensional (DGP-like) regime, over distances $r_c < r < R$.
 In this regime we introduce the quantity $N = [R/r_c] \gg 1$ (where $[...]$ denotes the integer part) and we can approximate the sum as follows:
\be \label{est9}
\sum_n  {1\over  1 + {r_c^2 n^2 \over4 R^2}} \approx \sum_{n=1}^{N} 1 + {4R^2\over r_c^2}\sum_{n =N}^{+\infty}{1 \over n^2} \approx \textrm{const.} {R\over r_c} \, .
\ee
In the last approximation, we have used the fact that the first term evaluates to $N \approx R/r_c$, and that  the sum in the second term is, for $N\gg 1$, the remainder of the sum of inverse squared integers, which behaves as $1/N$ asymptotically.  Dropping again factors of order one, we arrive at:
\be \label{est10}
\Delta Q \sim (U_H v_0)^2 {1 \over r_c} \, , \qquad {r_c \over R} \ll 1 \, .
\ee
\end{enumerate}
We can understand the extra suppression factor $R/r_c$ in case 1 compared to case 2 from the fact that, in the  regime $R/r_c \gg 1$ there is a large number of KK modes which contribute and are unsuppressed on the brane: those with masses above the compactification scale $1/R$ but below the cross-over scale  $1/r_c$.

\paragraph{Higgs mass term correction.}

The Higgs has a quadratic non-derivative coupling only to heavy bulk modes. We can incorporate this fact in our toy model by supposing that the bulk field which gives rise to the KK spectrum has a bulk mass $M_0$, and the spectrum is then:
\be \label{est11}
m_n^2 = M_0^2 + {n^2 \over R^2}.
\ee
This does not affect the KK normalization nor the value of the wavefunction on the brane, which are still as in (\ref{est3}).  The correction $\Delta M^2$ to the Higgs mass is then obtained from equation (\ref{tach3}):
\be \label{est12}
\Delta M^2 \simeq  { F_0^2 \over T_H} {1\over 2\pi R}  \sum_{n=1}^{\infty}{1\over \left( 1 + {r_c^2 n^2 \over4 R^2}\right) \left(M_0^2 + {n^2\over R^2}\right)},
\ee
where we have approximated again ${\cal N}_n \simeq 1$ and we have defined:
\be \label{est13}
F_0 = \left( 2 X_H'(\f_0) v_0 + 4  S_H'(\f_0) v_0^3\right).
\ee

There are two dimensionless parameters which control the result:
\be \label{est14}
{R\over r_c}, \qquad  R M_0.
\ee
\begin{enumerate}
\item ${r_c / R} \gg 1 $ and $M_0 \ll 1/R $.\\
In this case  the sum can be approximated by
\be \label{est15}
{R^4 \over r_c^2} \sum_{n=1}^{\infty}{1\over n^4} = \textrm{const.} {R^4 \over r_c^2} \, ,
\ee
and we obtain:
\be \label{est16}
\Delta M^2 \approx { F_0^2 \over T_H} {R^3 \over r_c^2} \, .
\ee
\item  ${r_c / R} \gg 1 $ and $M_0 \gg 1/R $. \\
In this case we can neglect the ``$1$''  only in the first factor the denominator of the general term of the sum in (\ref{est12}).  Therefore:
\be \label{est17}
\sum_{n=1}^{\infty}{1\over \left( 1 + {r_c^2 n^2 \over4 R^2}\right) \left(M_0^2 + {n^2\over R^2}\right)} \approx {R^2 \over r_c^2} \sum_{n=1}^{\infty}{1\over n^2\left(M_0^2 + {n^2\over R^2}\right)} \, .
\ee
Introducing $N = [M_0 R] \gg 1$ we can approximate the last expression as:
\be \label{est18}
  {R^2 \over r_c^2}\left({1\over M_0^2} \sum_{n=1}^{N}{1\over n^2} +  R^2 \sum_{n=N}^{\infty}{1\over n^4}\right) \approx \textrm{const.} {R^2 \over r_c^2 M_0^2} \, .
\ee
In the last approximation we have used the fact that the remainder of the series of $1/n^4$ from $N$ scales as $N^{-3}$ so the second term in equation (\ref{est18}) gives a  contribution $\sim 1/(r_c^2 M_0^4)$, which is  negligible with respect to the first term, $~ R^2/(r_c^2 M_0^2)$. From (\ref{est18}) we arrive at:
\be \label{est19}
\Delta M^2 \approx { F_0^2 \over T_H} {R \over r_c^2 M_0^2}.
\ee
\item ${r_c / R} \ll 1 $ and $M_0 \ll 1/R $.\\
In this case the roles of the first and second factor in the denominator of the summand in (\ref{est12}) are interchanged, and proceeding as in case 3 we obtain:
\be
\Delta M^2 \approx { F_0^2 \over T_H} R.
\ee
\item  ${r_c / R} \ll 1 $ and $M_0 \gg 1/R $ . \\
Define two large integers $N_1 = R/r_c$ and $N_2 = M_0 R$.
Proceeding as in the cases above, the dominant part of the series is the sum of the first $N_1$ or $N_2$  terms, whichever is smaller. We then find that the sum evaluates to:
\be
\sum_{n=1}^{\infty}{1\over \left( 1 + {r_c^2 n^2 \over4 R^2}\right) \left(M_0^2 + {n^2\over R^2}\right)} \approx \left\{\begin{array}{ll} {R\over M_0} & \quad M_0 < {1\over r_c} \, , \\
&\\
{R\over M_0^2 r_c} &  \quad M_0 > {1\over r_c} \, , \end{array} \right.
\ee
which leads to:
\be
\Delta M^2 \approx \left\{\begin{array}{ll}  { F_0^2 \over T_H} {1\over M_0} & \quad M_0 < {1\over r_c} \, , \\
&\\
 { F_0^2 \over T_H} {1\over M_0^2 r_c} &  \quad M_0 > {1\over r_c} \, . \end{array} \right. 
\ee
\end{enumerate}

\subsubsection{Continuous models (abridged)}
Bulk potentials with a softer behavior at infinity than (\ref{est1}) lead to geometries  with a continuous KK spectrum. A toy model which mimics this case is the single-brane RS model with induced gravity on the brane, where the bulk is a cut-off $AdS_5$ space-time. The two relevant parameters now are the bulk curvature scale $k$, and the crossover scale $r_c$, still given by $M_4^2/M_p^3$.
As in the previous subsection, the main difference between the toy model and the complete holographic setup is that the former contains a zero-mode which mediates four-dimensional interactions at large distances, while in the latter this mode is projected out. However, this is not important since we are interested in the effect of the (continuous) tower of massive modes.

This model was analyzed in \cite{ktt2} where one can find analytic expressions (in terms of Bessel functions) for the KK wave-functions evaluated on the brane. We  need the large-mass and small-mass asymptotic behavior:
\be \label{est20}
\psi_m(r_0) = \left\{\begin{array}{ll} {\sqrt{m/k} \over 1+ k r_c } & \quad m/k \ll 1 \, , \\
& \\
{1\over {\sqrt{1 + r_c^2 m^2/4}}} & \quad m/k \gg 1 \, , \end{array} \right.
\ee
where now $m$ is a continuous mass parameter.  The wave-functions are plane-wave normalized in the IR, and we still have ${\cal N}_{m} \simeq 1$.  We can set again $a_0=1$. With this choice,  all mass scales appearing in the equations  are to be understood as measured by brane observers.

The sums in the previous section turn into integrals over $m$. It turns out that,  when evaluated, these integrals lead exactly to the same estimates as in the previous subsection, with the substitution $R \to 1/k$. The reason is that   the suppression of the wave function on the brane for large mass in  (\ref{est20})  is the same as found for the flat  compactification  on a circle, equation  (\ref{est3}).  The fact that most of the effect comes from summing over a large number of modes makes the difference between sums and integrals negligible.

\subsection{General remarks}
Based on the  results we have found in  this appendix that, whether or not there are instabilities clearly depends on the details of the model parameters and superpotentials. However we can draw a few general conclusions.
\begin{itemize}
\item The KK corrections to the Higgs kinetic term and mass are always finite when we sum over the whole KK  tower.
\item The corrections are proportional to the Higgs vev and the various ``superpotential'' coupling the Higgs to the dilaton and the curvature.
\item  Depending on the model features, the effects scale with different combinations of $R$ and $r_c$ (kinetic term), or $R, r_c$ and $M_0$ (mass term). To have definite results, one has to look at a concrete model.  However, notice that in the regime which is likely the most pheno-friendly ($r_c \gg R$ or $r_c \gg 1/k$) the corrections are always suppressed by powers of the  of small numbers $R/r_c$ or $1/kr_c$. In this regime,  we can obtain more insight by rephrasing the estimates (\ref{est8}) and (\ref{est16}-\ref{est19}) in terms of the four-dimensional Planck scale $M_4$, using the relation:
\be
M_4^2 = M_p^3 r_c \, .
\ee
Recall also that the Higgs and its vev $v_0$ we are using are in units of  $M_p$ (the bulk Planck scale), and that $T_H = M_p^{-1}$. Using these relations, we can rewrite   (\ref{est8}) as
\be \label{est21}
{\Delta Q \over T_H} \approx U_H^2 v_{phys}^2 {R \over r_c} \left({1\over r_c M_4}\right)^{2/3},
\ee
where $v_{phys} \equiv v_0 M_p $ is the physical Higgs vev with dimension of energy.
The quantity (\ref{est21})  must be smaller than unity if we want the Higgs boson not to turn into a ghost. This can be easily achieved naturally for reasonable values of $U_H$, due to the suppression factor $R/r_c$ and especially to the huge suppression by $r_c M_4$, which is roughly  the ratio between the 4d Planck scale and a macroscopic (astrophysical or cosmological) scale.

Similarly, we can write the ratio between the Higgs mass and  the mass correction in equations (\ref{est16}) and (\ref{est19}) (limiting ourselves to the case $R\ll r_c$)  as:
\be \label{est22}
{\Delta M^2 \over m_H^2} \approx (X_0')^2 R^2 {v_{phys}^2 \over m_H^2} {R \over r_c}  \left({1\over r_c M_4}\right)^{2/3}, \qquad M_0 \ll 1/R \, ,
\ee
or
\be\label{est23}
{\Delta M^2 \over m_H^2} \approx {(X_0')^2 \over M_0^2} {v_{phys}^2 \over m_H^2} {R \over r_c}  \left({1\over r_c M_4}\right)^{2/3}, \qquad M_0 \gg 1/R \, .
\ee
In both cases the large  suppression factors $R/r_c$ and especially $(r_c M_4)^{2/3}$ make it very plausible that one does not need any fine tuning to make $\Delta M^2/m_H^2$ small, thereby preventing the Higgs from turning tachyonic.  In  writing equations  (\ref{est22}-\ref{est23}) we have neglected the term proportional to $S_0'$ in $F_0$ (see equation) because it is suppressed by even more powers of the four-dimensional Planck scale.

\item Given that the results are the same for the softest of  non-confining potential (i.e.~a cosmological constant)  as for ``steep'' confining  potentials, it is reasonable to assume that the same results will hold in the intermediate classes of general non-confining potentials
(i.e.~$0< b<\sqrt{2/3}$) as well as  ``soft'' confining potentials (i.e.~$b=\sqrt{2/3}$, $0<P<1$).
\end{itemize}

\end{appendix}


\end{document}